\documentclass[prx,showkeys,showpacs,footnoteinbib,twocolumn,
unsortedaddress]{revtex4}
\usepackage{amsmath,amssymb,epic,eepic}
\usepackage[binary-units]{siunitx}
\usepackage[utf8]{inputenc}
\usepackage{graphicx}
\usepackage{hyperref}
\usepackage[capitalise]{cleveref}
\usepackage{todonotes}
\usepackage{bbm}
\usepackage{tikz}
\usepackage{circuitikz}
\usepackage{braket}
\usepackage[caption=false]{subfig}
\usepackage{fontawesome}
\usetikzlibrary{%
        shapes.geometric,calc,intersections,%
        decorations.markings,decorations.text,decorations,%
        patterns,decorations.pathmorphing,fpu,%
		decorations.pathreplacing,%
        positioning,automata,shapes.multipart,circuits,%
        circuits.ee, circuits.ee.IEC, shapes.gates.ee,%
		arrows.meta,arrows,external
}

\newcommand{\mytodo}[1]%
{{\todo[inline,backgroundcolor=blue!10!white]{#1}
}}
\newcommand{\me}{\mathrm{e}}
\newcommand{\mi}{\mathrm{i}}
\newcommand{\md}{\mathrm{d}}
\DeclareMathOperator{\trace}{tr}
\DeclareMathOperator{\heaviside}{H}

\DeclareMathOperator{\sinc}{sinc}

\begin{document}

\title{Processing quantum signals carried by electrical currents}

\author{B. Roussel$^{1,2}$}
\author{C. Cabart$^1$}
\author{G. F\`eve$^3$}
\author{P. Degiovanni$^1$}

\affiliation{(1) Univ Lyon, ENS de Lyon, Université Claude Bernard
Lyon 1, CNRS,
Laboratoire de Physique, F-69342 Lyon, France}

\affiliation{(2) European Space Agency - Advanced Concepts Team, ESTEC,
Keplerlaan 1, 2201 AZ Noordwijk, The Netherlands.}

\affiliation{(3) Laboratoire
de Physique de l'Ecole Normale Sup\'erieure, ENS, Universit\'e
PSL, CNRS, Sorbonne Universit\'e, Universit\'e Paris-Diderot, Sorbonne Paris
Cit\'e, Paris, France}

\begin{abstract}
Recent developments in the coherent manipulation of electrons in
ballistic conductors include the generation of time-periodic
electrical currents involving one to few electronic excitations per period. 
However, using individual electrons as carrier of quantum
information for flying qubit computation or quantum metrology 
applications calls for a general method to unravel the
single-particle excitations embedded in a quantum electrical current
and how quantum information is encoded within it. Here, we
propose a general signal processing algorithm to extract the
elementary single-particle states, called electronic atoms of signal,
present in any periodic
quantum electrical current. These excitations
and their mutual quantum coherence describe the excess single-electron
coherence in the same way musical notes and score describe 
a sound signal emitted by a music instrument. This method, which is the
first step towards the development of signal processing of quantum
electrical currents is illustrated by assessing the quality of
experimentally relevant single
electron sources. The example of randomized quantum electrical
currents obtained by regularly clocked but randomly injected unit charge 
Lorentzian voltage pulses enables us to discuss how
interplay of
the coherence of the applied voltage and of the Pauli principle
alter the quantum coherence between the emitted single particle
excitations. 
	\\[5mm]
Version of \today
\end{abstract}



\maketitle

\section{Introduction}

These recent years have seen spectacular breakthrough in the
manipulation of quantum electric circuits.
On-demand single-electron sources
in quantum Hall edge channels
\cite{Feve-2007-1,Leicht-2011-1,Dubois-2013-1,Kataoka-2016-1},
2D electron gases using electron pumps~\cite{Fletcher-2013-1} or
surface acoustic
waves~\cite{Hermelin-2011-1} and in tunnel
junctions~\cite{Gabelli-2012-1}
enable us to engineer
time-dependent quantum electrical currents involving one to a few
elementary excitations per period. This emerging field, called
electron quantum optics, precisely aims at generating,
manipulating and characterizing such ``quantum beams of
electricity'' in metallic quantum conductors
\cite{Bocquillon-2014-1}. The latest advances
have given access to the single-particle wavefunctions carried by such
quantum electrical currents together with
their emission probabilities and
coherences
\cite{Bisognin-2019-2}
thereby demonstrating our ability to access electronic quantum states at an
unprecedented level. These achievements strongly suggest that this field
is now sufficiently mature for exploring its applications.

From a quantum technology point of view, quantum electrical currents
carry quantum information through their single-, two- and ultimately,
many-particle content. For example, single electrons delocalized on two
one-dimensional channels have been proposed as ``railroad flying
qubit''~\cite{Bertoni-2000-1,Ionicioiu-2001-1,
Bertoni-2007-1,Zibold-2007-1,Bertoni-2009-1} in which a 
qubit state is encoded in the quantum delocalization of an
electron on two copropagating 1D channels \cite{Yamamoto-2012-1}. This is a very promising line
of research towards the developpement of quantum spintronics
\cite{Bertrand-2016-1,Flentje-2017-1} and, in the longer term, of free
electron quantum computation
\cite{Beenakker-2004-2,Barnes-2000-1,Takada-2019-1}.

This information can be accessed through a 
hierarchy of electronic coherences
similar to the ones
introduced by Glauber \cite{Glauber-1963-1} for photons. These
coherences are the ``quantum signals''
carried by the quantum electrical current in a metallic conductor.
Because of
the parity super-selection rule 
\cite{Wick-1952-1,Friis-2016-1,
Johansson-2016-1},
the first
non-zero electronic quantum signal is the
single-electron coherence~\cite{Degio-2011-1,Haack-2012-2}
containing all information on
single-particle excitations within the system. The next one is the
second order electronic coherence~\cite{Moskalets-2014-1,Thibierge-2016-1} 
that describes
two-particle excitations within the beam.

Measuring these quantum signals requires quantum tomography protocols
for $n$-electron coherence. Such tomography protocols are all based
on the transformation of quantum signals into measurable quantities. For
exemple, 
Mach-Zehnder, Hong-Ou-Mandel (HOM)
and Franson electron interferometry
experiments 
realize 
``filtering'' or ``overlaps''
on electronic coherences~\cite{Roussel-2016-2}, thereby encoding
the results of these operations into experimentally
accessible quantities such as average current~\cite{Haack-2011-1} and current
correlations~\cite{Olkhovskaya-2008-1,Splettstoesser-2009-1}. Electronic HOM
interferometry~\cite{Bocquillon-2013-1} is at the core of
the recently demonstrated HOM single-electron
tomography~\cite{Bisognin-2019-2,Jullien-2014-1} whereas, for
higher-energy ($\si{\milli\electronvolt}$) electrons, a
time-dependent quantum point contact was used as a time-dependent energy
filter for reconstructing single-electron
coherence~\cite{Fletcher-2019-1}.

This however leaves open the question of decoding classical or quantum
information encoded within quantum electrical currents. This requires
finding appropriate
representations of electronic coherences. In the present context,
``appropriate'' means simple with respect to the reference
state which is a Fermi sea at a given chemical potentiel. We
therefore consider the excess 
single-particle coherence describing the single-particle
content in terms of electron and hole excitations with respect to the
reference Fermi sea. Ideally, we are looking for the simplest
possible description, requiring minimal data to encode this
description of the single-particle content.

\begin{figure*}
	\includegraphics{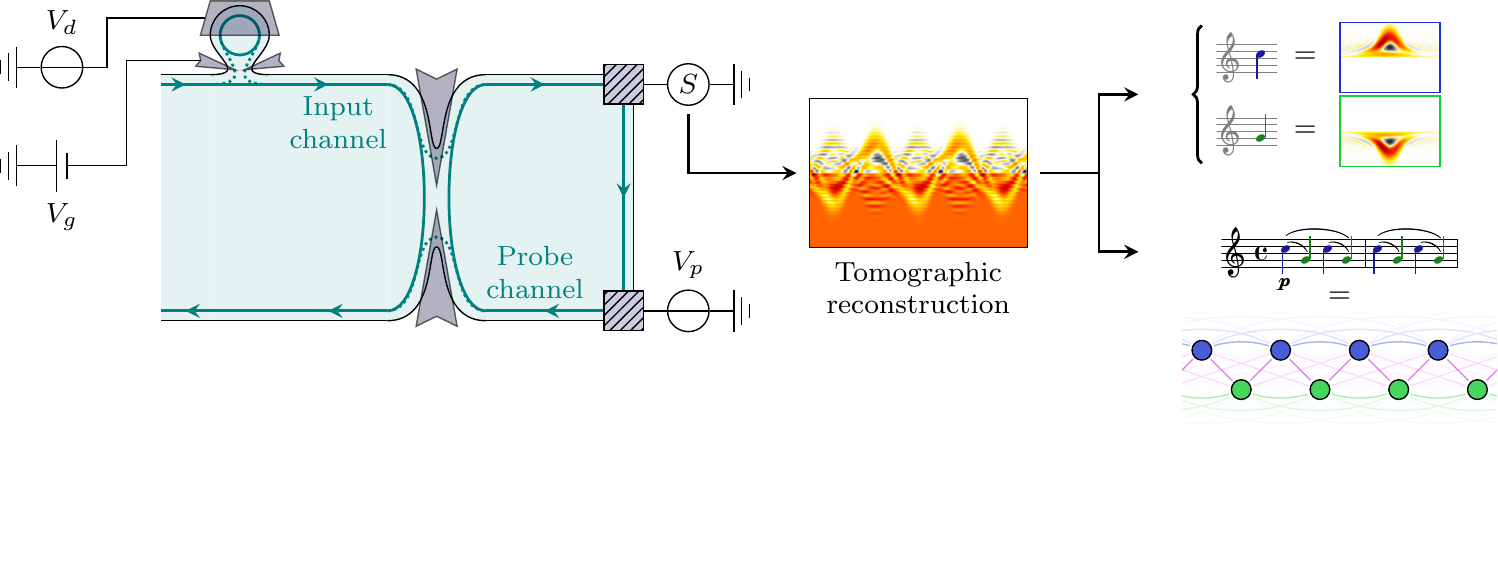}
\caption{\label{figure/visual-abstract} Schematic of the process for
extracting the single particle content from a quantum electrical
current. Left part: the Hong--Ou--Mandel interferometer uses
two-particle interferences to encode the overlap between the 
injected single electron coherences into the outgoing current noise
\cite{Roussel-2016-2}. Middle part: the single electron coherence is
reconstructed from current noise measurements 
\cite{Degio-2010-4,Jullien-2014-1,Bisognin-2019-1}. Right part: 
the result of the tomographic reconstruction, 
depicted here as the electronic Wigner distribution function \cite{Ferraro-2013-1}, is
processed by the algorithm described in the present paper to obtain a
description of single electron coherence 
in terms of electronic atoms of signal (counterparts of
musical notes) arranged according to a ``quantum coherence score''
(counterpart of the music score). 
	}
\end{figure*}

In this paper, we show in full generality
that such a description exists: any excess time-periodic single-electron coherence
admits a minimal description in terms of quasi-periodic single-electron
and single-hole excitations which are the time-domain counterparts of Bloch
waves in
solid state physics \cite{Bloch-1929-1}. This implies that only electron and hole Bloch wave emission
probabilities as well as electron/hole coherences between two different
Bloch waves are required to know the single electron coherence.
Considering the counterpart of Wannier
functions \cite{Wannier-1937-1}, which are localized wave-functions contrary to Bloch waves, 
the excess single-electron coherence can then be expressed in
terms of a set of mutually orthogonal single-particle states called
electronic atoms of signal~\cite{Roussel-2016-2} thereby
providing us with a discrete description of the electronic coherence.
We shall see that
electronic atoms of signals and the discrete representation of single-electron coherence 
can be viewed as the counterpart of music notes
and of a musical score as pictured on \cref{figure/visual-abstract}.
Therefore, the extraction of such a simple form of single-electron
coherence provides us with the appropriate toolbox to develop a full wave-packet based
approach on quantum transport envisioned in pioneering works
\cite{Martin-1992-1,Hassler-2008-1}. In a broader perspective, it
is a
crucial step in the development of
``quantum signal processing'' for quantum electrical currents that
extends the general paradigm of
signal processing~\cite{Moura-2009-1} to the quantum realm. It would
entitle us with
an enabling set of technologies and methods aiming at
encoding, transferring and retrieving quantum
information carried by these 
``quantum signals'', a crucial step for the
applications of electron-based quantum technologies.

Whenever interactions can be neglected, this description can be
used to describe the full many-body state of the electron fluid and
therefore to access many-particle quantities such as the electron/hole
entanglement entropies. This connexion can be made explicit using time
periodic single scattering theory and has been used to obtain the full
counting statistics of single particle excitations \cite{Yin-2019-1}.

The entanglement entropy inferres from this representation
of single-electron coherence can then be used
to assess the quality of experimentally relevant single electron sources
such as the mesoscopic capacitor. We are also able to obtain an explicit
description of the single-electron excitations emitted. Finally, in
order to
illustrate the possibility for modulating emission probabilities and
coherences, we apply our algorithm to 
the recently introduced
randomized trains of Lorentzian pulses \cite{Glattli-2018-1}, an
interesting example that enlightens the role of the Pauli exclusion principle
in electronic quantum signals.

This paper is structured as follows: in \cref{sec/questions}, we
introduce the problem of finding a simple representation of
single-electron coherence. Then, in \cref{sec/Floquet}, we present
our
algorithm for finding such a representation for any time-periodic excess
single-electron coherence. The relation of this representation to 
electron/hole entanglement is discussed in \cref{sec/many-body}. Finally,
we apply our method to the study of the mesoscopic capacitor to assess
its quality as a single-electron source, and to periodic and
randomized Leviton trains 
in \cref{sec:source-diagnostic}.

\section{Statement of the problem}
\label{sec/questions}

Let us now introduce and motivate the problem considered here by considering
simple trains of excitations used to model the emission by
experimentally demonstrated single to few
electron sources. These simple examples will enable us to write down a simple representation
of the excess single-electron coherence, a generalization of 
which will be shown to exist in \cref{sec/Floquet}.

\subsection{Electronic coherence}
\label{sec/questions/Ge}

The central concept of electron quantum optics are the electronic
coherences defined by analogy with Glauber's coherences of
photon quantum optics \cite{Glauber-1963-1}. They carry all the information on the
fermionic $n$-particle states propagating within the conductor. Here
we focus on single-electron coherence, which, at position $x$ along a single chiral
electronic channel, is
defined as \cite{Degio-2010-4,Degio-2011-1}
\begin{equation}
\mathcal{G}^{(e)}_{\rho,x}(t|t')=\trace(\psi(x,t)\,\rho\,
\psi^\dagger(x,t'))\,.
\end{equation}
where $\rho$ denotes the many-body reduced density operator for the
electron fluid and $\psi$ the fermionic field operator describing the
electrons.
When all the electronic sources 
are switched off, $\mathcal{G}^{(e)}_{x,\mathrm{off}}$
coincides with the equilibrium single-electron coherence characterized by
a chemical potential $\mu$ and an electronic temperature
$T_{\mathrm{el}}$. When sources are switched on, the excess
single-electron coherence defined by $\mathcal{G}^{(e)}_{x,\mathrm{on}}=
\mathcal{G}^{(e)}_{x,\mathrm{off}}+\Delta\mathcal{G}^{(e)}_x$ contain 
all the information on the single-particle excitations generated by the sources and drives
that are switched on. 

The single-electron coherence can be studied in the time domain as well
as in the frequency domain but is most conveniently visualized 
using a real-valued time/frequency representation called the Wigner
distribution function \cite{Ferraro-2013-1}
\begin{equation}
\label{eq/Wigner}
W_{\rho,x}^{(e)}(t,\omega)=\int_{\mathbb{R}}
v_F\mathcal{G}_{\rho,x}^{(e)}\left(
t+\frac{\tau}{2} \middle| t-\frac{\tau}{2}\right)\,\me^{\mi\omega\tau}\md\tau
\end{equation}

\subsection{Electron and hole trains}
\label{sec/questions/trains}

\begin{figure}
\centering
\includegraphics[width=8cm]{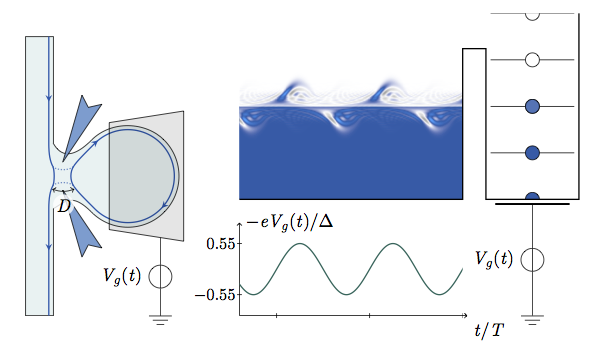}
\caption{\label{fig:capa-meso} Left panel: The mesoscopic capacitor is a ballistic
quantum conductor formed by connecting a quantum dot to a chiral edge
channel via a quantum point contact of transparency $D$. 
Right panel: Modelization as a driven quantum 
dot with level spacing $\Delta$ connected to an electronic reservoir.
The mesoscopic capacitor is driven by an a.c.\@ voltage drive
$V_g(t)$ applied to the top gate. Applying a d.c.\@ voltage bias to the top
gate shifts the energy levels of the dot. The mesoscopic capacitor
emits a stream of electron
and hole excitations whose Wigner distribution function
$W^{(e)}_S(t,\omega)$ is depicted as a density plot on the
right panel.
}
\end{figure}

An ideal periodic single-electron source is a periodically operated 
device that emits exactly one
single-electron excitation on top of the Fermi sea $\ket{F_{\mu=0}}$ during each period.
The corresponding many body state is an electron train of the form
\begin{equation}
|\Psi_{\text{SES}}\rangle = \prod_{l\in \mathbb{Z}}
\psi^\dagger[\varphi_{e,l}] |F_{\mu=0}\rangle
\end{equation}
where $\psi^\dagger[\varphi_{e,l}]$ creates a single particle excitation
in the electronic wavefunction $\varphi_{e,l}$. It differ from
$\varphi_{e,l=0}$ by translation by $lT$ in the time domain. Ideally, one
would like each of these electronic excitations to be perfectly
distinguishable from each other which means that $\varphi_{e,l}$ and
$\varphi_{e,l'}$ are orthogonal when $l\neq l'$. In this case, the
excess single-electron coherence is 
\begin{equation}
\label{eq:ideal-source:SES}
\Delta \mathcal{G}^{(e)}%
	(t|t')=\sum_{l=-\infty}^{+\infty}
\varphi_{e,l}(t)\,\varphi_{e,l}(t')^*
\end{equation}
For example, 
when driven by a square gate voltage $V_g(t)$ and for a suitable value of the
dot transparency $D=D_{\text{opt}}$ 
the mesoscopic capacitor depicted on \cref{fig:capa-meso}
ideally
generates one
electron excitation and one hole excitation per period
\cite{Feve-2007-1,Mahe-2008-1}:
\begin{equation}
	\label{eq/SES/MB-state}
|\Psi_{\text{MC-SES}}\rangle = \prod_{l=-\infty}^{+\infty}
\psi^\dagger[\varphi_{e,l}]\psi[\varphi_{h,l}]\,|F_{\mu=0}\rangle
\end{equation}
Here $\varphi_{e,l}$ and $\varphi_{h,l}$ are time translated by $lT$ from 
the emitted electron $\varphi_{e,0}$ and hole wavefunctions
$\varphi_{h,0}$ and are mutually orthogonal and normalized.
The excess single-electron coherence is then given by
\begin{equation}
\label{eq:ideal-source:LPA}
\Delta \mathcal{G}^{(e)}%
	=\sum_{l=-\infty}^{+\infty}
\varphi_{e,l}(t)\,\varphi_{e,l}(t')^*%
-\sum_{l=-\infty}^{+\infty}
\varphi_{h,l}(t)\,\varphi_{h,l}(t')^*
\end{equation}
where the hole contribution naturally comes with a minus sign. 

When
closing the dot, the time needed to emis an electronic (or hole)
excitation becomes larger than $T/2$. It was then argued 
\cite{Degio-2010-4} that the mesoscopic capacitor emits a quantum
superposition of no excitation and an elementary electron/hole pair
during each period. 
Such a state would be parametrized as
\begin{equation}
|\Psi_{\text{e/h}}(u,v)\rangle=\prod_{l=-\infty}^{+\infty}
\left(u+v\psi^\dagger[\varphi_{e,l}]\psi[\varphi_{h,l}]\right)|F_{\mu=0}\rangle
\end{equation}
where $|u|^2+|v|^2=1$. The resulting single-electron
coherence is then
\begin{subequations}
\label{eq:ideal-source:eh-pairs}
\begin{align}
\Delta\mathcal{G}^{(e)}(t|t')
&=
\sum_{l\in\mathbb{Z}}\big[ |v|^2
\varphi_{e,l}(t)\varphi_{e,l}(t')^*
-|v|^2
\varphi_{h,l}(t)\varphi_{h,l}(t')^*\big]
\label{eq:ideal-source:eh-pairs:1}
\\
	&+\sum_{l\in\mathbb{Z}}\big[u\,v^*
\varphi_{e,l}(t)\varphi_{h,l}(t')^*
+vu^*\varphi_{h,l}(t)\varphi_{e,l}^*(t')\big]
\label{eq:ideal-source:eh-pairs:2}
\end{align}
\end{subequations}
in which the r.h.s. of \cref{eq:ideal-source:eh-pairs:1} lives in the
quadrants of electron
and hole excitations (see \cref{fig/eh-quadrants})
whereas \cref{eq:ideal-source:eh-pairs:2} represents the electron/hole pair
coherence arising from $|\Psi_{\text{e/h}}(u,v)\rangle$ whenever $uv\neq
0$. \Cref{eq:ideal-source:LPA} is recovered for $(u,v)=(0,1)$ which
should therefore correspond to $D\simeq D_{\text{opt}}$ 
whereas for $(u,v)=(1,0)$ one recovers the Fermi sea, the result expected
when the dot is totally closed ($D=0$). The case where
$|u|^2=|v|^2\simeq 1/2$ could thus be viewed as 
the excess electronic coherence from an ideal source 
emitting a coherent superposition of ``nothing'' and of a single electron/hole pair
per period.
It corresponds to
maximal electron/hole entanglement \cite{Roussel-2016-2}. 
Note however that the r.h.s of \cref{eq:ideal-source:eh-pairs} does not involves
inter-period coherences (terms with $l\neq l'$).

\subsection{Electronic atoms of signal}
\label{sec/questions/electronic-atoms-signal}

\Cref{eq:ideal-source:SES,eq:ideal-source:LPA,eq:ideal-source:eh-pairs}
correspond to
ideal sources and have a simple expression in terms of a family of
single-electron wavefunctions called \emph{electronic atoms of
signal}~\cite{Roussel-2016-2}.
Electronic atoms of signal consists in a family of 
normalized mutually orthogonal single-electron wavefunctions
$\varphi_{a,l}$ which are translated by multiples of $T$~:
\begin{subequations}
\begin{align}
\varphi_{a,l}(t)&=\varphi_{a,0}(t-lT)\\
\langle \varphi_{a,l}|\varphi_{a',l'}\rangle
&=\delta_{l,l'}\delta_{a,a'}\,.
\end{align}
\end{subequations}
Although a decomposition of the form \eqref{eq:ideal-source:SES} is known
\cite{Moskalets-2015-1} for a $T$-periodic train of Lorentzian voltages
pulses of unit charge at zero temperature, 
realistic sources are, in general, not ideal.
Even the forms given by expressions~\labelcref{eq:ideal-source:LPA,eq:ideal-source:eh-pairs}
are too simple to
describe the excess single-electron coherence of all experimentally
realistic sources. First of all, even at
very low temperature, they
correspond to ideal
operating regimes which are only asymptotic with respect to the
experimental parameters as in the case of the mesoscopic capacitor at
$D\simeq D_{\text{opt}}$. Moreover, at non-zero temperature $T_{\mathrm{el}}$,
electron/hole pairs are
generated from thermal fluctuations and introduce an underlying thermal coherence
time $\hbar/k_BT_{\text{el}}$ a priori unrelated to the period $T$. It 
may lead to interperiod coherences not present in expression
\labelcref{eq:ideal-source:eh-pairs}. 
Last but not least, when electronic coherence
is measured at some distance from such a source, Coulomb interactions
alter the electronic coherence in a drastic way
\cite{Wahl-2013-1,Ferraro-2014-2,Marguerite-2016-1}, leading to 
extra-electron/hole pairs \cite{Cabart-2018-1}.

These remarks rise the question of
finding a way to express an arbitrary periodic 
single-electron coherence in terms of suitable electronic atoms of signals. We will
now present a systematic procedure for obtaining such an expression
together with the appropriate electronic atoms of signals from 
single-electron coherence. This procedure
can be applied to data obtained from
a numerical computation but also to experimental 
data obtained from an electronic
tomography protocol as recently demonstrated in
\cite{Bisognin-2019-2}. 

\section{Floquet--Bloch--Wannier analysis}
\label{sec/Floquet}

\subsection{Sketch of the method}
\label{sec:Floquet:method}

\Cref{eq:ideal-source:SES,eq:ideal-source:LPA,eq:ideal-source:eh-pairs}
have in common that their purely electron 
and purely hole parts are very simple. This
characteristic is at the heart of our signal processing algorithm
for analyzing single-electron coherence. 
The key idea, which is to exploit time periodicity of single-electron
coherence
\begin{equation}
	\mathcal{G}^{(e)}_{\rho,x}(t+T|t'+T)=\mathcal{G}_{\rho,x}^{(e)}(t|t')\,,
\end{equation}
lies at the heart of
Floquet theorem \cite{Floquet-1883-1}, the time-domain counterpart of 
Bloch's theorem for electronic waves in a periodic crystal
\cite{Bloch-1929-1}.

However, in the present situation, we are looking for a simple
description of $\mathcal{G}_\rho^{(e)}$ in terms of electron and hole
excitations with respect to a reference Fermi sea (here
$\ket{F_{\mu=0}}$). Consequently, Floquet's theorem has to be
adapted in order to be compatible with the decomposition of the
single-particle space of states into a direct sum
$\mathcal{H}=\mathcal{H}_+\oplus
\mathcal{H}_-$ of electron and hole excitations that have positive
(resp. negative) energy with respect to the $\mu=0$ Fermi level.
As we shall see,
this can be done and the corresponding 
eigenvalues have a transparent physical meaning as an occupation number. 
Finally, 
as in band theory of solids, localized single-particle states
\cite{Wannier-1937-1}
can then be constructed. We will show in \cref{sec:Floquet:Wannier} 
that these are the
electronic atoms of signals 
suitable for
describing the quantum electrical current under consideration.

\subsection{Floquet--Bloch analysis}
\label{sec:Floquet:diagonalization}

Introducting localized single-particle states $|t\rangle$
such that $\langle t|t'\rangle= v_F^{-1}\delta(t-t')$, the
dimensionless Hermitian operator $\mathbf{G}^{(e)}$ is defined by 
\begin{equation}
	\label{eq:Ge:definition}
	\mathbf{G}^{(e)}= v_F^2\int_{\mathbb{R}^2}
	\ket{t}\mathcal{G}^{(e)}(t,t')\bra{t'}\,\md t\,\md t'\,.
\end{equation}
in which the $(\rho,x)$ index has been dropped out for simplicity.
The conjugation relation
$\mathcal{G}^{(e)}(t|t')^*=\mathcal{G}^{(e)}(t'|t)$ 
for single-electron coherence translates into
$\mathbf{G}^{(e)}$ being Hermitian. Furthermore, if
we introduce the single-particle state $\ket{\varphi}$ corresponding to
an excitation described by a normalized wavefunction $\varphi$
\begin{equation}
	\ket{\varphi} = v_F \int_{\mathbb{R}} \varphi(t) \ket{t} \, \md t,
\end{equation}
its occupation probability is a real number between $0$ and $1$ given by
\begin{equation}
	p[\varphi]
	=
	\Braket{\varphi | \mathbf{G}^{(e)} | \varphi}.
\end{equation}
thereby ensuring that
$\mathbf{G}^{(e)}$ is a positive operator, bounded by $1$.
For a $T$-periodic source, time periodicity of single-electron coherence 
translates into the commutation of
$\mathbf{G}^{(e)}$  with the time-translation operator
$\mathbf{T}_T$ defined by $\mathbf{T}_T|t\rangle = |t+T\rangle$. 

As explained before, our analysis has to be performed separately on the electron
and hole subspaces $\mathcal{H}_\pm$.
The adapted Floquet-Bloch theorem proven in
\cref{appendix:Floquet} 
provides us with a basis of single-particle state
which partially diagonalizes the single-electron operator while being
compatible with the decomposition into electron and hole excitations. 

\begin{figure}
   \centering
   \includegraphics{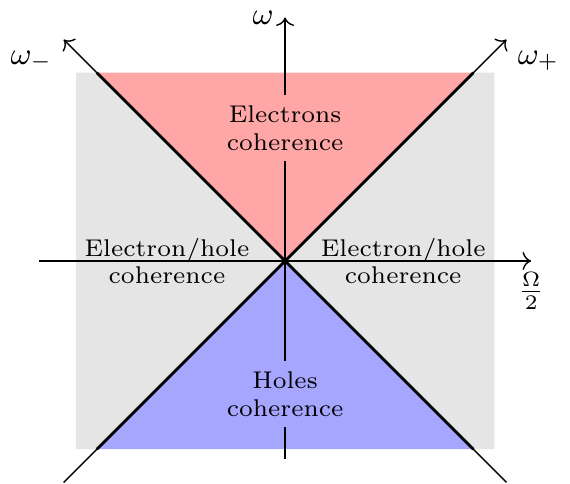}
    
	\caption{\label{fig/eh-quadrants} Frequency 
	domain quadrants for single-electron coherence: depending on the
	signs $(\varepsilon_+,\varepsilon_-)$ of $(\omega_+,\omega_-)$, 
	we are considering the matrix
	elements 
	$\braket{\omega_- | \mathbf{G}^{(e)} | \omega_+}$ of
	$\mathbf{G}^{(e)}_{\varepsilon_+,\varepsilon_-}$ in the
	$\ket{\omega}$ basis of plane waves (see
	\cref{appendix:conv:G:operator} for normalizations).
   The
	electronic quadrant (in red) defined by both
	$\omega_+=\omega+\Omega/2$ and $\omega_-=\omega-\Omega/2$ positive gives information
	about electronic excitations. The hole quadrant (in blue) defined by both $\omega_+$ and $\omega_-$
	being negative gives information about hole excitations.
The two electron/hole quadrants ($\omega_+ \omega_- < 0$, light grey)
contain information about
electron/hole coherences.
    }
\end{figure}

More precisely, this result states that there exists an orthonormal basis 
$\ket{\psi^{(e)}_{a,\nu}}$ of the
positive-energy Hilbert space $\mathcal{H}_+$ and an orthonormal basis 
$\ket{\psi^{(h)}_{b,\nu}}$
of the negative-energy
Hilbert space $\mathcal{H}_-$ which are respectively indexed by band
indexes $a$
(resp. $b$) and quasi-energies $0\leq \nu<2\pi f$
which 
are all eigenvectors of the time-translation
operator $\mathbf{T}_T$ with eigenvalue $\me^{-\mi \nu T}$ and
satisfy the normalization condition
\begin{equation}
\label{eq:Bloch:normalization}
    \langle \psi^{(e)}_{a,\nu}|\psi^{(e)}_{a',\nu'}\rangle =
2\pi\delta_{a,a'}\delta_{\mathbb{R}/2\pi f\mathbb{Z}}(\nu-\nu').
\end{equation}
where $\delta_{\mathbb{R}/2 \pi f \mathbb{Z}}$ is a Dirac comb of period
$2 \pi f$. A similar relation is obtained for the hole states
$\ket{\psi^{(h)}_{b,\nu}}$. In this basis, 
the projections of the single-electron operators on the electron and
hole quadrants (see \cref{fig/eh-quadrants}) 
are diagonalized and their eigenvalues can be expressed 
as the occupation numbers of the corresponding single-electron
states. 
Finally, the full
electronic coherence $\mathbf{G}^{(e)}$ also contains the
information on electron/hole coherences (see \cref{fig/eh-quadrants}) 
which also commutes with
$\mathbf{T}_T$. As explained in \cref{appendix:Floquet}, all this leads
to the following 
form
of single-electron coherence: 
\begin{widetext}
\begin{subequations}
	\label{eq:Bloch:result}
\begin{align}
\label{eq:Bloch:diagonal}
\mathbf{G}^{(e)}&=\int_0^{2\pi f}\left(\sum_a
g_a^{(e)}(\nu)\,|\psi^{(e)}_{a,\nu}\rangle \langle \psi^{(e)}_{a,\nu}|\,
  +\sum_b 
  (1-g_b^{(h)}(\nu))
  \ket{\psi^{(h)}_{b,\nu}} \bra{\psi^{(h)}_{b,\nu}}\right)
  \,\frac{\md\nu}{2\pi}\\
&+\sum_{a,b}\int_0^{2\pi
f}\left(g_{ab}^{(eh)}(\nu)\,
|\psi_{a,\nu}^{(e)}\rangle \langle \psi^{(h)}_{b,\nu}|
+g_{ba}^{(he)}(\nu)\,
|\psi_{a,\nu}^{(e)}\rangle \langle \psi^{(h)}_{b,\nu}|\right)
\, \frac{\md\nu}{2\pi},
	\label{eq:Bloch:off-diagonal}
\end{align}
\end{subequations}
\end{widetext}
where $g^{(eh)}_{ab}(\nu)=g^{(he)}_{ba}(\nu)^*$ in order to
ensure hermiticity of $\mathbf{G}^{(e)}$. 
Let us note that the Floquet-Bloch
wavefunctions being quasi-periodic, these are extended states which are
not localized on a specific period. This is not yet the description in
terms of electronic atoms of signals that will be discussed in the
forthcoming subsection. Before moving to this description, let us recall
that the outcome of the electronic tomography protocol originally
proposed in Ref.
\cite{Degio-2010-4} is an experimental determination of
$\Delta_0\mathbf{G}^{(e)}$ to which the diagonalization procedure can be
applied, therefore leading to an Floquet-Bloch electronic and hole
eigenstates and spectrum as was done in \cite{Bisognin-2019-2}.

As discussed in \cref{appendix:Floquet}, the restriction
$\mathbf{G}^{(e)}_{++}$ of the coherence operator $\mathbf{G}^{(e)}$ to
the electron quadrant, is also positive and bounded by~1.
This leads to $0 \le g^{(e)}_a(\nu) \le 1$ for all $(a,\nu)$, thus
showing that they can
interpreted as the occupation number for the Floquet-Bloch electronic
states $\ket{\psi^{(e)}_{a,\nu}}$.
In the same way, $0\leq g_b^{(h)}(\nu)\leq 1$ since $1-g^{(h)}_b(\nu)$ is the
occupation number of the hole electronic state
$\ket{\psi^{(h)}_{b,\nu}}$. %

\subsection{Electronic atoms of signal}
\label{sec:signal-processing:Wannier}
\label{sec:Floquet:Wannier}

\subsubsection{Floquet--Wannier states}

Since we are interested into finding a description of the excess
single-electron coherence in terms of electronic atoms of signal
\cite{Roussel-2016-2} which are
normalized localized single-electron states, we consider 
Floquet-Wannier states which are the analogous of localized
orbitals in solid-state band theory \citep{Wannier-1937-1}. 
They are defined for $l\in\mathbb{Z}$ as
\begin{equation}
	\label{eq:Wannier:definition}
	\Ket{\varphi_{a,l}} = \frac{1}{\sqrt{f}}\int_0^{2\pi f} 
	\me^{-\mi\nu lt}
	\Ket{\psi_{a,\nu}}\,\frac{\md\nu}{2\pi},
\end{equation}
and form an orthonormal family as implied by \cref{eq:Bloch:normalization}.
Moreover, for a given band, all the states
$(|\varphi_{a,l}\rangle)_{l\in\mathbb{Z}}$ are related by
time translation since $\mathbf{T}_T\ket{\psi_{a,\nu}}=\me^{-\mi\nu
T}\ket{\psi_{a,\nu}}$ and \cref{eq:Wannier:definition} imply that:
\begin{equation}
\label{eq:Wannier:time-translation}
\mathbf{T}_T|\varphi_{a,l}\rangle = |\varphi_{a,l+1}\rangle.
\end{equation}
Exactly as in solid-state band theory \citep{Marzari-2012-1}, there 
are ambiguities
in the determination of
electronic atoms of signals coming 
from the possibility to redefine 
the Floquet-Bloch eigenvectors at a given quasi-energy $\nu$ within
each degenerate subspace of the projection of $\mathbf{G}^{(e)}$ on the electron
or the hole subspace. 
These ambiguities
are extensively discussed in
\cref{appendix:Floquet:Wannier:ambiguities}.
To circumvent these difficulties,
we will focus here on the
electronic atoms of signal that have the smallest spreading in time.
This minimal spreading principle~\cite{Marzari-2012-1}, detailed in
\cref{appendix:Floquet:Wannier:spreading}, has the
advantage of producing maximally localized electronic atoms of signal.
This provides a clear wiew of single-electron coherence within the electronic
fluids in terms of single-particle states that are most visibly
associated with a given period. 

To understand the meaning of such a
description, a musical analogy is convenient: the sound signal
associated with a music instrument can be described in terms of 
elementary units wich are ``music notes'' arranged along a 
``music score'' which specifies the notes to be emitted at a given time.
The electronic atoms of signal can indeed be viewed as the electron
quantum optics counterparts of ``notes'' and the expression of the excess
single-electron coherence in the basis of ``notes'' can be viewed as its
``quantum coherence score''. We will now discuss the specific form of
the ``quantum coherence score'' of a $T$-periodic single electron
coherence.

\subsubsection{Quantum coherence score}

The single-electron coherence
restricted to the electronic quadrant $\mathbf{G}^{(e)}_{++}$ can then be rewritten as
\begin{equation}
	\label{eq:Wannier:Ge}
	\mathbf{G}^{(e)}_{++}
	=
	\sum_a\sum_{l_+,l_-}
	g^{(e)}_a(l_+-l_-)
	|\varphi^{(e)}_{a,l_+}\rangle
	\langle\varphi^{(e)}_{a,l_-}|
\end{equation}
where 
\begin{equation}
\label{eq:Wannier:pa-ee}
	g^{(e)}_a(l)
	=
	\int_0^{2\pi f} g^{(e)}_a(\nu) \, 
	\me^{\mi\nu Tl} \frac{\md\nu}{2\pi f}\,.
\end{equation}
For $l\neq 0$, $g^{(e)}(l)$ represents the interperiod coherence over
$|l|$ periods whereas
$g^{(e)}_a(0)$ is the emission probability for the $\varphi_{a}$
electronic atom of signal at each period. 
Note that there is no coherence between electronic atoms of
signals associated with different bands. However, electronic coherence may
extend over more than one time period:
a flat band ($g_a^{(e)}(\nu)$ constant) won't
lead to inter-period coherences 
whereas a non-flat band
will.
The typical scale over which $g^{(e)}_a(\nu)$
varies is nothing but the inverse timescale over which inter-period
coherence exists. The same considerations apply to hole bands.
Finally, using these Floquet-Wannier states,
the electron/hole coherences $g_{ab}^{(eh)}(l_+-l_-)=
\bra{\varphi_{b,l_-}^{(h)}}\mathbf{G}^{(e)}
\ket{\varphi_{a,l_+}^{(e)}}$ in this basis are given by
\begin{equation}
\label{eq:Wannier:pab-eh}
g_{ab}^{(eh)}(l)=
\int_0^{2\pi f}g_{ab}^{(eh)}(\nu)\,\me^{\mi l\nu T}
\frac{\md\nu}{2\pi f}\,.
\end{equation}
Because electron/hole coherence couples different bands,
different choices of electronic atoms of signal lead to different values
for $g_{ab}^{(eh)}(\Delta l)$. This is not the case for the coherence
between purely electronic or purely hole  wavepackets given by
\cref{eq:Wannier:pa-ee}.

Note that inter-period and electron/hole coherences make the ``quantum
coherence score''
richer than an ordinary (classical) music score which only specifies 
the note that has to be played at a given time and its intensity. The
electronic atoms of signal and the associated ``quantum coherence
score'' are the natural language to describe an arbitrary excess
single-electron coherence. The ``quantum coherence score'' could in
principle be used to encode some quantum information within a quantum
electrical current.

Exactly as a tight-binding quadratic 
Hamiltonian in a
specific Wannier orbital basis is a natural way to describe electron
hoping
within a condensed-matter system, the ``quantum coherence score'' is
the first step in characterizing the many-body state of the electronic
system. The excess second-order electronic coherence
\cite{Thibierge-2016-1} can also been
expressed in terms of electronic atoms of signal, thereby providing a
view of the first non-trivial electronic correlations within the
electronic fluid. Understanding the many-body state of the electronic
fluid in terms of these discrete representations of first- and
higher-order excess electronic coherences
is a very interesting perspective for electron quantum optics. Although
its simplest aspect will be discussed in \cref{sec/many-body}, a full
discussion would go way beyond the scope of the present paper.

Positivity of the electronic and hole coherences $\mathbf{G}^{(e)}$ 
and $\mathbf{G}^{(h)}$
leads to
Cauchy-Schwarz inequalities.
Within the electron and hole quadrants,
it leads to
\begin{subequations}
\begin{align}
	\left|g_a^{(e)}(l)\right| &\leq \min(g^{(e)}_a,1-g^{(e)}_a) \\
	\left|g_b^{(h)}(l)\right| &\leq \min(g^{(h)}_b,1-g^{(h)}_b)
\end{align}
\end{subequations}
with $g^{(e)}_a=g_a^{(e)}(l=0)$ and
$g^{(h)}_b=g_b^{(h)}(l=0)$
denoting the respective averages of 
$g^{(e)}_a(\nu)$  and $g^{(h)}_b(\nu)$
over $0\leq \nu < 2\pi f$. 
In the electron/hole quadrants, the Cauchy-Schwarz
inequalities bound the electron/hole coherences:
\begin{subequations}
	\begin{align}
	\left| g_{ab}^{(eh)}(l)\right|^2
		&\leq
	g^{(e)}_a\,
	\left(1-g_b^{(h)}\right)\\
		\left| g_{ab}^{(eh)}(l)\right|^2
		&\leq
    g^{(h)}_b\,
		\left(1-g_a^{(e)}\right)\,.
	\end{align}
\end{subequations}
These inequalities immediately show that,
in the absence of electron
($g_a^{(e)}(\nu)=0$ for all $\nu$ and $a$)
or hole ($g_b^{(h)}(\nu)=0$) excitations, there are no
electron/hole coherences ($g_{ab}^{(eh)}(l)=0$) as well as no
coherence
between the missing excitations as noted in Ref.~\cite{Ferraro-2013-1}.

\subsubsection{Martin-Landauer wavepackets}
\label{sec:Floquet:examples:stationary}

\begin{figure}
	\includegraphics{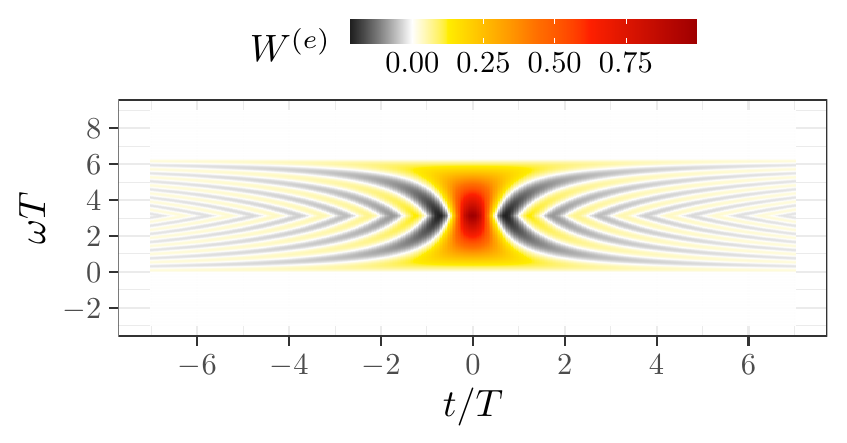}
	\caption{\label{fig/ML-Wigner} Wigner representation of the
	Martin-Landauer wavepacket $\text{ML}_{0,0}$ as a function of $t/T$ and
	$\omega T$.}
\end{figure}

Let us illustrate these ideas on the example of a stationary 
electronic state. In this case, the
single-electron coherence only depends on $t-t'$ and is the Fourier
transform of the electronic distribution function $f_e(\omega)$. 
Such a state can be
viewed as $T$-periodic for any period $T$ so, let us chose one and perform the
corresponding Floquet--Bloch analysis. The excess single-electron
coherence being already diagonal in the plane-wave basis, 
the Floquet--Bloch waves are plane waves
$\psi_{m,\nu}(t)=v_F^{-1/2}
\me^{-\mi(\nu+2\pi mf)t}$ ($0\leq \nu<2\pi f$). 
The bands are then indexed by an integer $m\in
\mathbb{N}$ and the corresponding eigenvalues are given
by:
\begin{equation}
\label{eq/Floquet-Bloch/Martin-Landauer/eigenvalues}
	g_m^{(e)}(\nu)=f_e(\nu+2\pi m f)\,.
\end{equation}
The corresponding
electronic atoms of signal are obtained by summing plane waves over an energy band of
width $hf$, centered at energies $(m+1/2)hf$ with $m$ integer. These are
the
Martin--Landauer wavepackets \cite{Martin-1992-1}:
\begin{equation}
\label{eq/Floquet/examples/Martin-Landauer}
\text{ML}_{m,0}(t)=\frac{1}{\sqrt{v_FT}}\,
\frac{\sin{(\pi ft)}}{\pi ft}\,\me^{-2 \mi \pi (m+1/2)ft}
\end{equation}
which are known in
the signal-processing community as the Shannon wavelets. Their Wigner
representation $W_{\text{ML}_{m,l}}(t,\omega)=W(t,\omega-2\pi(m+1/2)f)$,
defined for a single electron wavepacket $\varphi$ 
by using $\varphi(t+\tau/2)\,\varphi(t-\tau/2)^*$ 
instead of $\mathcal{G}^{(e)}(t+\tau/2,t-\tau/2)$
in Eq. \eqref{eq/Wigner}, are
time and frequency translated 
from
\begin{equation}
 W(t,\omega) = 
	\Theta\left[\pi f-|\omega|\right]\left(1-\frac{|\omega|}{\pi f}\right)\sinc(2(\pi f-|\omega|)t)
\end{equation}
The Wigner representation $W_{\text{ML}_{0,0}}$, depicted on
\cref{fig/ML-Wigner}, is clearly localized in the $0\leq \omega \leq
2\pi f$ energy band and decays as $\sim 1/t$ in time.

Because of \cref{eq/Floquet-Bloch/Martin-Landauer/eigenvalues}, the bands
are generically not flat thereby implying the existence of interperiod coherences
and therefore, of an associated coherence time. The idea is then to
choose the period $T$ so that, the electron distribution function is as
flat as possible over energy bands of width $hf$. 

For example,
at zero temperature, the non-equilibrium distribution function generated
by a 
d.c.\@ biased QPC with
$V_{\mathrm{dc}}<0$, 
is a step function that jumps from $1$ for $\omega< 0$ to the
transmission probability $0<\mathcal{T}<1$ from the biased incoming electrode to the 
outgoing one we are considering
for $0\leq \omega \leq
-eV_{\mathrm{dc}}/\hbar$. It then abrubtly falls to zero
$\omega > -eV_{\mathrm{dc}}/\hbar>0$. The natural choice of $T$ is then 
$T=h/e|V_{\mathrm{dc}}|$. The excess coherence has only one non trivial
band corresponding to $g^{(e)}_0(\nu)=\mathcal{T}$ for $0\leq \nu<2\pi
f=
|eV_{\mathrm{dc}}|/\hbar$. 
The excess 
electronic coherence is then naturally
described in terms of Martin-Landauer wavepackets associated with period
$T=h/e|V_{\text{dc}}|$:
\begin{equation}
	\label{eq/ML/biased-QPC}
	\Delta_0\mathbf{G}^{(e)}=
	\mathcal{T}\sum_{l\in\mathbb{Z}}\ket{\text{ML}_{0,l}}\bra{\text{ML}_{0,l}}\,.
\end{equation}
Consequently, this excess single-electron coherence corresponds to a
train of
Martin-Landauer wavepackets 
without interperiod coherences and each of them
being
emitted with probability $\mathcal{T}$. Stationarity is
visible through the invariance of $\Delta_0\mathbf{G}^{(e)}$ through
translation by $\mathbf{T}_{\Delta t}$ for any $\Delta t$: timeshifting all
the Martin-Landauer wavepackets in the r.h.s. of \cref{eq/ML/biased-QPC}
$\ket{\text{ML}_{m,l}}\mapsto \mathbf{T}_{\Delta
t}\ket{\text{ML}_{m,l}}$ still gives $\Delta_0\mathbf{G}^{(e)}$.

At finite temperature $T_{\mathrm{el}}>\SI{0}{\kelvin}$, the electronic distribution
function is 
smeared over a scale $k_BT_{\mathrm{el}}/\hbar$, thus leading to a
non-flat band spectrum. Therefore, there
are always inter-period coherences over the thermal coherence time
$h/k_BT_{\mathrm{el}}$. It might seem surprising that when
$T_{\text{el}} = \SI{0}{\kelvin}$, the inter-period coherences go to
zero whereas the thermal coherence time goes to infinity. This comes
from the fact that when decreasing the temperature, as the off-diagonal
coherences spread over more and more period, their modulus decreases
and ultimately vanishes at zero temperature. 

\subsection{Relation to experimentally-relevant quantities}
\label{sec:Floquet:signals:detection}

Let us now explain how experimental signals are related to these
spectral quantities. We will first discuss the value of the dip in an HOM experiment,
a simple HOM based repeated detection
scheme of a given electronic excitation 
and finally a time-dependent energy filter
based on a driven QPC \cite{Locane-2019-1}. 

\subsubsection{The Hong--Ou--Mandel dip}
\label{sec:Floquet:signals:HOM}

In the case of an HOM experiment with two identical sources $S_1$ and
$S_2$ on the
incoming channels of a beam splitter with reflexion and transmission
probabilities $\mathcal{R}$ and $\mathcal{T}$,
the depth of the HOM dip, obtained by synchronizing the sources, 
can be related to the Floquet-Bloch
spectral properties of single-electron coherence. This comes from the
expression of 
the two-particle interference contribution to low-frequency noise in an
HOM experiment as \cite{Degio-2010-4}:
\begin{equation}
\label{eq:HOM:total:time-resolved}
	\mathcal{Q}(t,t')
=-e^2v_F^2\mathcal{RT}
\left(\mathcal{G}^{(e)}_1(t,t')\mathcal{G}^{(h)}_2(t,t')
+\left[1\leftrightarrow 2\right]
\right)\,.
\end{equation}
We consider the low frequency
noise defined by integrating over $\tau=t-t'$ and 
averaging over $\bar{t}=(t+t')/2$. Expanding both contributions in
the r.h.s. of \cref{eq:HOM:total:time-resolved} in terms of
$\Delta_0 \mathbf{G}^{(e)}$
leads to three distinct contributions. Two of them involve only
one of the incoming excess single-electron coherences and
correspond to the partitioning of single-particle excitations from
one of the two incoming channels at
the QPC (HBT contribution) whereas the third one involves the excess single electron
coherence of both sources and accounts for two-particle interferences
between them (HOM contribution).

At zero temperature and with identical and synchronized sources on two incoming channels
the excess noise in the HBT (only one source {\it on}) and HOM
experiments (both sources {\it on}) are obtained as sums of a background
which comes from the
transmitted or reflected excess noise $\Delta S_S$ of the sources 
and of
two-excitation interference contributions denoted by $\Delta
S_{\text{HBT}}$ and $\Delta S_{\text{HOM}}$:
\begin{subequations}
	\begin{align}
		\label{eq:HBT:total1}
		\Delta S_{11}^{(\text{HBT}_1)}&=\mathcal{R}^2\Delta
	S_{\text{S}}+\Delta S_{\text{HBT}}\\
		\label{eq:HBT:total2}
		\Delta S_{11}^{(\text{HBT}_2)}&=\mathcal{T}^2\Delta
    S_{\text{S}}+\Delta S_{\text{HBT}}\\
		\label{eq:HOM:total}
		\Delta S_{11}^{(\text{HOM})} &=
		(\mathcal{R}^2+\mathcal{T}^2)\Delta S_S
		+\Delta S_{\text{HOM}}\,.
	\end{align}
\end{subequations}
As shown in \cref{appendix/partition-noise}, 
$\Delta
S_{\text{HBT}}$ and $\Delta S_{\text{HOM}}$
are given by:
\begin{subequations}
\begin{align}
	\label{eq:HBT:excess}
\Delta S_{\text{HBT}} &=
	e^2\mathcal{RT}\int_0^{2\pi f}
	\left(\sum_ag_a^{(e)}(\nu)+\sum_bg_b^{(h)}(\nu)\right)
	\,\frac{\md \nu}{2\pi}\\
	\Delta S_{\text{HOM}}
	&= 2e^2\mathcal{RT}\left[\int_0^{2\pi f}\sum_a
	(1-g_a^{(e)}(\nu))
g_a^{(e)}(\nu)\,\frac{\md\nu}{2\pi}
	\right. \nonumber
	\\
	&+\int_0^{2\pi f}
	\sum_bg_b^{(h)}(\nu)(1-g_b^{(h)}(\nu))\,
\frac{\md\nu}{2\pi}
	\nonumber
	\\
	& \left. -2\int_0^{2\pi f}\sum_{a,b}\left|
	g^{(eh)}_{ab}(\nu)\right|^2\frac{\md\nu}{2\pi}\right].
	\label{eq:HOM:excess}
\end{align}
\end{subequations}
In \cref{appendix/partition-noise/HOM-dip}, we show that the depth of
the HOM dip, which is the difference $\Delta S_{\text{dip}}=2\Delta
S_{\text{HBT}}-\Delta S_{\text{HOM}}$ at this operating point can be
expressed simply in terms of the fluctuation of the total charge emitted
per period $(\Delta
Q)^2_{\text{w}}$:
\begin{equation}
    \frac{\left[\Delta
    S_{\text{dip}}\right]}{\left[\Delta
    S_{\text{dip}}^{(\text{max})}\right]} =
    1-\frac{(\Delta
    Q)^2_{\text{w}}}{\overline{N}_{\mathrm{tot}}}\,.
\end{equation}
in which $\bar{N}_{\mathrm{tot}}$ is the sum of the average number of
electron and hole excitations (see \cref{eq/Ne-Nh-definition}) and
$(\Delta
Q)^2_{\text{w}}$ is
computed from first order coherences using Wick's theorem (see \cref{appendix/partition-noise/HOM-dip}). 
If the many-body state does satisfy Wick's
theorem, which is the case whenever interactions can be neglected, then
this corresponds to the actual vanishing of charge fluctuations.
Under this hypothesis, 
an ideal single electron (see \cref{eq:ideal-source:SES}) or single electron and single
hole (see \cref{eq:ideal-source:LPA}) source would
reach this bound and therefore, under the assumption that interactions
can be neglected, would lead to a maximally deep HOM dip at zero
temperature. Another important 
example is the state with a single coherent electron/hole pair
obtained by the action of an operator of the form
\begin{equation}
	\sqrt{1-g}\,\mathbf{1}+\me^{\mi\Theta}\sqrt{g}\,\psi^\dagger[\varphi_e]\psi[\varphi_h]
\end{equation}
where $0<g<1$, acting on the Fermi sea $\ket{F_{\mu=0}}$.
In this case, the average number of electron (as well as hole)
excitations is $g$ but the charge fluctuation is exactly zero:
$(\Delta Q)^2_\text{w}=0$. Consequently, the dip does to its maximum
value. As we shall see in \cref{sec/many-body}, this is the case for all states
obtained by acting with a $T$-periodic time-dependant scatterer on
$\ket{F_{\mu=0}}$.

However, let us recall that this is not true when interactions, for example between the
sources and the beam-splitter are present \cite{Marguerite-2016-1}: the
depth of the dip is decreased by electronic decoherence. The dip may also
not be maximally deep, at zero temperature, when the emission process involves
some classical randomness, one example being the randomized train of
levitons considered in \cref{sec/Levitons/random}.

\subsubsection{Repeated HOM detections}

Because electronic atoms of signals are localized in time, they are
suitable single-particle states to discuss repeated detection protocols.
Let us discuss such a protocol based on two-particle
interferometry using an ideal beam splitter with energy-independent transmission
probability $\mathcal{T}$ (HOM interferometry).

On one incoming channel, we consider a $T$-periodic source $S$
whereas on the other incoming channel, we have
a specific ideal
electronic source $S_a$ which emits a periodic trains of electronic atoms of signals
$\ket{\varphi_{a,l}}$, not necessarily related to the ones present
emitted by $S$. Its excess single-electron coherence is thus
\begin{equation}
\Delta \mathcal{G}_{S_a}^{(e)}(t,t')=\sum_{l=0}^{N}
\varphi_{a,l}(t)\,\varphi_{a,l}^*(t')\,.
\end{equation}
The resulting outgoing current noise contains an HOM contribution proportional to the overlap between
$\Delta\mathcal{G}_S^{(e)}$ and $\Delta\mathcal{G}_{S_a}^{(e)}$ \cite{Ferraro-2013-1}. Using
the $T$ periodicity of $\Delta\mathcal{G}_S^{(e)}$, the experimental
signal scales as $N\gg 1$ which quantifies the total acquisition time $NT$:
\begin{equation}
\label{eq/HOM/signal}
	\int_{[-\frac{NT}{2},\frac{NT}{2}]^2} 
\Delta\mathcal{G}_{S_a}^{(e)}(t,t')^*\Delta\mathcal{G}^{(e)}_S(t,t')\,\md t\,\md t'\sim
N\,\bar{p}_a\,.
\end{equation}
This 
overlap counts the number of times an electron in the
single-particle state $|\varphi_{a,l}\rangle$ is scattered against an
electronic excitation in the same single-particle state for 
$N$ periods of duration $T$. Since $S_a$ is an ideal source sending a train of $N$
identical excitations shifted by multiples of $T$, the quantity
$\bar{p}_a$ should be
interpreted as the average number of times, the
single-particle state
$\varphi_{a}$ is emitted per period. If the $\varphi_{a,l}$ are among
the electronic atoms of signal emitted by $S$, then when the emission of
$S_a$ is synchronized with the emission of these atoms of signal by $S$, 
$\bar{p}_a=\overline{g}_a$ is the probability of emission of these
electronic atoms of signal by $S$.

\subsubsection{Time-frequency filtering}
\label{sec:Floquet:signals:filtering}

A repeated detection protocol can also be realized by scattering the electron
flow through a periodically driven energy-dependent scatterer. Recently,
such a time-frequency
filtering has been demontrated and used for single-electron
tomography~\cite{Fletcher-2019-1}. It relies on
a quantum point contact
with an energy-dependent transmission probability $T(\omega)$ 
equipped with a top electrostatic gate driven by a time dependent voltage
$V_d(t)$. The signal collected by such a device is the total
charge transmitted through the QPC which can be rewritten as a linear filtering of 
the incident single-electron excess coherence \cite{Locane-2019-1}:
\begin{equation}
	\label{eq/time-energy-filter/filtering}
	Q = -e\int_{\mathbb{R}^2} 
	v_F\Delta\mathcal{G}^{(e)}\left(t+\frac{\tau}{2},t-\frac{\tau}{2}
	\right)\mathcal{F}_d(t,\tau)^*
	\,\md t\,\md \tau
	\end{equation}
with the filter's kernel being given by
\begin{equation}
	\label{eq/time-energy-filter/filter}
	\mathcal{F}_d(t,\tau)=\int_{\mathbb{R}} T(\omega)\,
	\me^{\mi \omega\tau
	+
	\frac{\mi e}{\hbar}\int_{t-\tau/2}^{t+\tau/2}
	\mathcal{V}_d(t')\md t'}
	\frac{\md \omega}{2\pi}
\end{equation}
in which $\mathcal{V}_d(t')$ is proportional to the driving voltage
$V_d$\footnote{More precisely, $\mathcal{V}_d(t)=\int_{-\infty}^x
\left(\partial_xU\right)(x,t-x/v_F)\md x$ in which $U(x,t)=V_d(t) u(x)$
denotes the voltage seen by the electrons as they fly across the QPC
($u(x)$ being non zero close to the constriction of the QPC and rapidly
vanishing away from it).}.

For a time-periodic driving at frequency $f=1/T$, the linear filter is
also $T$-periodic.
Since $\mathcal{F}_d(t,\tau)^*=\mathcal{F}_d(t,-\tau)$, the  
Floquet-Bloch analysis can be applied to the filter. 
Besides the example
considered in Ref. \cite{Locane-2019-1}, the case of a driven quantum
dot \cite{Yamahata-2019-1,Wenz-2016-1} corresponding
to a Lorentzian transmission probability $T(\omega)$ centered at 
$\omega_0 >0$ with width $\gamma_0\ll \omega_0$ is
worth considering since it corresponds to a dot filtering mostly
electronic excitations around the energy $\hbar \omega_0$. Provided the
drive is such that $\omega_0-e\mathcal{V}_d(t)/\hbar\gg \gamma_0$, we expect that
only purely electronic excitations are transmitted.
We should therefore be able to diagonalize the
(dimensionless)
filtering operator 
\begin{equation}
	\mathbf{F}_d =v_F\int_{\mathbb{R}^2}
	\ket{t_+}\,\mathcal{F}_d\left(\frac{t_++t_-}{2},t_+-t_-\right)\,
	\bra{t_-}\,\md t_+\md t_-\,
\end{equation}
within the electronic quadrant, thus leading to:
\begin{equation}
	\mathbf{F}_d=\sum_\alpha\int_0^{2\pi f} 
	\mathcal{F}_{\alpha}(\nu)\,\ket{\psi^{(d)}_{\alpha,\nu}}
	\bra{\psi^{(d)}_{\alpha,\nu}}
	\frac{\md \nu}{2\pi}\,
\end{equation}
in which the eigenstates $\ket{\psi^{(d)}_{\alpha,\nu}}$ are electronic
Floquet-Bloch waves of the filter.
The eigenvalues $\mathcal{F}_{\alpha}(\nu)$ are
real but (as far as we know) are not restricted. 
In practice, acquisition of the experimental signal requires
a $T$-periodic single-electron excess coherence
$\Delta_0\mathbf{G}^{(e)}$ and measurement over $N\gg 1$ periods. Then,
the total transmitted charge increases linearily with time. The average
charge transmitted per period $Q_T$, and thereby the transmitted dc current $\langle
I_{\text{tdc}}\rangle = Q_T/T$, can then be expressed in terms of the
electronic atoms of signals $\ket{\varphi^{(d)}_{\alpha,l}}$ arising from the
Floquet-Bloch waves of the filter. Decomposing
\begin{equation}
	\mathbf{F}_d=\sum_\alpha\sum_{(l_+,l_-)\in\mathbb{Z}^2}
	\mathcal{F}_\alpha(l-l')\,\ket{\varphi^{(e)}_{\alpha,l_+}}
	\bra{\varphi^{(e)}_{\alpha,l_-}}
\end{equation}
in which $\mathcal{F}_{\alpha}(l_+-l_-)$ is related to
$\mathcal{F}_\alpha(\nu)$ by \cref{eq:Wannier:definition}, 
leads to the average transmitted
dc current:
\begin{equation}
	\langle I_{\text{tdc}}\rangle =-ef
	\sum_\alpha \sum_{l\in\mathbb{Z}}
	\mathcal{F}_{\alpha}(l)\,
	\braket{\varphi^{(d)}_{\alpha,0}
	| \Delta_0\mathbf{G}^{(e)} | \varphi^{(d)}_{\alpha,l}}\,.
\end{equation}
Therefore, the driven QPC studied in \cite{Locane-2019-1} appears as
a linear filter acting on linear coherence which generalizes to the
time-dependent case, the quantum dot energy filter originally used to study 
electronic relaxation in quantum Hall edge channels
\cite{Altimiras-2010-1}. 
The average current is then directly proportional to
the overlap between $\Delta_0\mathbf{G}^{(e)}$ and the time-dependent
filter's
``quantum coherence score'' introduced in
\cref{sec:signal-processing:Wannier}.
Generically, several bands may be present and therefore
several electronic atoms of signal may be needed. However, we can hope that suitable
drives may lead to filtering by mostly one band, or equivalently one
type of
electronic atom of signal thereby enabling us to probe the presence of a specific atom of signal 
within $\Delta_0\mathbf{G}^{(e)}$. 

\section{Many-body properties}
\label{sec/many-body}

Until now, we have focused on the properties of the electronic fluid at the
single-particle level, assuming nothing more than $T$-periodicity. 
However, when interactions within the electronic
fluid can be neglected, the single-particle description actually gives
us access to the whole many-body state. This is notably the case when the
single-electron source is modelled by single-particle scattering
processes.

In this section, we will explain how the Floquet-Bloch analysis
allows us to give a simple many-body description and
unravel some of the symmetries hidden within the band structure. We will
also be able to rederive the single particle scattering operator
leading to such a single-electron coherence, thereby exploring the path
followed in Ref. \cite{Yin-2019-1} the other way around.
Furthermore, by giving a direct insight on electron/hole entanglement,
the Floquet--Bloch analysis is well
suited to quantify the quality of electron
sources. We will use it in \cref{sec:source-diagnostic} to
identify the best operating experimental
parameters for a given source.

\subsection{Many-body state at zero temperature}

For the sake of simplicity but without loss of generality, we shall focus
on a $T$-periodic coherence corresponding to a vanishing average dc
current so that the chemical potential of the electron fluid is exactly
zero, a specific case also considered in \cite{Yue-2019-1}.
We assume that single electron coherence is the result of 
$T$-periodic 
single-particle scattering $\mu=0$ Fermi sea
($T_{\mathrm{el}}=\SI{0}{\kelvin}$)
$\ket{F}$ which thereby generally describes a $T$-periodic ac source
whenever interactions can be neglected.

In order to derive the many-body state at zero temperature, the method
consists into 
finding an expression of the many-body Floquet operator from the
Floquet-Bloch decomposition (see \cref{appendix/manybody}) thereby
inverting the procedure described in Refs. \cite{Yin-2019-1,Yue-2019-1}. 
Applying this operator to the Fermi sea leads to the general form of many-body
state $\ket{\Psi}$ emitted by the source:
\begin{equation}
	\label{eq:manybody:zeroTstate}
	\prod_{\substack{\nu \in [0, 2 \pi f[\\ a \in \mathbb{N}}}
	\left(
		\sqrt{1 - g^{(e)}_a(\nu)}
		+
		\sqrt{g^{(e)}_a(\nu)}\,
		\psi^\dagger\left[\psi^{(e)}_{a,\nu}\right]
		\psi\left[\psi^{(h)}_{a,\nu}\right]
	\right)
	\ket{F}.
\end{equation}
From this expression, we notice an important symmetry on
the spectrum: because electrons and holes 
are emitted in pairs, we have 
$g_a^{(e)}(\nu) = g_a^{(h)}(\nu)$. It is worth noting that
this relation is different from electron-hole symmetry, which reverses
frequencies and, as such, would be $g_a^{(e)}(\nu) = g_a^{(h)}(2\pi f-\nu)$.
Consequently, 
when electron/hole symmetry is satisfied in state
\eqref{eq:manybody:zeroTstate}, the Floquet--Bloch spectrum
exhibits the symmetry:
$g_a^{(e)}(\nu) = g_a^{(e)}(2\pi f-\nu)$.

\subsubsection{The case of flat bands}

When the bands are flat, we can go further in the analysis and reexpress
the many-body state in terms of Floquet-Wannier wavefunctions. The flat
band case ($g_a(\nu) = g_a$) happens for a purely a.c.\@ voltage
drive at zero temperature. The case of a.c.\@ voltage drives have been
studied previously in
\cite{Vanevic-2008-1,Vanevic-2016-1,Vanevic-2017-1}. In this case,
as shown in \cref{appendix/manybody},
\cref{eq:manybody:zeroTstate} can be rewritten as
\begin{equation}
	\ket{\Psi}
	=
	\prod_{a \in \mathbb{N}}
	\prod_{l \in \mathbb{Z}}
	\left(
		\sqrt{1-g_a}
		+
		\sqrt{g_a}\,
		\psi^\dagger\left[\varphi^{(e)}_{a,l}\right]
		\psi\left[\varphi^{(h)}_{a,l}\right]
	\right)
	\ket{F}.
\end{equation}
This is the formula for a classical voltage drive found
in~\cite{Vanevic-2016-1}. 
Since there is no relative phase between $\sqrt{1-g_g}$ and
$\sqrt{g_a}\,\psi^\dagger[\varphi^{(e)}_{a,l}]\psi[\varphi^{(h)}_{a,l}]$, 
once a determination for the electronic
Floquet-Wannier wavefunctions has been chosen, it determines also the
wavefunctions for holes, up to a global phase. As such, we do not expect
the Floquet-Wannier wavefunctions to be minimally spread for both
electrons and holes.
This also allows us to come back to the ansatz guessed in
\cref{eq:ideal-source:eh-pairs}. This ansatz works only in the case of
flat bands. While this is the case for a classical drive, we will see
that it
is usually not the case for the mesoscopic capacitor.

\subsection{Electron/hole entanglement entropy}
\label{sec:eh-entanglement:entropy}

Accessing the many-body state allows us to quantify the quality
of a single-electron (or more generally, a $n$-electron) source. Such a
source would emit electrons and holes independantly, without
correlations besides Fermi statistics. Furthermore, because our sources are described as
noiseless single-particle scattering from an equilibrium state 
quantifying
outgoing correlations gives the amount of correlation generated during
the scattering process. When the global state is
pure ($T_{\mathrm{el}}=\SI{0}{\kelvin}$), the correlations only come from entanglement.

Although the question of entanglement is a complicated problem in a
many-body system \cite{Amico-2008-1}, 
the very definition of electron and hole provides us
with a natural way to split the many-body Hilbert space in two
orthogonal components, thereby enabling us to fall back on a more
familiar description. Thanks to the parity super-selection rule for
fermions \cite{Wick-1952-1,Friis-2016-1,Johansson-2016-1} and to the
absence of superconducting correlations in a metallic conductor, the
many-body density operator is block-diagonal, only exhibiting coherences
between states having the same number of electron and hole excitations
with respect to a reference Fermi sea. Quantifying the electron/hole
entanglement could, in principle, be done by looking separately into all
these superselection sectors but this would require knowing electronic
coherences to all orders. 

Fortunately, when Wick's theorem is satisfied, the full many-body state
depends only on first-order
coherence. This is also true for the partial trace on positive or
negative energy states since higher-order correlations
functions expressed in frequency basis are just correlation functions of
the whole state taken in the simplex of positive frequencies and
thereby, they also obey Wick's theorem. From this and the superselection
rule follows that 
the many-body state associated to the
electronic quadrant corresponds to filling non-coherently each
Floquet-Bloch mode $\ket{\psi^{(e)}_{a,\nu}}$ with its probability
$g^{(e)}_{a}(\nu)$:
\begin{align}
	\widehat{\rho}_{++}
	=&
	\bigotimes_{\substack{\nu \in [0, 2 \pi f[\\ a \in \mathbb{N}}}
	\left(
		\left(
			1 - g^{(e)}_a(\nu)
		\right)
	\ket{0}\bra{0}\right.
	\nonumber\\
	&+\left.
		g^{(e)}_a(\nu)\, \psi^\dagger[\psi^{(e)}_{a,\nu}]
		\ket{0}\bra{0}
		\psi[\psi^{(e)}_{a,\nu}]
	\right).
	\label{eq:many-body:2levels}
\end{align}
Its form is reminiscent of a the thermal state with a mode-dependent
temperature. 

An important property of an ideal $n$-electron source is that there are no
correlations between the electron and hole excitations it emits. Namely,
we expect the full many-body state
associated to positive and negative frequencies to factorize as
	$\widehat{\rho}_{\text{SES}}
	=
	\widehat{\rho}_{++} \otimes \widehat{\rho}_{--}$.
In the present case of a pure many-body state for the whole electronic
fluid, the departure from such a factorized form is measured by the
von~Neumann entanglement entropy of the electrons (or the holes), a quantitative
measure of
entanglement in this case \cite{Bennett-1996-2}. 
Starting from the
expression~\labelcref{eq:many-body:2levels} for the
many-body state, is it given by
\begin{align}
\label{eq:eh-entanglement:entropy}
	S_{\text{vN}}
	=
	&-
	\sum_{a \in \mathbb{N}}
	\int_0^{2\pi f} \left(
		g^{(e)}_a(\nu) \log_2 (g^{(e)}_a(\nu))\right.\nonumber\\
		&+\left. (1-g_a^{(e)}(\nu)) \log_2 (1 - g^{(e)}_a(\nu))
	\right) \, \frac{\md \nu}{2 \pi f}\,.
\end{align}
Therefore, the entanglement entropy can be inferred from the properties
of the Floquet-Bloch spectrum for electrons (or holes) which thereby
appears as an entanglement spectrum \cite{Li-2008-1}. This connexion has
been exploited to quantify entanglement between spatially separated
regions in many-body fermionic systems or generated by a quantum point
contact \cite{Francis-Song-2012-1} but can indeed be applied to more
general decompositions of the full single-particle state in a sum of
two orthogonal components 
\cite{Klich-2006-1}. 

Finally, at non-zero temperature, the von Neumann entropy is no longer
the sole measure of entanglement since the global state is not pure
anymore. For practical use in
experiments \cite{Bisognin-2019-2},
the departure of the full many-body state from a pure state
can be quantified using a purity indicator which can also expressed in terms of the
Floquet-Bloch electron and hole spectra
$(g^{(e)}_a(\nu),g^{(h)}_b(\nu))$ and of electron hole coherences
$g^{(eh)}_{ab}(\nu)$ for $0\leq \nu<2\pi f$ as detailed in
\cref{appendix/purity}.

\section{Electron source analysis}
\label{sec:source-diagnostic}

We now apply our signal-processing technique to numerical
data coming from a Floquet modelization of periodic electron sources.
Our goal is to use the Floquet-Bloch spectrum to assess the quality
of sources as single-electron sources, along the lines discussed in
the previous section. Because of their experimental importance, the 
mesoscopic capacitor \cite{Feve-2007-1} and the Leviton source
\cite{Dubois-2013-2} will be discussed.
The former offers the possibility to emit single-electron excitations
well separated from the Fermi surface. 
The
Leviton source exploits the fact that a suitable rearrangement of an
infinite Fermi sea can lead to the generation of purely electronic
excitations~\cite{Levitov-1996-1} in a simple way.

\subsection{The mesoscopic capacitor}
\label{sec/capacitor}

\subsubsection{Model}
\label{sec/capacitor/model}

The mesoscopic capacitor depicted on
\cref{fig:capa-meso}
is modeled using Floquet scattering theory
\cite{Moskalets-book}, since, in most
experimentally relevant regimes, interaction effects within the
capacitor itself can be neglected. 
Within this framework~\cite{Moskalets-2008-1,Degio-2010-4}, the mesoscopic capacitor is
characterized by the level spacing of
the dot $\Delta$, the transparency $D$ of
the QPC (see \cref{fig:capa-meso}) as well as by the voltage drive
$V_g(t)$. The Floquet scattering matrix relating the outgoing fermionic
field to the incoming one is then expressed as
\cite{Moskalets-2008-1,Degio-2010-4}
\begin{equation}
S(t,t')=\exp{\left(\frac{\mi e}{\hbar}\int_{t'}^tV_g(\tau)\,\md\tau\right)}
\mathcal{S}_0(t-t'),
\end{equation}
where $\mathcal{S}_0$ denotes the scattering matrix of the dot which, in
the frequency domain, is given by
\begin{equation}
\mathcal{S}_0(\omega)= \frac{\sqrt{1-D}-\me^{2\mi\pi
\hbar(\omega-\omega_0)/\Delta}}{1-\sqrt{1-D}\,\me^{2\mi\pi
\hbar(\omega-\omega_0)/\Delta}},
\end{equation}
where $\omega_0$ comes from a dc bias applied to the dot. Adjusting it
so that $\omega_0=0$ ensures that a peak
in the density of states of the dot is located at the Fermi level in the
absence of external drive.

We will now discuss the operating regimes of the mesoscopic
capacitor operated by a sinusoidal drive $V_g(t)=V\sin{(2\pi ft)}$ at frequency $f$
by computing the electron/hole entanglement from the
Floquet-Bloch spectrum for the electronic excitations at fixed $\Delta$
and driving frequency $f$ in terms of the
experimentally controlled parameters $D$ and $V$, the latter being the
amplitude of the drive applied to the mesoscopic capacitor. 
In the present case, we work at fixed drive frequency and dot
geometry so that $\Delta/hf\simeq 20$, which corresponds to experimentally
realistic conditions. Having in mind the original experiments
\cite{Feve-2007-1,Marguerite-2016-2}, 
results for
a square drive are given in
\cref{appendix/capacitor/square-drive}.

\subsubsection{Electron/hole entanglement entropy}
\label{sec/capacitor/sinusoidal/entropy}

\begin{figure}
	\centering
	\includegraphics{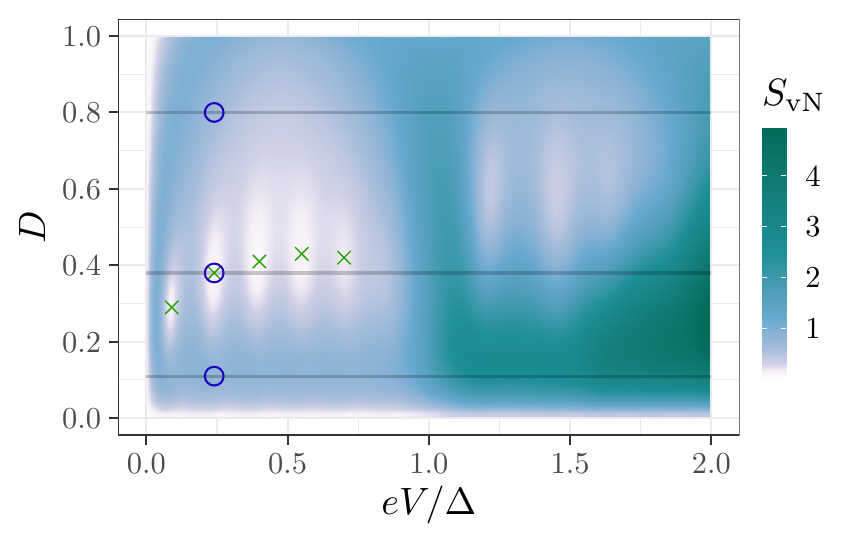}
\caption{\label{fig:capa-meso:examples:sinus:entropy-map} 
Density plot of the electron/hole
entanglement entropy at zero temperature for the mesoscopic capacitor
operated with a sine drive at frequency $f$ such that 
$\Delta/hf=20$ as a function of $eV/\Delta$ and $D$. Crosses correspond
	to the five local minima of the entropy (see
	\cref{table:entropy:cos:pos}) where the source is the closest to an
	ideal single-electron source. For the second local minimum, we have
	chosen three values of $D$: the optimal one, one above and one below
	(round points) to discuss the effect of varying $D$ for a fixed drive.
}
\end{figure}

\begin{figure}
	\centering
	\includegraphics[width=8.6cm]{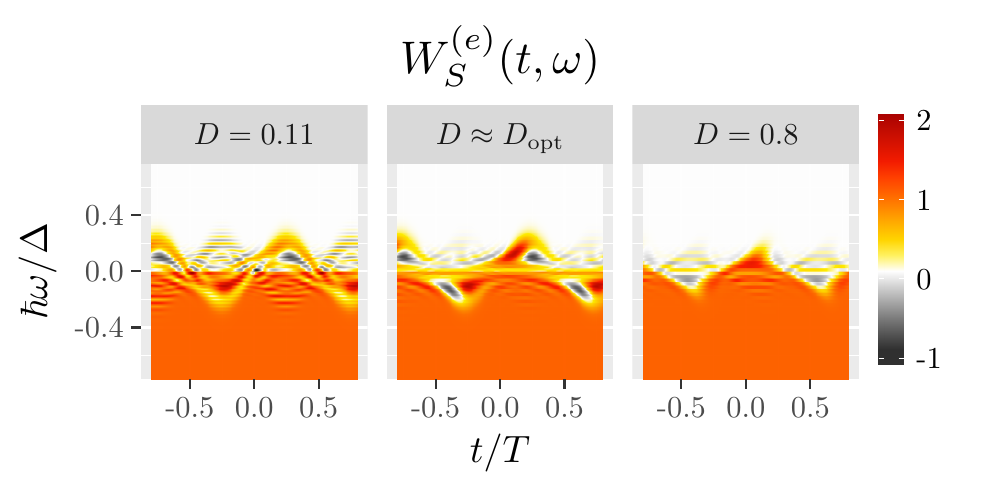} 
	\caption{\label{fig:capa-meso:examples:sinus:wigners} 
	Density plots of
	the full Wigner distribution function $W_S^{(e)}(t,\omega)$ for the
	sine-drive case as a
	function of $t/T$ and $\hbar\omega/\Delta$ for the three round
	points appearing on \cref{fig:capa-meso:examples:sinus:entropy-map}.}
\end{figure}

\Cref{fig:capa-meso:examples:sinus:entropy-map} presents a density plot of
the entropy defined by \cref{eq:eh-entanglement:entropy} as a function of $D$ and
$eV/\Delta$ at fixed $\Delta/hf=20$. There are shallow zones with minima in
each square $eV/\Delta\in ]n,n+1]$ ($n\in\mathbb{N}$) and $0<D\leq 1$.
A global minimum can be found at
$eV_{\text{opt}}/\Delta$ slightly less than $\num{0.24}$ and $D_{\text{opt}}
\approx \num{0.38}$ 
and the corresponding entropy is
very low: $\SI{0.06}{\bit}$. As we shall see in Secs. \ref{sec/capacitor/sinusoidal/entropy}
and \ref{sec/capacitor/sinusoidal/Wannier}
this is a regime where the
mesoscopic capacitor behaves almost ideally, emitting exactly one 
electron and one hole excitation per period. Decreasing $D$ from this
value leads to a
local maximum of the entropy (for $D\simeq 0.11$) before a decrease to
zero when
$D\rightarrow 0$ corresponding to a regime where the source emits
nothing. 
For each of the three points 
located at the same value of $eV/\Delta$ and corresponding to $D=0.8$,
$D=D_{\text{opt}}$ and $D\simeq 0.11$, the full electronic
Wigner distribution function is depicted on 
\cref{fig:capa-meso:examples:sinus:wigners}. As for the square drive
case, interference fringes, characteristic from inter-period electronic
coherence as well as
for electron/hole coherences, are visible for $D\simeq 0.11$ and to a
lesser extent for $D=0.8$ whereas they are much more discrete for
$D=D_{\text{opt}}$.

\begin{figure}
	\centering
	\includegraphics{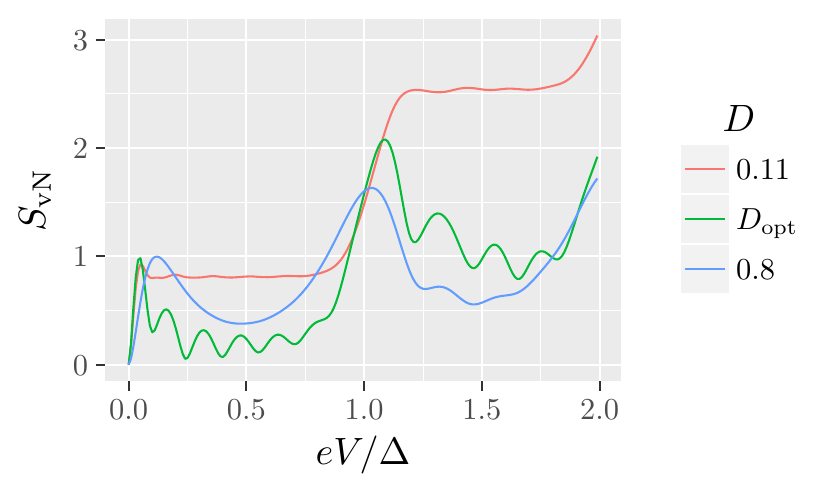}
	\caption{\label{fig:capa-meso:examples:sinus:entropy-cuts} Cuts of
	the entropy $S_{\text{vN}}$ in the 
	sine-drive case for the three horizontal lines corresponding to
	$D=0.11$, $D=D_{\text{opt}}$ and
	$D=0.8$ on \cref{fig:capa-meso:examples:sinus:entropy-map} 
	as functions of $eV/\Delta$.}
\end{figure}

The local minima on \cref{fig:capa-meso:examples:sinus:entropy-map} correspond to 
quite low values of the electron/hole entanglement entropy. They can
also be seen on \cref{fig:capa-meso:examples:sinus:entropy-cuts}
presenting cuts for fixed value of $D$ of $S_{\text{vN}}$ as functions
of $eV/\Delta$. By running a simplex minimization algorithm, we can find
position and entropy value at each minimum as summarized on
\cref{table:entropy:cos:pos}.

\begin{table}[h]
\caption{Positions in the $(D,eV/\Delta)$ plane 
and values of $S_{\text{vN}}$ (in bits) for the entropy minima --
	crosses on \cref{fig:capa-meso:examples:sinus:entropy-map} -- in the
sine-drive case for $eV/\Delta
\le 1$ (when about one electron per period is emitted).
At each of these operating points, the source emits a single electronic
atom of signal per period whose Wigner representation
is depicted \cref{fig:capa-meso:examples:sinus:wannier-minimas}. The
associated hole atom of signal is charge conjugated and shifted by a
half-period.
}

\label{table:entropy:cos:pos}
\setlength{\tabcolsep}{7pt}
\renewcommand{\arraystretch}{1.2}
\begin{tabular}{@{\hskip 2pt}cccc@{\hskip 2pt}}
	\toprule
	& $D$ & $eV/\Delta$ & $S_{\text{vN}}$ \\
	\colrule
	1 & 0.29 & 0.09 & 0.10 \\
	2 & 0.38 & 0.24 & 0.06 \\
	3 & 0.41 & 0.40 & 0.06 \\
	4 & 0.43 & 0.55 & 0.10 \\
	5 & 0.42 & 0.70 & 0.18 \\
	\botrule
\end{tabular}
\end{table}
There are also local minima in the second square where $1<eV_D/\Delta\leq 2$ but
the corresponding entropy values are higher (above $\SI{0.3}{\bit}$). 
In this zone we send three electrons and three holes
per period. As such, it is not surprising that the purity of the source
is lower, since we expect to excite more electron/hole pairs.

\subsubsection{The Floquet-Bloch spectrum}
\label{sec/capacitor/sinusoidal/spectrum}

\begin{figure}
	\includegraphics{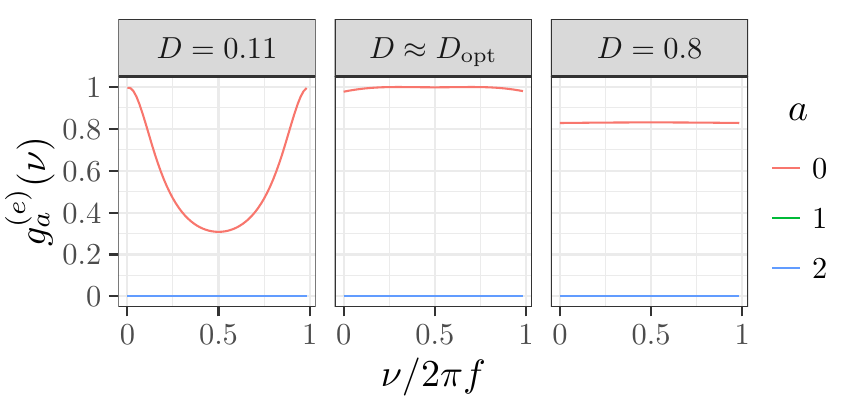}
\caption{\label{fig:capa-meso:examples:sinus:spectrum}
The Floquet--Bloch spectra for the three selected round
points in the sine-drive case appearing on \cref{fig:capa-meso:examples:sinus:entropy-map}.
Only the first three bands are represented, all the other ones 
being even closer to zero. 
}
\end{figure}

Let us review the Floquet-Bloch spectra for the three round points 
marked on \cref{fig:capa-meso:examples:sinus:entropy-map}.
The middle and right panels of \cref{fig:capa-meso:examples:sinus:spectrum}
depict flat bands. 
The middle panel
corresponds to the absolute minimum of the entropy, shows one band with
average very close to one. This corresponds to the
best operating point as a single-electron source. 
Opening the dot ($D=0.8$, right panel) also leads to flat bands as
expected but we note that the eigenvalues for the first band (which is
the only one that is non negligible) is only $0.83$.

Going to a closed dot ($D=0.11$, left
panel) leads to a curved first band with average $0.5$. 
This point
corresponds to  the local maximum of the entropy
between $D=0$ and $D=D_{\text{opt}}$ along
$eV=eV_{\text{opt}}$. 
At this point, the entropy is equal to $\SI{0.85}{\bit}$.
Starting from the optimal point, decreasing $D$ increases the escape time
of the electron and hole excitations. In previous publications
\cite{Degio-2010-4,Roussel-2016-2}, we had argued that, in a specific regime,
the mesoscopic capacitor emits
a quantum superposition of nothing and of an
elementary electron/hole pair on top of the Fermi sea. Decreasing $D$
would
increase the amplitude of the emission of the electron/hole pair
from modulus very close to one to modulus zero and this explains the
behavior of the entropy with decreasing $D$ at fixed $eV/\Delta$.
However, when $D$ is decreased, inter-period
coherences (or equivalently band curvature) appear due electron and hole delocalization 
over more than one half-period.
This shows how
our analysis unravels what happens more precisely than the  previously used simple 
picture.

\subsubsection{Electronic atoms of signal and coherences}
\label{sec/capacitor/sinusoidal/Wannier}

\begin{figure}
	\includegraphics{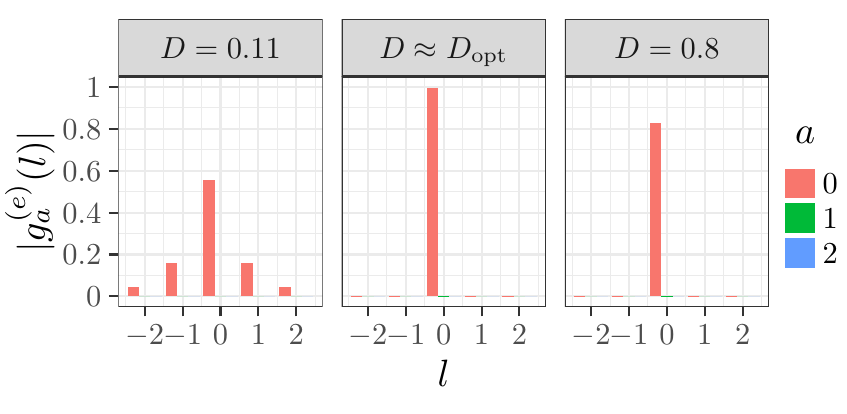}
	\caption{
		\label{fig:capa-meso:examples:sinus:e-coherences}
		Modulus of the interperiod coherences $|g^{(e)}_a(l)|$ between the electronic atoms
		of signal of the $a=0$, $1$ and $2$ Floquet-Bloch bands given by
		\cref{eq:Wannier:pa-ee} as a function of $l$ for the three
		round points on \cref{fig:capa-meso:examples:sinus:entropy-map} 
		(sine-drive case).
	}
\end{figure}

Let us now discuss the electronic atoms of signal as well
as their coherence properties at the same three round points. As shown on 
\cref{fig:capa-meso:examples:sinus:e-coherences}, for a
widely open dot, there is still one type of electronic atom of signal
with no inter-period correlations that is
emitted per half period, although it is emitted with a probability less than one.
When closing the dot, we first encounter an optimal point ($D\sim
D_{\text{opt}}$) where only one
is emitted almost certainly: the mesoscopic capacitor behaves like an
almost ideal single-electron source(see \cref{eq/SES/MB-state}) and there are no inter-period
electronic coherences (see \cref{eq:ideal-source:LPA}). Finally, when 
closing the dot, the electronic escape time
from the dot
increase beyond $T/2$ and, consequently,
the elementary electron and hole
excitations emitted by the capacitor tends to delocalize over more than
one period. Moreover, electron hole coherences are generated and 
we encounter a point with a local maximum of
electron/hole entanglement ($D\simeq 0.11$). 
Analyzing the shape of the Wannier wavepackets
confirms that closing the dot leads to longer wavepackets.

\begin{figure}
	\includegraphics[width=8.6cm]{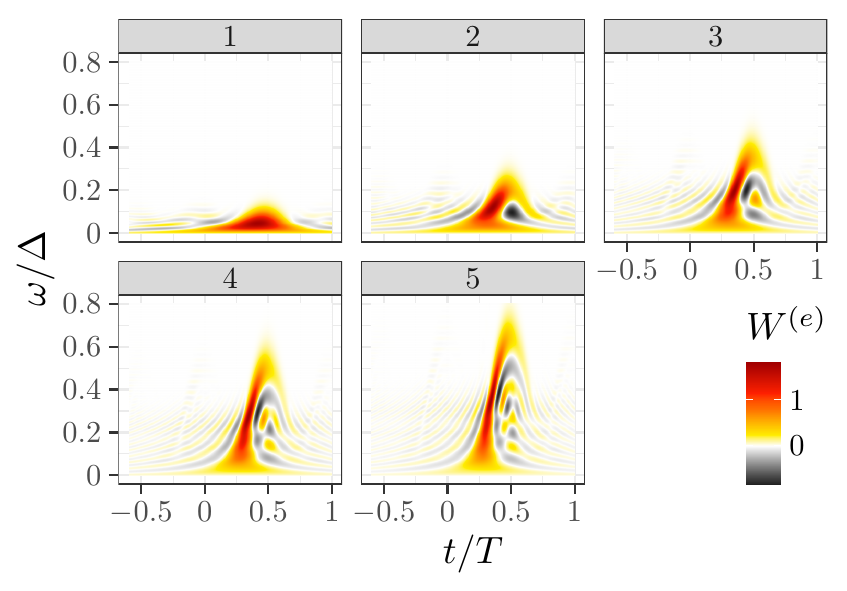} 
	\caption{\label{fig:capa-meso:examples:sinus:wannier-minimas}
		Dominant electronic atoms of signals emitted by the mesoscopic
		capacitor for the local minima of $S_{\text{vn}}$ appearing on
		\cref{fig:capa-meso:examples:sinus:entropy-map} (crosses) in the domain
		$0<D<1$ and $0<eV/\Delta<1$.
	}
\end{figure}

\Cref{fig:capa-meso:examples:sinus:wannier-minimas} presents the
dominant electronic atoms of signal for the local optimal points in the
quadrant $0<D<1$ and $0<eV/\Delta<1$. As we raise the drive amplitude,
the Wannier wavefunctions explore higher energies. For each minimum,
their is a corresponding number of negative bumps in the Wigner
representation. This suggests that these optimal regimes correspond to a 
resonance between the rising time of the drive voltage, the period and
the energy gap of the cavity.

In conclusion, this analysis demonstrates how the Floquet--Bloch
analysis can be used to find optimal operating points of single electron
sources and, more generally, to characterize what is emitted by the
source and to optimize wavepackets shaping strategies
\cite{Misiorny-2018-1}. As such, it can help improving the quantitative modeling of many electron
quantum optics experiments such as, for example, electronic decoherence experiments
\cite{Marguerite-2016-1}. 
Because the electronic excitations emitted by
the mesoscopic capacitor are quite sensitive to this phenomenon, and
also for practical reasons, the Leviton sources built from a suitably
driven Ohmic contact \cite{Dubois-2013-2}. We shall now apply our
analysis to this very important source.

\subsection{Leviton trains}
\label{sec/Levitons}

Let us now consider an Ohmic contact driven by time dependent
voltage which is a $T$-periodic train of Lorentzian pulses of width
$\tau_0$, each of them carrying an electric charge $q=-\alpha e$. 
The resulting time-dependent voltage 
\begin{equation}
V(t)=\frac{\alpha hf}{2e}\,
\frac{\sinh{(2\pi f\tau_0)}}{\sinh^2(\pi f\tau_0)+\sin^2(\pi ft)}
\end{equation}
has a d.c.\@ component
$V_{\mathrm{dc}}=\alpha hf/e$ and an a.c.\@ part
$V_{\text{ac}}(t)=V(t)-V_{\text{dc}}$ \cite{Dubois-2013-1}.

To understand the underlying physics, let us remember what happens in
the case of a single Lorentzian pulse of duration $\tau_0$ and
integer charge $\alpha=n>0$ at zero temperature. In this case,
the emitted many-body state is a Slater determinant built by adding on the
Fermi sea $n$ mutually orthogonal electronic single-electron excitations
whose wavefunctions are given in the frequency domain
by~\cite{Grenier-2013-1}:
\begin{equation}
\label{eq/Levitons/one-pulse}
\varphi_n(\omega)=\sqrt{4 \pi v_F \tau_0}\,\heaviside(\omega)\,
L_{n-1}(2\omega\tau_0)\,
\me^{-\omega\tau_0},
\end{equation}
where $L_n$ denotes the $n$th Laguerre polynomial and $\heaviside$ is
the Heaviside distribution. 
In the limit where
Lorentzian pulses are well separated $(f\tau_0\ll 1)$, we expect the
electronic atoms of signal, which we call Levitonoids, to have a strong overlap with these mutually orthogonal
wavefunction. 

For a Leviton train ($\alpha=1$), one could naively expect
each Lorentzian voltage pulse to carry exactly one
Levitonoid excitation. Although, this Levitonoid may tend to the isolated
Leviton in the limit $\tau_0\ll T$, 
in the case where the Lorentzian pulse start to overlap
($f\tau_0 \gtrsim 1$), the relation between the Levitonoid and the
Leviton is non-trivial because 
of the Pauli principle
since single-Leviton wavefunctions of width $\tau_0$ separated by
$T\ll \tau_0$ are not orthogonal.

To gain a better understanding of the way information is encoded in such
a compact electronic train, besides the
periodic train of Levitons, we shall consider in
\cref{sec/Levitons/random} a randomized train of 
Lorentzian voltage pulses \cite{Glattli-2018-1}, obtained by randomly choosing 
wether or
not each Lorentzian pulse is present in the drive or not. The single-electron coherence
associated with this statistical ensemble of voltage drives is still
$T$-periodic and our analysis can be applied.

\subsubsection{Levitonic atoms of signals}
\label{sec/Levitons/charge-1}

\begin{figure}
	\includegraphics{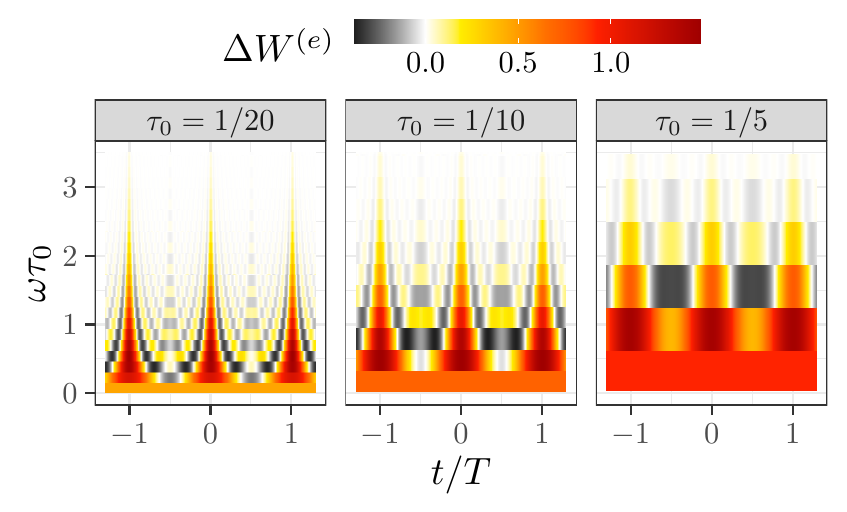}
	\caption{
		Wigner distribution function of a Leviton train for different ratio
		$\tau_0/T$ for increasing values of $\tau_0/T$.
		As we raise $\tau_0/T$, the duration of each Leviton becomes
		longer and longer and, compared to the energy scale
		$\hbar/\tau_0$, $hf$ becomes larger. Once $\tau_0$ is greater
		than
		$T$, the first band of width $2\pi f$, which has no
		time dependance, is the only one to remain. We are thus left
		with
		an almost stationary situation due to the raise of chemical
		potential by $\delta\mu=hf$.
	}
	\label{fig:signal-processing:levitons:1:all}
\end{figure}

\Cref{fig:signal-processing:levitons:1:all} shows the full Wigner
distribution function
of the $T$-periodic train of unit charge Lorentzian pulses 
for different values of $f \tau_0$. Varying this parameter swipes
from a dilute train in which each
Leviton is well separated from each other, to a compact train in which
the pulses are so spread over multiple periods that we only see
the variation of the chemical potential due to the d.c.\@ part.

\begin{figure}
	\includegraphics{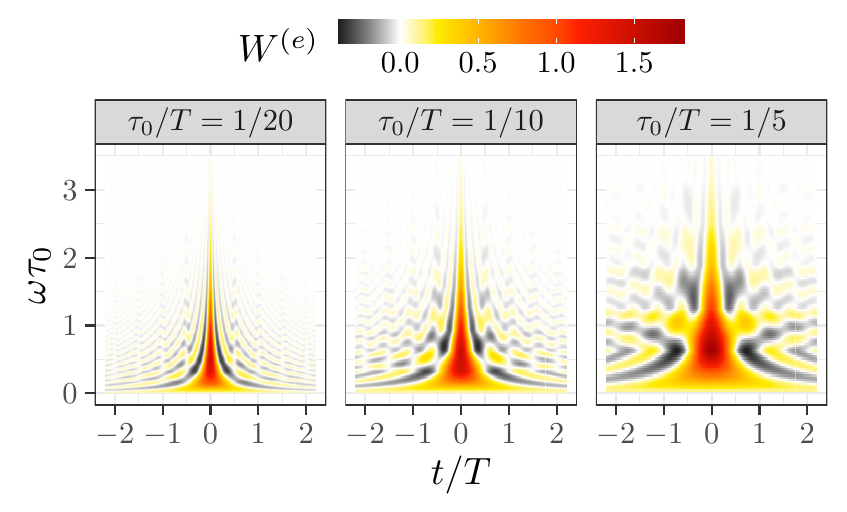}
	\caption{
		The electronic atoms of
		signal of a train of charge $\alpha$ Lorentzian pulses for $\alpha=1$
		and different values of $f\tau_0$. When $f\tau_0 \ll 1$, the
		wavepacket we obtain is very similar to a Leviton. When $f
		\tau_0 \simeq 1$, we recover a Martin-Landauer wavepacket. The
		atoms of signal found by our algorithm (lower panel) fit
		perfectly the ones predicted analytically (upper panel).
	}
	\label{fig:signal-processing:levitons:1} 
\end{figure}

In the $\alpha=1$ case, Moskalets has obtained explicit expression
for electronic atoms of signal associated with such a Leviton
train \citep{Moskalets-2015-1}. 
Each of them leads to a
Lorentzian current pulse of width $\tau_0$. This is
manifestly not the case for the electronic atoms of signal obtained
numerically whose Wigner representations are depited on
\cref{fig:signal-processing:levitons:1}.
Our numerical
algorithm produces wavepackets having the smallest spreading
in time whereas the analytical expressions obtained by Moskalets
possess a Lorentzian current pulse of width $\tau_0$. As shown in
\cref{appendix/levitonoids}, an analytical expression for the
minimally spread
wavepackets can be obtained:
\begin{equation}
	\label{eq/sigproc/levitonoid/minimal}
	\varphi_{\text{Lev}}(\omega) = \frac{1}{\sqrt{\mathcal{N}}}
	\heaviside(\omega)
	\me^{-\omega_{\text{int}} \tau_0},
\end{equation}
where $\mathcal{N}$ is a normalisation factor and $\omega_{\text{int}} =
2 \pi f \lfloor \omega/2\pi f \rfloor$ is the frequency counted in
multiple of $2 \pi f$. This minimally spread Levitonoid is the following
linear combination of Martin--Landauer's wavepackets
\begin{equation}
\ket{\text{Lev}}=\sqrt{1-\me^{-4\pi f\tau_0}}\sum_{n=0}^{+\infty}
	\me^{-2\pi nf\tau_0}\ket{\text{ML}_{n,0}}\,.
\end{equation}
The details of this derivation can be found in
\cref{appendix/levitonoids}. We have checked 
that this analytical form and
the one found by the algorithm match perfectly.

As can be seen from \cref{fig:signal-processing:levitons:1},
the minimally-spread Levitonoid tends to the
single-Leviton state $\ket{\varphi_1}$ obtained from
\cref{eq/Levitons/one-pulse} in the $f\tau_0\rightarrow 1$ limit. 
When we lower $f \tau_0$, the steps of width $2\pi f$
in \cref{eq/sigproc/levitonoid/minimal}
become smaller, thereby corresponding to an increasingly closer
staircase approximation of the decaying exponential. A measure of the
distance between
the minimal Levitonoid $\ket{\text{Lev}}$ and the single-Leviton 
state
$\ket{\varphi_1}$
is given by the overlap between these two single-particle states:
\begin{equation}
	\label{eq/Levitonoids/Leviton-overlap}
	\left|
		\Braket{\text{Lev} | \varphi_{1}}
	\right|^2
	=\frac{\tanh(\pi f\tau_0)}{\pi f\tau_0}
\end{equation}
which, for $f \tau_0 \ll 1$, departs quadratically from
unity.

In the regime where $f\tau_0\simeq 1$, the overlap between the minimal
Levitonoid and the single Leviton tends to zero as $1/\pi f\tau_0$. 
In this regime, it seems natural to compare our Levitonoids
to the Martin-Landauer wavepacket $\ket{\text{ML}_{1,0}}$
(compare the right panel of
\cref{fig:signal-processing:levitons:1} to
\cref{fig/ML-Wigner}). 
This overlap goes
exponentially to one as $f \tau_0$ goes to infinity
\begin{equation}
	\label{eq/Levitonoids/ML-overlap}
	\left|
		\Braket{\text{Lev} | \varphi_{\text{ML}_{0,0}}}
	\right|^2
	=
	1 - \me^{-4 \pi f \tau_0}.
\end{equation}
For the examples discussed above, when
$f \tau_0 = 1/5$, the overlap is around $\SI{92}{\percent}$. At $f
\tau_0 = 1$, the overlap is unity up to the sixth significative digit,
making the differentiation between a minimally-spread Levitonoid and a Martin-Landauer
impossible in practice.  

These behaviors shed light on the difference in terms of
typical temporal width between the minimally-spread Levitonoids and the wavepackets
introduced by Moskalets. For Moskalets' wavepackets, the typical
duration is always $\tau_0$. In our case, the typical duration is
$\tau_0$ when $\tau_0 \lesssim T$. However, when
$\tau_0\gtrsim T$, the minimal Levitonoid will have a
duration of the order of $T$ and, ultimately, in the $f\tau_0\rightarrow
0$ limit, tend  to
a Martin-Landauer wavepacket.

\subsubsection{The random train}
\label{sec/Levitons/random}

In order to distinguish between the electronic
wavepackets used to carry the information and the way they are injected,
we elaborate on the recent idea \cite{Glattli-2018-1} of randomizing the
emission process itself. We consider non-periodic trains of
electrons associated with infinite random binary chains $b_k$
($k\in\mathbb{Z}$) which determines whether a
Lorentzian pulse centered at $t_k=kT$ is added to the driving voltage
($b_k=1$ with probability
$p$) or not ($b_k=0$ with probability $1-p$): 
\begin{equation}
\label{eq/Levitons/random/voltage-drive}
V_d(t)=\sum_{k\in\mathbb{Z}}b_kV_{\text{Lev}}(t-kT)\,
\end{equation}
in which $V_{\text{Lev}}(t)$ corresponds to a Lorentzian pulse of width
$\tau_0$ carrying a charge $-e$ centered at $t=0$. Even if pulse
emission is randomized, the ensemble average properties of the electron
stream are still $T$-periodic
because emission is still centered on times $t_k=kT$ for
$k\in\mathbb{Z}$.
The average single-electron
coherence in the time domain has been evaluated
as~\cite{Glattli-2018-1}
\begin{widetext}
\begin{equation}
\label{eq/Levitons/random/coherence}
\mathcal{G}^{(e)}_{\text{R}_p}\left(t+\frac{\tau}{2}\big| t-\frac{\tau}{2}\right)
=\frac{\sin\left[\pi (ft-\theta_p(f\tau))\right]\,
\sin\left[\pi (ft+\theta_p(f\tau))\right]}{\sin\left[\pi f(t-\mi\tau_0+\frac{\tau}{2})\right]\,
\sin\left[\pi f\left(t+\mi\tau_0-\frac{\tau}{2}\right)\right]}\,\mathcal{G}_F^{(e)}(\tau)
\end{equation}
\end{widetext}
in which the index $\text{R}_p$ stands for ``randomly emitted with probability $p$'',
$\mathcal{G}_F^{(e)}$ denotes the Fermi sea's single-electron
coherence and
\begin{equation}
	\label{eq/Levitons/random/coherence/theta}
	\theta_p(x)=\sqrt{\frac{x^2}{4}-(f\tau_0)^2-\mi(1-2p)f\tau_0 \, x}\,.
\end{equation}
Eqs. \eqref{eq/Levitons/random/coherence} and
\eqref{eq/Levitons/random/coherence/theta} are the starting point for applying our
Floquet--Bloch analyzis for finding the electronic atoms of signals
underlying the randomized train of Lorenztian pulses. 
Remarkably, the analysis can be performed numerically but also
analytically, as explained in \cref{appendix/random}.

The first important result is that
the excess single-electron coherence can be described in terms of 
minimal Levitonoids which are the appropriate
electronic atoms of signal for the non-random $T$-periodic train. 
This illustrates quantitatively the motivation put forward in \citep{Glattli-2018-1}:
randomization enables us to separate what is emitted from the way
it is emitted. Single electron coherence is thus described in terms of the same
electronic atoms of signals but with a different ``quantum coherence
score''. 

\begin{figure}
	\includegraphics{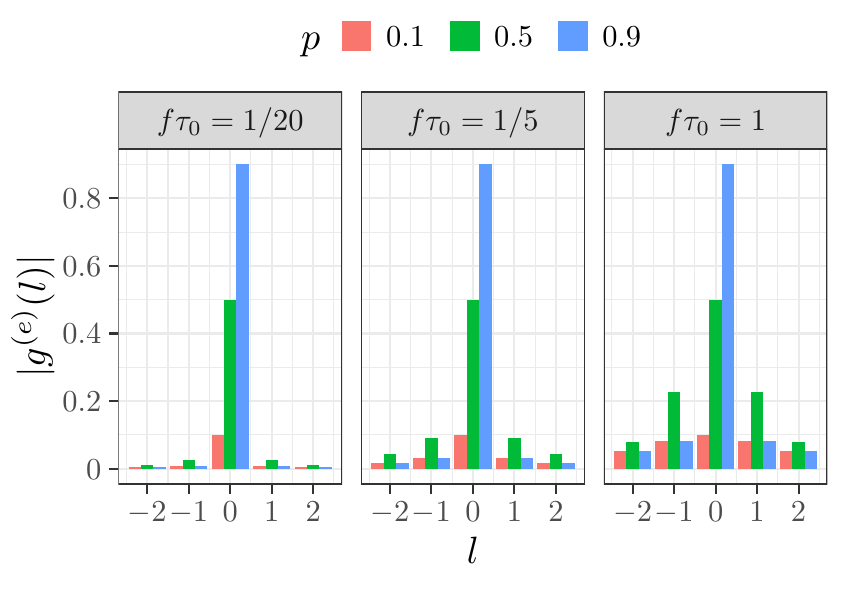}
	\caption{\label{figure/random/coherences}
		Coherence between time-shifted Levitonoids for the random
		train with emission probabilities
		$p = 0.1, 0.5, 0.9$ and a width
		$f\tau_0 = 1/20, 1/5, 1$. The central peak has value $p$ and
		we clearly see the increase of inter-period coherences when
		increasing $f\tau_0$ at fixed $p$ and their spreading
		when decreasing $p$ at fixed $f\tau_0$.
	}
\end{figure}

More precisely,
when the
pulses are widely spaced ($f\tau_0\ll 1$), we can associate a
single-electron excitation (the minimal Levitonoid) which is very close
to the
Leviton wavepacket (see \cref{eq/Levitonoids/Leviton-overlap}) 
with each Lorentzian pulse. Lowering $p$ then
just lowers the emission 
probability of the corresponding single-electron excitation (see the
left panel of \cref{figure/random/coherences}). In this
regime,
randomization just lowers the intensity of the emission, as would
be expected with a classical ensemble of musicians choosing to play, or
not to play, one of the periodically repeated note from the ``I Gotta
Feeling'' music score. 

However, for the random Leviton train, 
lowering the emission probability of each Lorentzian pulses can lead to
subtle effects when the Pauli principle starts to enter the game, in the
$f\tau_0\sim 1$ regime or above. This is the second important result
from our detailed analysis:
although, the excess single-electron
coherence is still
described in terms of minimal Levitonoids, lowering $p$ also introduces
inter-period coherences depicted on
\cref{figure/random/coherences}. At fixed $p$, they increase with 
$f\tau_0$ as seen by comparing the three 
panels of \cref{figure/random/coherences}. In this regime, the
modification of the ``quantum coherence score'' induced by lowering $p$
is not naively classical as in the $f\tau_0\ll 1$ regime: interperdiod
coherences are revealed. 
This can be understood as follows: the
limit of a dense $T$-periodic train is recovered for $p\rightarrow 1$: 
one Levitonoid is then
emitted per period without any inter-period coherences. But then, decreasing $p$
opens some space on the adjacent periods: among all the classical drives 
building the statistical ensemble underlying $\text{R}_p$, the weight of
those containing pulses separated by more than $T$ increases and this
contributes to the increasing weight of inter-period coherences. 
In the limit
$p\rightarrow 0$, we thus expect to recover an excess electronic coherence
spreading over $|\tau|\lesssim \tau_0$, very similar to the one of an
isolated Leviton because, in this limit, the weight of trajectories for
which an emitted Lorentzian pulse is separated from the nearest other
emitted pulses by more
than $\tau_0$ goes to unity. This explains the increasing inter-period
coherences in the low $p$, high $f\tau_0$ regime.

\begin{figure}
	\includegraphics{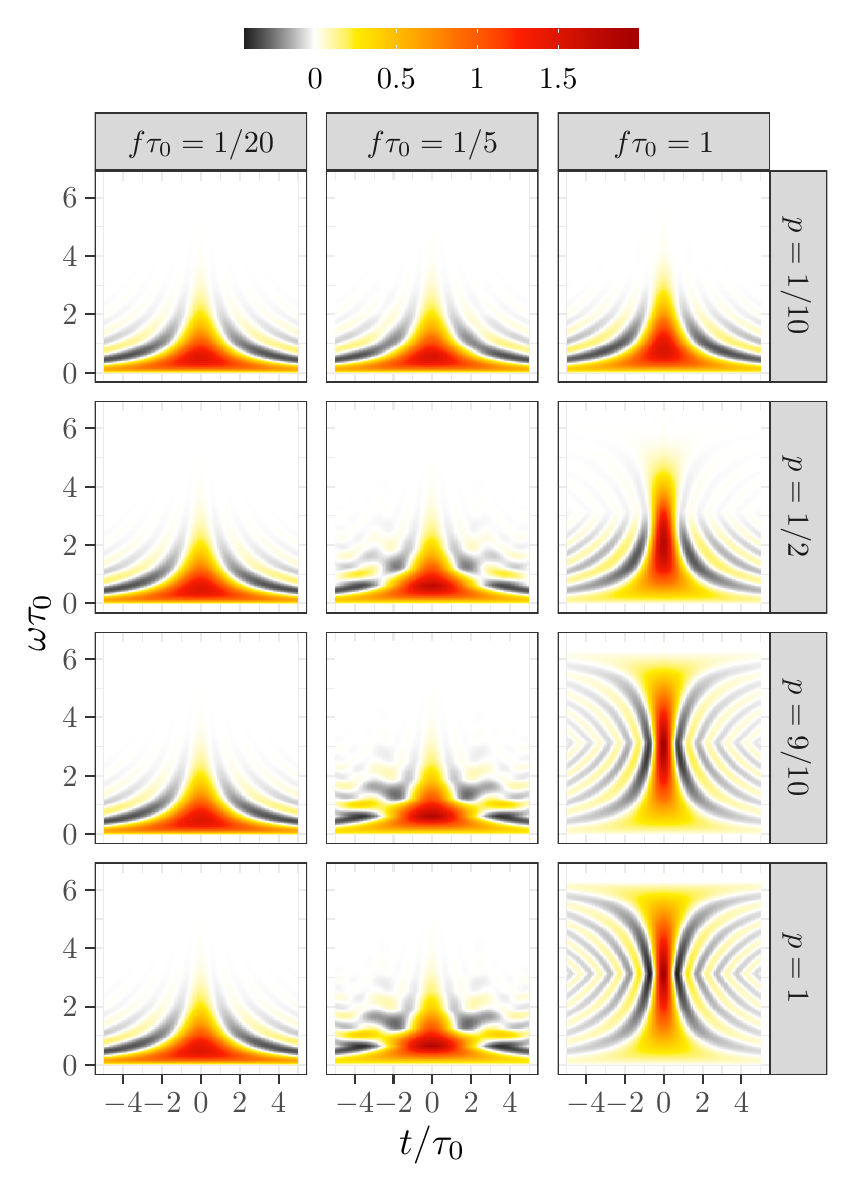}
	\caption{\label{fig/random/Wigner-of-Glattlions}
		Wigner representations of $p$-Glattlions for
		$p = 0.1$, $0.5$, $0.9$, $1$ and a width
		$f\tau_0 = 1/20$, $1/5$ and $1$. The case
		$p = 1$ corresponds to the minimal Levitonoids introduced 
in \cref{sec/Levitons/charge-1}. 
Note that, on this figure, time is counted in units of $\tau_0$ so that
we clearly see
how the $p$-Glattlion
interpolates between a Leviton-like Wigner representation of width $\tau_0$ for $p=1/10$ to 
a minimal Levitonoid one for $p=9/10$. 
	}
\end{figure}

Remarkably, and this is the third result from our in-depth analysis,
these interperiod coherences can be
recasted in terms of normalized
single-electron states, which we call the
$p$-Glattlion and which are 
\begin{equation}
	\label{eq/Levitons/random/Glattlion}
	\ket{\text{Gla}_p}
    =\int_0^{+\infty} %
    \sqrt{\frac{1-\me^{-4 \pi f \tau_0}}{p f}}
    \me^{-\omega_{\text{int}}\tau_0}
    \sqrt{g^{(e)}(\omega)}\,\ket{\omega}\,\md\omega\,.
\end{equation}
Using these single-particle states, the excess single-electron coherence
can be written as
\begin{equation}
	\label{eq/Levitons/random/Glattlion-resummation}
	\Delta_0\mathbf{G}_{\text{R}_p}^{(e)}
    =
    p
    \sum_{l \in \mathbb{Z}}
    \ket{\text{Gla}_{p,l}} \bra{\text{Gla}_{p,l}}
\end{equation}
in which $\ket{\text{Gla}_{p,l}}=\mathbf{T}_T^l\ket{\text{Gla}_p}$ is
the $p$-Glattlion translated by
$l T$. This rewriting resums the interperiod coherences in the
minimal Levitonoid basis into pure single-electron states. Ultimately,
such a rewriting reflects the coherence of the voltage pulses trains used to
build the random ensemble $\text{R}_p$. The price to
pay is that these states cannot be viewed as electronic atoms of signals
since they are not mutually orthogonal between different periods (see
\cref{eq/Glattlions/scalar-products}).
The Wigner representations of $\text{Gla}_p$ for various $(p,f\tau_0)$ are
plotted on \cref{fig/random/Wigner-of-Glattlions}.
These single-particle states interpolate 
between Leviton wavefunctions of width $\tau_0$ in the limit 
$f\tau_0\ll 1$ or
$p\ll 1$ and the minimal Levitonoid obtained for $p=1$.

\section{Conclusion and perspectives}
\label{sec:signal-processing:conclusion}

In this work, we have introduced a representation of the single-electron
coherence of a periodic electron source in terms of perfectly
distinguishable normalized single-particle
wavefunctions associated with each period which we call ``electronic
atoms of signals'' \cite{Roussel-2016-2}. This description, which is
the counterpart of the Karhunen-Loève decomposition for classical
signals~\cite{Book-Papoulis-Signal}, enables us to obtain a simple
description of the single-particle content emitted by the source in
discrete terms. The electronic
atoms of signal are the building blocks of the single-electron coherence
which are emitted according to their emission
probabilities and quantum coherences. Such a decomposition is very
reminiscent of the way
music can be described in terms of notes
arranged along a specific score: the emission probability being the
analogous of the strength at which the note is played whereas the 
coherences are specifically quantum. This type of
decomposition, generalized to non-periodic quantum electrical currents
is a convenient way to represent general
single-electron quantum signals exactly as a music score represents
a generically non-periodic piece of music. 
This is the appropriate framework to discuss the encoding 
and decoding of the quantum
information embedded within a quantum electrical current.

Being able to access the single-particle content of a quantum electric
current suggests that a very high degree of control 
may be envisioned in the near future. This is
particularily important for the potential
applications of electron quantum optics to quantum sensing of
electromagnetic fields on a sub-micrometric space 
and sub-nanosecond time scale.

In particular, our study of electron sources also shows that, generically, an electron
source emits several electron or hole wavefunctions. The multiplicity of
emitted excitations is enhanced by non-zero temperature as shown in the
recent experimental study \cite{Bisognin-2019-2}. From a signal
processing point of view, this means that in general, electron quantum
optics sources and detectors respectively emit and detect
many different electron and hole excitations. In
this sense, they are quantum counterparts to
MIMO (Multiple-Input Multiple-Output)
classical microwave devices such as advanced radars and Wi-Fi routers, which 
make use of many (spatial) modes to improve
transmission or detection performances. 
In the long run, the representation of electonic coherence in terms of
electronic atoms of signal will be instrumental for characterizing and
improving the 
performances of quantum sensing devices based on quantum electric
currents, exactly as MIMO is now used in radar
technology \cite{Bliss-2003-1}. It may 
as well help quantifying and maybe improving quantum information flow
within these
devices, as was done in classical signal
processing \cite{Raleigh-1998-1}.

This decomposition may also bring new insights on
physical phenomena such as
electron fractionalization~\cite{Grenier-2013-1,Freulon-2015-1}, the
effect of temperature on trains of multiparticle states
\cite{Moskalets-2018-1} and interaction-induced electronic
decoherence \cite{Marguerite-2016-1,Rodriguez-2020-1}. Since
the Floquet-Bloch decomposition provides a zero-order guess 
for the many-body state from
single-particle coherence in the absence of interactions, it is the
perfect starting point for a more refined description of the many-body
states based, for example, on the adaptation to electron quantum optics of the unitary coupled
cluster method now used in variational quantum eigensolvers
\cite{Romero-2018-1}. Such an ansatz would reproduce 
deviations
from Wick's theorem at higher
and higher order coherences, thereby providing a clear insight of the electronic
coherences in terms of the many-body state.

Finally, our quantum analyzer may
also offer a way to access to the recently studied
electron/hole
entanglement~\cite{Dasenbrook-2015-1,Hofer-2016-2,Dasenbrook-2016-1}
and,
supplemented by other measurements \cite{Thibierge-2016-1}, to quantify
more precisely the importance of interaction-induced
higher order quantum correlations as well as of thermal
fluctuations~\cite{Moskalets-2017-1,Moskalets-2018-3}. 

The general
quantum signal processing method presented here is also directly relevant for
electron quantum optics in other systems such as topological insulators
\cite{Ferraro-2014-1} and, with some adaptation, in strongly correlated
1D quantum edge channels such as fractional quantum Hall edges
\cite{Ferraro-2017-1} where it might shed some light on recently
predicted correlation effects within trains of Lorentzian pulses
\cite{Ronetti-2018-1}.
Finally, it can establish a bridge between electron and
microwave quantum optics~\cite{Gasse-2013-1,Forgues-2014-1,Grimsmo-2016-1,Virally-2016-1}, by probing 
the electronic content of
microwave photons injected from a transmission line into
a quantum conductor. However, this requires establishing a bridge between
the coherence properties of electrons and the quantum optical coherence of the emitted
radiation extending the work of Ref. \cite{Ferraro-2018-1}. 

\acknowledgments{We thank E. Bocquillon,
P. Borgnat and P. Flandrin for useful discussions.
This work is supported in part by the ANR grant ``1shot reloaded''
(ANR-14-CE32-0017), the ERC Consolidator grant ``EQuO'' and by the Joint Research Project “SEQUOIA” (17FUN04)
within the European  Metrology  Programme  for Innovation and 
Research (EMPIR) co-financed by the Participating States and from the European Union’s 
Horizon 2020 research and innovation programme.
}

\appendix
\section{Normalizations}
\label{appendix:conv:G:operator}

The single-particle states $|t\rangle$ and $|\omega\rangle$, normalized as
\begin{subequations}
\begin{align}
\langle t|t'\rangle &= v_F^{-1}\delta(t-t')\\
\langle \omega|\omega'\rangle &= \delta(\omega-\omega')
\end{align}
\end{subequations}
are related by
\begin{subequations}
\begin{align}
|t\rangle &= \frac{1}{\sqrt{2\pi v_F}}\int \md\omega\,\me^{\mi\omega
t}|\omega\rangle\\
|\omega\rangle &= \sqrt{\frac{v_F}{2\pi}}\int \md t\,\me^{-\mi\omega
t}|t\rangle
\end{align}
\end{subequations}
Using the expression of the fermion field operator
\begin{equation}
\psi(t)=\int_{\mathbb{R}} c(\omega)\,\me^{-\mi\omega
t}\frac{\md\omega}{\sqrt{2\pi v_F}}\,
\end{equation}
in terms of fermionic annihilation and creation operators $c(\omega)$
$c^\dagger(\omega)$ obeing the canonical anticommutation relations
$\{c(\omega),c^\dagger(\omega')\}=\delta(\omega-\omega')$, 
the $\mathbf{G}^{(e)}$ operator is expressed in
the $\ket{\omega}$ base as
\begin{equation}
\mathbf{G}^{(e)} 
= \int_{\mathbb{R}^2}  
|\omega_+\rangle\, \langle
c^\dagger(\omega_-)\,c(\omega_+)\rangle_\rho \langle
\omega_-|\,\md\omega_+\,\md\omega_-
\,.
\end{equation}

\section{Floquet-Bloch theory}
\label{appendix:Floquet}

\subsection{Diagonalizing the electron part}
\label{appendix:Floquet:electron}

Let us introduce
the projectors
$\mathbf{\Pi}_\pm$ on the space of
positive (resp.\@ negative) energy single-particle states. The
projections
$\mathbf{G}^{(e)}_{\varepsilon,\varepsilon}\ =
\mathbf{\Pi}_\varepsilon\,\mathbf{G}^{(e)}\,\mathbf{\Pi}_\varepsilon$ ($\varepsilon=\pm 1$)
of the single-electron coherence operator contain
information on 
electronic excitations for $\varepsilon=+$ and
on hole excitations for $\varepsilon=-$. These correspond to the
electron and hole quadrants of \cref{fig/eh-quadrants}. 
In the same way, the off-diagonal parts 
$\mathbf{G}^{(e)}_{\varepsilon,-\varepsilon}=\mathbf{\Pi}_\varepsilon\,\mathbf{G}^{(e)}\,\mathbf{\Pi}_{-\varepsilon}$
couple the electron and hole parts of the
single-particle state and encode electron/hole coherences. 

Note that
$\mathbf{G}^{(e)}_{++}$ contains all the electronic excitations, even
the thermal ones that are present at non-zero temperature when the
source is switched off. Keeping these is essential for having positive
operators to diagonalize. Denoting by $\Delta_0\mathbf{G}_{(e)}=
\mathbf{G}^{(e)}-\mathbf{G}_F^{(e)}=\mathbf{G}^{(e)}-\mathbf{\Pi}_-$,
we have
\begin{equation}
	\mathbf{G}^{(e)}=\mathbf{\Pi}_-+\mathbf{G}^{(e)}_{++}+\Delta_0\mathbf{G}^{(e)}_{--}+
	\mathbf{G}^{(e)}_{+-}+\mathbf{G}^{(e)}_{-+}
\end{equation}
where
$\mathbf{G}^{(e)}_{--}=\mathbf{\Pi}_-+\Delta_0\mathbf{G}^{(e)}_{--}$ and
$\mathbf{G}^{(e)}_{++}=\Delta_0\mathbf{G}^{(e)}_{++}$ (same for $+-$ and
$-+$).

The first step consists in diagonalizing the electron part of the excess
single-electron coherence $\mathbf{G}^{(e)}_{++}$. 
Since
$\mathbf{G}^{(e)}$ is hermitian as well as $\mathbf{\Pi}_+$,
$\mathbf{G}^{(e)}_{++}$ is also hermitian. Since
$[\mathbf{\Pi}_+,\mathbf{T}_T]=0$, $\mathbf{G}^{(e)}_{++}$ commutes with 
$\mathbf{T}_T$ and we also know that it is a positive operator bounded 
by $1$. Therefore, $\mathbf{G}^{(e)}_{++}$ and $\mathbf{T}_T$ 
can be diagonalized simultaneously.
Exactly as in solid state theory, the diagonalization is performed on
each of the eigenspaces of $\mathbf{T}_T$ which consist in
quasi-periodic single-particle states associated with a quasi-energy
$0\leq \nu<2\pi f\mathbb{Z}$ ($f=1/T$) and corresponding to the
eigenvalue $\me^{-\mi \nu T}$ of $\mathbf{T}_T$. The spectrum for 
$\mathbf{G}_{++}^{(e)}$ has a band structure with eigenvalues
$g^{(e)}(\nu)\in [0,1]$ for $0\leq 0<2\pi f$. We can therefore find an orthogonal
basis of eigenvectors
$\ket{\psi^{(e)}_{a,\nu}}\in\mathcal{H}_+$ such that
\begin{subequations}
	\begin{align}
		\mathbf{T}_T\ket{\psi^{(e)}_{a,\nu}}&=\me^{-\mi \nu
		T}\ket{\psi^{(e)}_{a,\nu}}\\
		\mathbf{G}^{(e)}_{++}\ket{\psi^{(e)}_{a,\nu}}&=g^{(e)}_a(\nu)
		\ket{\psi^{(e)}_{a,\nu}}
	\end{align}
\end{subequations}
These eigenvectors are called the electronic Floquet-Bloch vectors and
we can choose them to
satisfy the normalization conditions
\begin{equation}
	\braket{\psi^{(e)}_{a,\nu} | \psi^{(e)}_{a',\nu'} }
	=2\pi \delta_{a,a'}\delta(\nu-\nu')
\end{equation}
which is the same as the $\sqrt{2\pi}\ket{\omega}$ states. The explicit
form of the eigenvalue equations used in the numerical computation is 
discussed in \cref{appendix:Floquet:spectrum:eigenvalue-equation}. It relies on the
decomposition of each Floquet-Bloch state as a sum of plane waves
whose energies differ by a multiple of $hf$:
\begin{equation}
	\ket{\psi^{(e)}_{a,\nu}}=\sum_{n=0}^{+\infty} u^{(n)}_{a,\nu}
	\ket{\nu+2\pi nf}\,.
\end{equation}
The main difference with the usual Bloch theory in solid state physics
comes from the fact that, here, the sum is restricted to $n\in
\mathbb{N}$ because we are considering electronic excitations.

\subsection{Hole excitations and electron/hole coherences}
\label{appendix:Floquet:hole-and-eh}

Having discussed the electronic part of the single-electron coherence,
let us discuss the hole part as well as the electron/hole part. 
We can introduce a hole operator $\mathbf{G}^{(h)}$
defined by replacing $\mathcal{G}^{(e)}_{\rho,x}(t,t')$ in \cref{eq:Ge:definition} by 
\begin{equation}
\mathcal{G}^{(h)}_{\rho,x}(t,t')=\trace\left(
\psi^\dagger(x,t)\,\rho\,\psi(x,t')\right).
\end{equation}
This operator satisfies the same mathematical properties as
$\mathbf{G}^{(e)}$. This can be easily shown by using the anticommutation
relations of fermionic operators to relate electron and hole coherence
operators:
\begin{equation}
	\mathbf{G}^{(h)}=\mathbf{1}-\mathbf{C}\mathbf{G}^{(e)}\mathbf{C}^\dagger,
\end{equation}
where $\mathbf{C}$ is the anti-unitary involution that transforms
electrons in holes and vice-versa, defined in the time basis by
complex conjugation: $\braket{t | \mathbf{C}\psi}= \braket{t | \psi}^*$.
We then have $\mathbf{G}^{(h)} = \boldsymbol{\Pi}_{-} -
\mathbf{C} \Delta_0 \mathbf{G}^{(e)} \mathbf{C}^\dagger$ and therefore, 
holes can be dealt with along the same lines as electronic excitations.

However, rather than
focusing on the restriction of $\mathbf{G}^{(h)}$ to the positive frequencies quadrant, 
it turns out to be more convenient to focus on
$\Delta_0
\mathbf{G}^{(e)}_{--}$ 
defined as the restriction to the negative frequencies quadrant of $\Delta_0 \mathbf{G}^{(e)} =
\mathbf{G}^{(e)} - \boldsymbol{\Pi}_-$. Taking differences with respect
to the $\mu=0$ Fermi sea ensures that all exctitions,
including thermal ones, are taken into account. Then 
$\Delta_0
\mathbf{G}^{(e)}_{--}$ 
contains eigenfunctions of holes at negative
frequencies, with eigenvalues that are the opposite of 
hole occupation numbers. 

Exactly as $\mathbf{G}^{(e)}_{++}$, $\mathbf{G}^{(e)}_{--}$ can be
diagonalized simultaneously with $\mathbf{T}_T$. We thus introduce
an eigenbasis of hole single-particle states
$|\psi_{b,\nu}^{(h)}\rangle\in\mathcal{H}_-$
such that
\begin{equation}
	\label{eq:Bloch:diagonal:hole}
	\Delta_0 \mathbf{G}^{(e)}_{--}
	=
	- \sum_b \int_0^{2\pi f}
	g_b^{(h)}(\nu)
	\ket{\psi^{(h)}_{b,\nu}} \bra{\psi^{(h)}_{b,\nu}}
	\,\frac{\md\nu}{2\pi}\,.
\end{equation}
Using the completion relation
\begin{equation}
	\mathbf{\Pi}_{-}=\sum_b\int_0^{2\pi f}
	\ket{\psi^{(h)}_{b,\nu}}\bra{\psi^{(h)}_{b,\nu}}\,
	\frac{\md\nu}{2\pi}\,,
\end{equation}
the hole part $\mathbf{G}^{(e)}_{--}$
is then diagonal in the $\ket{\psi^{(h)}_{b,\nu}}$ basis with
respective eigenvalues $1-g_b^{(h)}(\nu)$:
\begin{equation}
	\mathbf{G}^{(e)}_{--}=\sum_b \int_0^{2\pi f}
	\left(1-g_b^{(h)}(\nu)\right)
    \ket{\psi^{(h)}_{b,\nu}} \bra{\psi^{(h)}_{b,\nu}}
    \,\frac{\md\nu}{2\pi}\,.
\end{equation}
This convention for the hole Floquet-Bloch spectrum ensures
that $\mathbf{G}^{(h)}_{++}$ is diagonalized by the eigenvectors
$\mathbf{C}|\psi^{(h)}_{b,\nu}\rangle$ with respective eigenvalue
$g_b^{(h)}(\nu)$. Let us notice that, for $0\leq \nu<2\pi f$, 
the hole eigenstate decomposition
into plane wave takes the form
\begin{equation}
	\label{eq/Floquet/periodic-function/hole}
	\ket{\psi^{(h)}_{b,\nu}}=\sum_{n=1}^{+\infty}
	v^{(n)}_{b,\nu}\ket{\nu-2\pi fn}
\end{equation}
in order to include only negative energy plane waves.
The final step for deriving \cref{eq:Bloch:result} is to introduce the 
electron/hole coherences in the basis of electronic and hole Floquet-Bloch
eigenstates:
\begin{subequations}
	\begin{align}
		\braket{\psi^{(e)}_{a,\nu} | \mathbf{G}^{(e)}
		|\psi^{(h)}_{b,\nu'}}
		&=2\pi\delta(\nu-\nu')\,g^{(eh)}_{ab}(\nu)\\
		\braket{\psi^{(h)}_{b,\nu} | \mathbf{G}^{(e)} |\psi^{(e)}_{a,\nu'}}
		&=2\pi\delta(\nu-\nu')\,g^{(he)}_{ba}(\nu)
	\end{align}
\end{subequations}

\subsection{Floquet--Bloch eigenvalues as occupation numbers}
\label{appendix:Floquet:spectrum:interpretation}

The normalization condition~\labelcref{eq:Bloch:normalization} for the eigenstates
$|\psi^{(e)}_{a,\nu}\rangle$ is the same as the one of plane waves except
for the fact that, in the present case, $\nu$ is a quasi-momentum
living in $\mathbb{R}/2\pi f\mathbb{Z}$. The destruction operator
associated with such an excitation is thus defined by direct analogy
with the operator $c(\omega)$:
\begin{equation}
	\psi[\psi^{(e)}_{a,\nu}]= \frac{v_F}{\sqrt{2\pi}}\int_{-\infty}^{+\infty}
	\psi^{(e)}_{a,\nu}(t)^*\,
\psi(t)\,\md t,
\end{equation}
where the normalization factor ensures the canonical anticommutation
relation 
\begin{equation}
\{\psi[\psi_{a,\nu}],
\psi^\dagger[\psi_{a',\nu'}]\}=\delta_{a,a'}\delta(\nu-\nu').
\end{equation}
It then follows that
\begin{equation}
	\langle \psi^\dagger[\psi^{(e)}_{a',\nu'}] \psi[\psi^{(e)}_{a,\nu}]\rangle = 
	\delta_{a,a'}\delta(\nu-\nu')\,g_a^{(e)}(\nu).
\end{equation}
This equation is analogous to the expression of the single-electron
coherence of a stationary state in the basis of fixed energy single
particle states $\ket{\omega}$ in terms of the electron distribution 
function $f_e(\omega)$:
\begin{equation}
	\langle c^\dagger(\omega')\,c(\omega)\rangle =
	\delta(\omega-\omega')\,f_e(\omega)\,.
\end{equation}
The eigenvalues $g_{a}^{(e)}(\nu)$ can thus be interpreted as the occupation numbers
of the single-particle states $|\psi^{(e)}_{a,\nu}\rangle$. We can
therefore interpret the 
spectrum of $\mathbf{G}^{(e)}_{++}$ as bands of
occupation numbers for the
Floquet-Bloch states $|\psi^{(e)}_{a,\nu}\rangle$ as a function of
their quasi-energy $\nu\in\mathbb{R}/2\pi
f\mathbb{Z}$. In the same way, the bands $\nu\mapsto g^{(h)}(\nu)$ can
be interpreted as giving the occupation numbers for the hole
excitations $\mathbf{C}\ket{\psi^{(h)}_{b,\nu}}$ which are quantum
superpositions of single particle states with energies
$hf-\hbar\nu$ shifted by positive multiples of $hf$. 

Of course, this raises the question of the band structure that can occur
in this type of problem. In order to get a hint on this question, we
must have a closer look at the underlying eigenvalue problems.

\subsection{Eigenvalue equations}
\label{appendix:Floquet:spectrum:eigenvalue-equation}

\begin{figure}
	\includegraphics{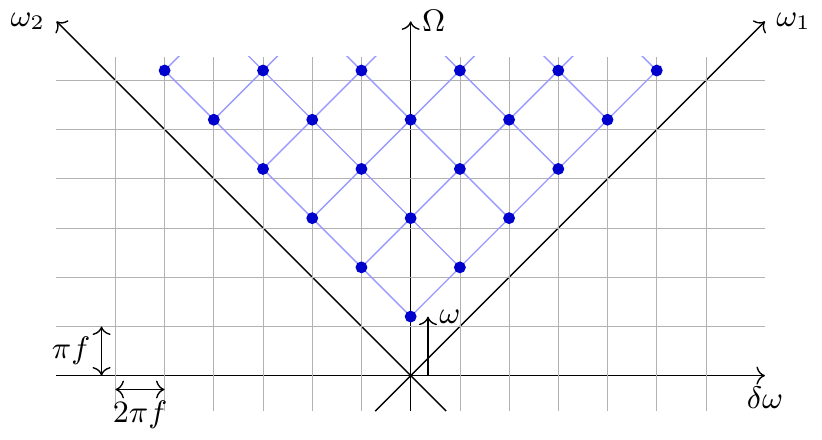}
	
	\caption{
		Graphical representation of the matrix $M(\nu)$.
		The first-order coherence in energy representation takes values
		for $\delta \nu$ being an integer multiple of $2 \pi f$. The
		matrix we extract at a given frequency $\nu$ is the one given
		by the value of the blue dots, that are spaced by $2 \pi f$ in
		both vertical and horizontal direction. Shifting the frequency
		$\nu$ by $\delta\nu$ corresponds to vertically translating all blue dots by
		$\delta \nu$.
	}
	\label{fig:bloch:matrix}
\end{figure}

The diagonalization problem that leads to the spectrum
$(g_a(\nu))_{a,\nu}$ and to the Floquet-Bloch eigenfunctions is
best expressed in the frequency
domain \cite{Degio-2010-4,Ferraro-2013-1}.
Exactly as in Bloch's theory, we introduce $T$-periodic
dimensionless functions $u_{a,\nu}$ such that 
$\psi_{a,\nu}(t)=\me^{-\mi\nu t}v_F^{-1/2}u_{a,\nu}(t)$.
Choosing a representiative of the quasi energy $\nu \in [0,2\pi f[$, 
we decompose $u_{a,\nu}(t)$ in Fourier series
\begin{equation}
\label{eq/Floquet/periodic-function}
u_{a,\nu}(t)=\sum_{n=0}^{+\infty} u_{a,\nu}^{(n)}\,
\me^{-2\mi\pi n f t}.
\end{equation}
where the sum goes from $n=0$ to $n=+\infty$ since we are
looking for purely electronic wavefunctions so each $\nu+2\pi nf$ is
positive.
The eigenvalue equation $\mathbf{G}^{(e)}_{++}|\psi_{a,\nu}\rangle =
g^{(e)}_a(\nu)|\psi_{a,\nu}\rangle$ can then be rewritten in terms of 
the single-electron coherence projected onto the electron quadrant.
With our choice of a
representative $\nu \in [0,2\pi
f[$ for the quasi-energy, the eigenvector equation for $g^{(e)}_a(\nu)$ is
\begin{equation}
\label{eq:Floquet:eigenvalue-equation}
\sum_{p\in\mathbb{N}} 
	W^{(e)}_{++,n-p}(\nu+\pi f(n+p))\,u_{a,\nu}^{(p)}=
	g^{(e)}_a(\nu)\,u_{a,\nu}^{(n)}.
\end{equation}
where, because of $T$-periodicity, we have decomposed the Wigner
distribution
function $W^{(e)}_{++}(t,\omega)$ associated with
$\mathbf{G}^{(e)}_{++}$ as a Fourier series:
\begin{equation}
	W^{(e)}_{++}(t,\omega)=\sum_{n\in\mathbb{Z}}\me^{-2\pi \mi
	nft}W^{(e)}_{++,n}(\omega)
\end{equation}
Note that, because we are considering the projection onto the electronic
quadrant, $W_{++,n}^{(e)}(\omega)=0$ for $\omega -\pi |n|f<0$ and equal to
$W^{(e)}_n(\omega)$ the $n$-th harmonic of the full Wigner function, for
$\omega -\pi |n|f\geq 0$.
\cref{eq:Floquet:eigenvalue-equation} is solved numerically to
determine the spectrum of the single-electron coherence restricted to
the electronic quadrant. We can also see it as the diagonalization of
the matrix $M(\nu)$, defined for each $\nu \in [0, 2 \pi f[$ as
\begin{equation}
	M_{np}(\nu)
	=
W^{(e)}_{++,n-p}(\nu+\pi f(n+p))	
\end{equation}
for $(n,p)\in\mathbb{N}^2$
This matrix is thus derived from the energy representation of the
first-order coherence as graphically pictured on
\cref{fig:bloch:matrix}.

The eigenvalues for the hole Floquet-Bloch matrix are obtained in the
same way starting from \eqref{eq/Floquet/periodic-function/hole} and
following the same step very precisely. This leads to the
eigenvalue equation ($0\leq \nu<2\pi f$ and $n\in\mathbb{N}^*$):
\begin{equation}
	\sum_{p=1}^{+\infty}
	W^{(e)}_{--,p-n}(\nu-\pi f(n+p))\,v^{(p)}_{b,\nu}=(1-g_b^{(h)}(\nu))
	\,v^{(n)}_{b,\nu}\,
\end{equation}
in which 
\begin{equation}
	W^{(e)}_{--}(t,\omega)=\sum_{n\in\mathbb{Z}}
	\me^{-2\pi \mi nft}W^{(e)}_{--,n}(\omega)
\end{equation}
is the Wigner distribution function associated with $\mathbf{G}^{(e)}_{--}$
and therefore $W^{(e)}_{--,n}(\omega)=0$ for $\omega+\pi |n|f>0$ and is
equal to $W^{(e)}_n(\omega)$ as soon as $\omega+\pi |n|f\leq 0$.

\subsection{Case of a voltage drive at zero temperature}
\label{appendix:Floquet:voltage-drive}

\begin{figure*}%
	\includegraphics{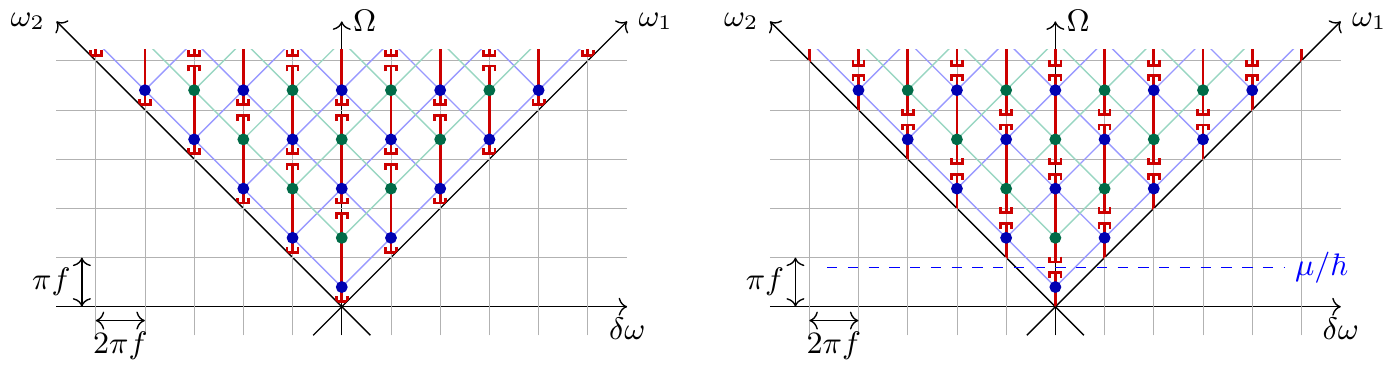}

	\caption{
		Matrices for a voltage drive at zero temperature.
		On the left, the case of an a.c.\@ voltage drive. In this case,
		the coherence is constant for all $\nu \in [0, 2\pi f[$. The
		eigenvalue problem does not depend anymore on the quasi-energy.
		On the right, we consider that there is a d.c.\@ part on top of
		the a.c.\@ voltage. In this case, the matrix $M(\nu)$ will be
		piecewise constant, with a step at $\nu = \mu/\hbar \pmod{2
		\pi f}$.
	}
	\label{fig:matrixdiag:vdrive}
\end{figure*}

A specific
feature of the case of a time-dependent classical drive is that,
at zero temperature, the energy coherence is piecewise constant, the
width of each step being $2 \pi f$. If we consider a purely a.c.\@ drive, 
the discontinuities does not appear when we extract the matrix
$M(\nu)$ for $\nu \in [0, 2 \pi f[$ (see \cref{fig:matrixdiag:vdrive},
left). As such, the eigenvalues will be independent on the
quasi-energy, and the eigenvectors of different quasi-energy can be
deduced by a frequency translation. It implies notably that it is
possible to find a set of Floquet-Wannier functions that are piecewise constant
in energy, with discontinuities happening every $2 \pi f$. Consequently,
the electronic atoms of signal are linear combinations of
Martin--Landauer wavepackets, a point already noticed in Ref.
\cite{Dubois-2013-1}, and there are no inter-period coherences due to
band flatness.

If we add a d.c.\@ part to the voltage, then it will shift the whole
energy coherence by $\mu/\hbar$ (see \cref{fig:matrixdiag:vdrive},
right). In this case, there are two possibilities:
\begin{itemize}
	\item If $\mu/h f$ is an integer, we are back to the a.c.\@ case,
		since the discontinuity will not happen for $\nu \in [0, 2
		\pi f[$.
	\item If $\mu/hf$ is not an integer, then the matrix $M(\nu)$ will be
		piecewise constant, with a step at $\nu = \mu/\hbar \ [2 \pi
		f]$. Similarly, the eigenvectors for $\nu \in [0, \omega_s[$ can be
		deduced by translating the eigenvectors at $\nu = 0$ in
		energy. The eigenvectors for $\nu \in [\omega_s, 2 \pi f[$
		can be deduced by translating the eigenvectors at $\omega_s$. In
		this case, we can find a set of Floquet-Wannier functions that are
		piecewise constant in energy, with steps happening at $2 \pi n
		f$ and $2 \pi n f + \omega_s$.
\end{itemize}

If we consider a small, non-zero temperature, such that $k_B
T_{\mathrm{el}} \ll h f$,
the steps will be smoothed out over a scale $k_B T_{\mathrm{el}}/\hbar$. We can thus
expect that the property mentioned above remains true, except at the
neighborhood of discontinuities.

This example demonstrates that, contrary to the case of bands in solid
state physics, the Floquet-Bloch bands we are considering here may
exhibit discontinuities that, indeed, may play a crucial role. For
example, this is the case when applying a dc-voltage bias corresponding
to a non-integer multiple of $-ef$ dc current  to an ac voltage drive or, more generally,
to a purely
ac-source. An example is a periodic train of Lorentzian voltage pulses
carrying a non integer charge in units of $-e$.

\section{Floquet-Wannier function ambiguities}
\label{appendix:Floquet:Wannier:ambiguities}

In this Appendix, we discuss the ambiguities
in the determination of
electronic atoms of signals and propose a minimal-spreading principle
for selecting a specific choice of electronic atoms of signals. 

\subsection{Origin of the ambiguities}
\label{appendix:Wannier:ambiguities}

Ambiguities in the choice of Floquet-Wannier wavefunctions can always be traced back
to degenerate common eigenspaces for $\mathbf{G}^{(e)}_{++}$ and 
$\mathbf{T}_T$. Let us introduce a unitary tranformation
$\mathbf{U}$ that
keeps $\mathbf{G}_{++}^{(e)}$ 
eigenspaces stable: $[\mathbf{U},\mathbf{G}_{++}^{(e)}]=0$,
then using the $\mathbf{U}|\psi_{a,\nu}\rangle$ states
in~\labelcref{eq:Wannier:definition}, we
obtain a new orthonormal family of 
Wannier functions which we denote by
$\left|\varphi^{[U]}_{a,l}\right\rangle$.
\Cref{eq:Wannier:time-translation} becomes
\begin{equation}
	\label{eq:Wannier:translation-action}
	\mathbf{T}_T \Ket{\varphi_{a,l}^{[\mathbf{U}]}}
	=
	\Ket{\varphi_{a,l+1}^{[\mathbf{T}_T\mathbf{U}\mathbf{T}_T^\dagger]}}.
\end{equation}
In order to satisfy the time-translation property of Wannier
wavefunctions~\labelcref{eq:Wannier:time-translation}, we require that
$\mathbf{U}$ preserves each eigenspace of $\mathbf{T}_T$ and we will then discuss 
what happens depending on the structure of the common eigenspaces of
$\mathbf{T}_T$ and
$\mathbf{G}^{(e)}_{++}$.

Preserving the eigenspaces of $\mathbf{T}_T$ immediately implies that
$\mathbf{U}$ preserves quasi-energy eigenspaces. Assuming that it leaves
each of them invariant, this means that it reduces
to a unitary transformation operating on the space generated by all the
Floquet-Bloch states at a given quasi-energy.
Let us now analyze what happens depending on the eigenspaces of
$\mathbf{G}^{(e)}_{++}$ at fixed quasi-energy.

In the case where the Floquet-Bloch bands
are non-degenerate, injective ($g_a(\nu)\neq g_a(\nu')$ for
$\nu\neq \nu'$) and do not cross, each common eigenspace is one
dimensional and the only possibility for redefining the Floquet-Bloch
eigenstates is to introduce quasi-energy dependent phases:
\begin{equation}
\label{eq:Wannier:phase-ambiguity}
|\psi_{a,\nu}\rangle \mapsto
\me^{\mi\theta_a(\nu)}|\psi_{a,\nu}\rangle.
\end{equation}
Such quasi-energy dependent phases $\theta(\nu)$ fall into
different topological sectors which are labeled by the winding number
\begin{equation}
n_\text{w}=\frac{1}{2\pi}\int_0^{2\pi f}\frac{\md
	\theta(\nu)}{\md\nu}\,\md\nu.
\end{equation}
For example $\theta_n(\nu)=nT\nu$ has winding number
$n\in\mathbb{Z}$ and \cref{eq:Wannier:definition} implies that
\begin{equation}
	\Ket{\varphi_{a,l}^{[\me^{\mi\theta_n}]}} = |\varphi_{a,l+n}\rangle.
\end{equation}
Consequently, 
a topologically non-trivial phase has the same effect as combining
a translation by an integer number of periods with a topologically
trivial energy-dependent phase. 

In the case of $n$ degenerate Floquet-Bloch bands over the whole 
quasi-energy interval, the above
phases are replaced by a quasi-energy dependent
unitary transformation $U(\nu)\in \mathrm{U}(n)$ for $0\leq \nu<2\pi f$ so
that, considering $A_\alpha$ the set of $n$ band indexes, the new
Wannier functions are defined by:
\begin{equation}
\label{eq:Wannier:unitary-ambiguity}
	\Ket{\varphi_{a,l}^{[U]}}
	=
	\frac{1}{\sqrt{f}}
	\int_0^{2\pi f} \sum_{b\in A_\alpha}
	U_{a,b}(\nu)\ket{\psi_{b,\nu}}
	\,\frac{\md\nu}{2 \pi}.
\end{equation}
Such transformations are directly relevant when a source emits $n$
single-electron excitations on top of the Fermi sea.
In this case $g_a(\nu)=1$ for several values
of $a$. The topological sectors of such quasi-energy dependent
unitaries are classified by the topological sectors of the overall phase
since all groups $\mathrm{SU}(n\geq 2)$ are simply connected.
Let us now discuss in more detail the properties of the Floquet-Bloch
band structure to see wich situation is more likely to be encountered.

A first observation is
that bands may have discontinuities. From our observations for classical
voltage drives and for the mesoscopic capacitor driven by a sinusoidal or square 
voltage, it
seems that such discontinuities appear when temperature is non-zero, for
purely a.c.\@ sources.

Finally, since physical states are defined up to a phase, 
we have an extra possibility for defining electronic atoms of signals in 
the case of flat bands. For example, 
one could replace \cref{eq:Wannier:time-translation} 
by its projective version, that is
introducing a phase in front of $|\varphi_{a,l+1}\rangle$. Combining this
with \cref{eq:Wannier:translation-action} 
leads to 
\begin{equation}
U_\Omega|\psi_{a,\nu}\rangle = |\psi_{a,\nu+\Omega}\rangle,
\end{equation}
where the addition is considered modulo $2\pi f$ ($\Omega\in
\mathbb{R}/2\pi f\mathbb{Z}$). Substituting this into \cref{eq:Wannier:definition} leads to 
\begin{equation}
	\Ket{\varphi_{a,l}^{[U_\Omega]}}
	=
	\me^{\mi \Omega T}\Ket{\varphi_{a,l+1}}.
\end{equation}
The time translation property~\labelcref{eq:Wannier:time-translation} is
satisfied up to a phase.

\subsection{Minimal-spreading principle}
\label{appendix:Floquet:Wannier:spreading}

Let us now discuss the general method used to determine suitable
electronic atoms of signals.
Exactly as in solid-state physics, a natural idea is to look for 
maximally-localized Wannier functions \cite{Marzari-2012-1}. 
Let us consider $\varphi_a$ such
a wave-function, the spreading $\langle (\Delta
t)^2\rangle _{\varphi_a}$  is defined as
\begin{equation}
	\label{eq:Wannier:spreading}
	\begin{aligned}
		\langle (\Delta
		t)^2\rangle _{\varphi_a}
		&=v_F\int_{\mathbb{R}} t^2|\varphi_{a,0}(t)|^2\,\md t\\
		&- \left(v_F\int_{\mathbb{R}} t\,|\varphi_{a,0}(t)|^2\md
		t\right)^2.
	\end{aligned}
\end{equation}
Let us consider directly the case of $n$ degenerated
bands
$g_a(\nu)=p_\alpha(\nu)$ for all $0\leq \nu <2\pi f$ and $a\in A_\alpha$. 
We then have a quasi-energy dependent unitary 
transformation ambiguity described by \cref{eq:Wannier:unitary-ambiguity}. 
Maximally localized Wannier wavefunctions are now found by
minimizing the quadratic functional 
\begin{equation}
\label{eq:Wannier:functional}
	\mathcal{S}[U]
	=
	\sum_{a\in A_\alpha}
	\left\langle
		(\Delta t)^2
	\right\rangle_{\ket{\varphi_a^{[U]}}}
\end{equation}
over $U(\nu)\in \mathcal{U}(n)$ for $0\leq \nu < 2\pi f$.
Note that the right-hand side of \cref{eq:Wannier:spreading} may be divergent due to the
large time behavior of $|\varphi_{a,0}(t)|^2$ as, for example, in the
case of a Leviton train. In such a case, we should therefore regularize
it by subtracting the same quantity for a reference choice of the
unitary operator such as
$U(\nu)=\mathbf{1}$. 

Numerically, the implementation of the minimization process is
straightforward in the case of a non-degenerate band. Since there is a
natural cut-off for the length of the wavepacket, in this case it is
easy to compute the functional~\labelcref{eq:Wannier:spreading} from an
arbitrary phase~\labelcref{eq:Wannier:phase-ambiguity}. More
importantly, it is also easy to compute the gradient, giving access to
all efficient gradient-based minimization algorithms. In our case, we
rely on the GSL implementation of the Fletcher-Reeves algorithm
\cite{Fletcher-1964-1}. It consists in a succession of line
minimizations. We begin at a given point (which can either be random
phase or a null phase), and the first direction of minimization is given
by the gradient. Then, at each iteration, a new direction is chosen,
depending on the previous search direction, the gradient of current
iteration and the norm of the gradient of previous iteration. The
iteration ends when the gradient is orthogonal to the line of search.

For the degenerate case \labelcref{eq:Wannier:unitary-ambiguity}, there
are several difficulties. First, we need to parametrize the unitary
matrices $U(\nu)$. For this, we introduce $\Theta(\nu)$, Hermitian
matrices such that
\begin{equation}
	U(\nu) = \exp(\mi \Theta(\nu)).
\end{equation}
The main difficulty here is that, since $\mathrm{U}(n \le 2)$ is a
non-commutative group, it becomes hard to compute the gradients of the
functional $\mathcal{S}[U]$. However, it is still easy to compute them
if we consider a starting point at $U = \mathbbm{1}$. In the following,
we will denote $\ket{\psi_{A,\nu}}$ the vector containing every
wavefunctions $\ket{\psi_{a,\nu}}$ with $a \in A$, $A$ being the
degenerate set of bands we want to minimize on. the matrix $U(\nu)$
acts on this vector space, mixing wavefunctions. At each iteration $n
> 1$ of the algorithm, we now replace the wavefunctions
$\ket{\psi_{A,\nu}^{(n-1)}}$ by the wavefunctions
\begin{equation}
	\Ket{\psi_{A,\nu}^{(n)}}
	=
	\me^{\mi x_n H_n(\nu)} \Ket{\psi_{A,\nu}^{(n-1)}},
\end{equation}
$H_n$ being the search direction and $x_n$ the real parameter that
minimize this search direction. This allows us to always start the line
minimization process from $U = \mathbbm{1}$. To determine the minimum,
we check whether our search direction is orthogonal to the local
gradient computed by shifting $\me^{\mi x_n H_n(\nu)}$ to identity.
What makes everything work is that all quantities needed to compute the
new direction of minimization are either invariant on the point of the
$\mathrm{U}(n)$ group we consider them (norm of the previous gradient),
computed locally (the new gradient) or trivially transported (previous
search direction, which is parallel to the transport). After $N$
iterations, we end up with 
\begin{equation}
	\Ket{\psi_{A,\nu}^{(N)}}
	=
	\me^{\mi \Theta_N(\nu)}
	\cdots
	\me^{\mi \Theta_1(\nu)}
	\Ket{\psi_{A,\nu}^{(0)}}.
\end{equation}
emphasing the non-commutative character of the group we are minimizing
on.

\section{HBT and HOM current noise}
\label{appendix/partition-noise}

\subsection{Explicit expressions}

The outgoing current correlation $S_{11}^{(\mathrm{out})}(t,t')=\langle
i_{1\text{out}}(t)\,i_{1\text{out}}(t')\rangle 
-\langle i_{1\text{out}}(t)\rangle\langle i_{1\text{out}}(t')\rangle $
after a quantum point contact whose
scattering matrix is
$$\begin{pmatrix}
	\sqrt{\mathcal{R}} & \mi\sqrt{\mathcal{T}} \\
	\mi\sqrt{\mathcal{T}} & \sqrt{\mathcal{R}} 
\end{pmatrix}
$$
has been computed in Ref. \cite{Degio-2010-4} in terms of the incoming
current correlators and single electron coherences:
\begin{equation}
	\label{eq/HOM/full-noise}
		S_{11}^{(\mathrm{out})}(t,t') =\mathcal{R}^2
		S_{11}^{(\text{in})}(t,t') +\mathcal{T}^2
        S_{22}^{(\text{in})}(t,t') 
		+ \mathcal{RT}\mathcal{Q}(t,t')\,
\end{equation}
in which 
\begin{equation}
		\label{eq/HOM/Q-definition}
		\mathcal{Q}(t,t')= e^2v_F^2\left(\mathcal{G}^{(e)}_1(t',t)\,
		\mathcal{G}^{(h)}_2(t',t)+\left[1\leftrightarrow
		2\right]\right)\,
\end{equation}
encodes two-particle
interferences effects between the two incoming channels. The excess
current noise is therefore given by
\begin{subequations}
	\label{eq/HOM/excess-noise}
    \begin{align}
		\label{eq/HOM/excess-noise/input-noise}
        \Delta S_{11}^{(\mathrm{out})}(t,t') &=\mathcal{R}^2
        \Delta S_{11}^{(\text{in})}(t,t') +\mathcal{T}^2
		\Delta S_{22}^{(\text{in})}(t,t') \\
        &+ \mathcal{RT}\Delta \mathcal{Q}(t,t')\,.
		\label{eq/HOM/excess-noise/Q-contribution}
    \end{align}
\end{subequations}
Hanbury Brown
and Twiss (HBT) experiments correspond to one of the sources being switched
{\it on}
and the other one being switched {\it off}. From now on, let us assume that
both sources $S_1$ and $S_2$ are identical and synchronized.
Under this hypothesis, $\Delta S_{11}^{(\text{in})}=\Delta
S_{22}^{(\text{in})}=\Delta S_S$ is the excess noise generated by the
source.

At zero temperature, the HBT
excess zero frequency current noise can be expressed using the Floquet--Bloch spectrum
as $\mathcal{RT}\Delta \mathcal{Q}_{\text{HBT}}$ where
\begin{equation}
	\label{eq/HOM/Q-HBT}
	\Delta \mathcal{Q}_{\text{HBT}}
	=e^2\int_0^{2\pi f}\left[\sum_a
	g^{(e)}_{a}(\nu)+\sum_b
	g^{(h)}_{b}(\nu)\right]\,\frac{\md \nu}{2\pi}\,
\end{equation}
in which $\Delta \mathcal{Q}_{\text{HBT}}$ (see
\cref{eq/HOM/excess-noise/Q-contribution}) 
arises from the partitioning of electron and hole excitations at the
QPC not contained in the partitioning of the incoming current noises
(r.h.s of \cref{eq/HOM/excess-noise/input-noise} at zero
frequency). This leads to \cref{eq:HBT:excess}.

\begin{figure}
	\includegraphics{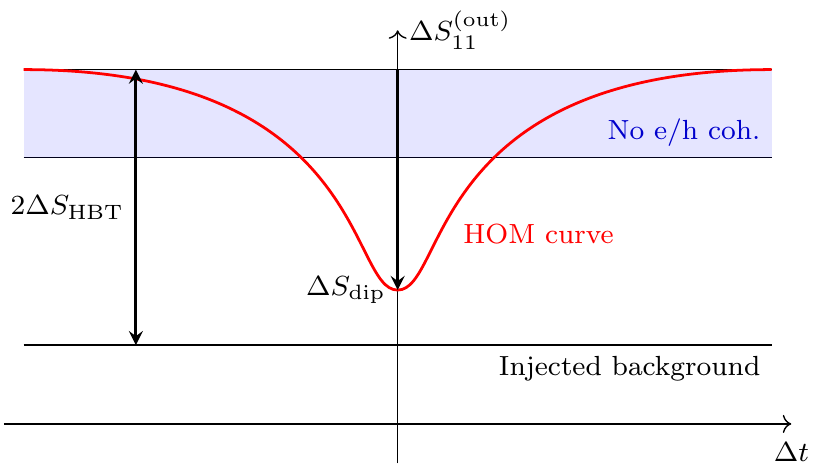}
	
	\caption{\label{fig/HOM/HOM-dip}
	Depth of the HOM dip at zero temperature: in an HOM experiment, the excess low frequency
	current noise is above the background noise 
	$(\mathcal{R}^2+\mathcal{T}^2)\,\Delta S_S$ injected by the sources.
	When the sources, which emit localized excitations, 
	are sufficiently desynchronized, the
	current noise is expected to reach $\Delta
	S_{\text{HBT}}=e^2f\mathcal{RT}(\overline{N}_e+\overline{N}_h)$
	exceeding
	this background noise by $2\Delta S_{\text{HBT}}$. 
	At fixed Floquet--Bloch spectra $(g^{(e)}_a(\nu),g_b^{(h)}(\nu))$, the depth of
	the HOM dip has a lower bound which translates into an upper bound
	for the excess noise equal to $2e^2f\mathcal{RT}((\Delta
	N_e)^2_{\text{w}}+(\Delta N_h)^2_{\text{w}})$ above the injected background
	noise (light blue zone). This bound is reached for
	vanishing electron/hole coherences. 
	}
\end{figure}

When both sources are switched {\it on}, an Hong--Ou--Mandel experiment is
performed. Using \cref{eq/HOM/excess-noise}, 
the corresponding excess noise is the sum of the excess
noise of the two possible HBT experiments 
\begin{equation}
    \label{eq/HOM/HOM-excess-noise}
        \Delta S_{11}^{(\mathrm{out})} =\Delta S_{1}^{(\mathrm{HBT})}
		+\Delta S_{2}^{(\mathrm{HBT})}
		+\mathcal{RT}\Delta\mathcal{Q}_{\text{HOM}}
\end{equation}
and of a two-excitations interference 
contribution involving the two sources $S_1$ and $S_2$ which 
can be expressed as
\begin{align}
	\label{eq/HOM/Q-HOM}
	\Delta\mathcal{Q}_{\text{HOM}}
	&=-2e^2\int_0^{2\pi f}\left(\sum_a g^{(e)}_a(\nu)^2+
	\sum_bg^{(h)}_b(\nu)^2\right)\,
	\frac{\md \nu}{2\pi}\nonumber \\
	&-4e^2\sum_{a,b}\int_0^{2\pi f}\left| 
	g^{(eh)}_{ab}(\nu)\right|^2
	\frac{\md \nu}{2\pi}
\end{align}
using the Floquet--Bloch analysis of the single electron coherence
emitted by the source $S_1$, identical to $S_2$ here. Adding
twice the r.h.s. of
\cref{eq/HOM/Q-HBT} (one for each source) to the r.h.s. of \cref{eq/HOM/Q-HOM} 
leads to the total contribution $(\mathcal{R}^2+\mathcal{T}^2)\,\Delta
S_S$ to the zero
frequency current noise that comes on top of the partitioning of the
sources intrinsic excess current noise $\Delta S_S$.
In the end, as shown on \cref{fig/HOM/HOM-dip},
the final
expression \eqref{eq:HOM:total} for the excess zero frequency current noise at the HOM dip is
the sum of the excess current noise of the sources transmitted by the
two sources (which is always positive), 
to which is added the total two-excitation interference
contribution given by \cref{eq:HOM:excess}. 

\subsection{The HOM dip}
\label{appendix/partition-noise/HOM-dip}

Let us now use this to discuss the depth of the HOM dip defined as
the difference between the HOM excess noise given by
\cref{eq/HOM/HOM-excess-noise}
and the sum of the two HBT noises.
Counting the dip's
depth
positively, its expression is
\begin{equation}
	\left[\Delta S_{\text{dip}}\right]=
-\mathcal{RT}
\,\Delta
Q_{\text{HOM}}
\end{equation}
where $\Delta\mathcal{Q}_{\text{HOM}}$ is given by \cref{eq/HOM/Q-HOM}.

It is interesting to rewrite these expressions in terms of physically
more appealing quantities.
For this, let us introduce an infrared
regularization to compute traces of $T$-periodic operators. Let
\begin{equation}
	\mathbf{X}=\sum_a\int_0^{2\pi f}X_a(\nu)\,\frac{\md\nu}{2\pi}
\end{equation}
a diagonal operator in the basis of electronic Floquet--Bloch
eigenstates.
The trace of this operator, which is defined and acts on $\mathcal{H}_+$
is divergent. Nevertheless, we can regularize it. Inverting
\cref{eq:Wannier:definition}, its expression in the basis formed by the electronic
atoms of signals is 
\begin{equation}
	\mathbf{X}=\sum_a\sum_{(l_+,l_)\in\mathbb{Z}^2}
	X_a(l_+-l_-)\,\ket{\varphi_{a,l_+}}\bra{\varphi_{a,l_-}}\,.
\end{equation}
in which $X_a(l)$ is related to $X_a(\nu)$ by \cref{eq:Wannier:pa-ee}.
Taking the trace over a subspace generated by the electronic atoms of
signal over a range of $N$ periods gives
\begin{equation}
	\mathrm{Tr}_N(\mathbf{X})=N\sum_a\int_0^{2\pi f}
	X_a(\nu)\,\frac{\md\nu}{2\pi f}\,.
\end{equation}
This leads to the definition of the per-period regularized trace
as 
$\overline{X}=\mathrm{Tr}_N(\mathbf{X})/N$.
With this definition, the average number of electronic and hole
excitations emitted per period are given by
\begin{subequations}
	\label{eq/Ne-Nh-definition}
	\begin{align}
		\overline{N}_e &=\sum_a\int_0^{2\pi f}
		g_a^{(e)}(\nu)\,\frac{\md \nu}{2\pi f}\\
\overline{N}_h &=\sum_b\int_0^{2\pi f}
        g_b^{(h)}(\nu)\,\frac{\md \nu}{2\pi f}\,.
	\end{align}
\end{subequations}
The quadratic terms in the eigenvalues or in the electron/hole
coherences appearing in \cref{eq/HOM/Q-HOM} 
correspond to what would be obtained, assuming Wick's theorem
be
valid. Therefore, we denote these quantities with a ``$\mathrm{w}$'' index.
Using Wick's theorem with $\mathbf{G}^{(e)}$ as single electron
coherence,
the second moments of the numbers of excitations emitted per periods
would be given by
\begin{subequations}
	\label{eq/HOM/Wick-second-moments}
	\begin{align}
		(\Delta N_e)_\mathrm{w}^2 &= \sum_a\int_0^{2\pi
		f}g_a^{(e)}(\nu)(1-g_a^{(e)}(\nu))
		\,\frac{\md \nu}{2\pi f}\\
		(\Delta N_h)_\mathrm{w}^2 &=\sum_b\int_0^{2\pi
		f}g_b^{(h)}(\nu)(1-g_b^{(h)}(\nu))
        \,\frac{\md \nu}{2\pi f}\\
		\mathrm{Cov}(N_e,N_h)_\mathrm{w} &= 
		\sum_{a,b}\int_0^{2\pi f}
		\left|g_{ab}^{(eh)}(\nu)\right|^2\,\frac{\md\nu}{2\pi f}\,.
	\end{align}
\end{subequations}
Let us stress that, in the presence of interactions, these are not the
actual moments of the numbers of electronic and hole excitations. The
actual values differ
from these Wick values by a contribution arising from the difference between
the intrinsic excess second order coherence introduced in 
Ref. \cite{Thibierge-2016-1} and its expected value from Wick's theorem. 

The absolute upper bound of the depth of the HOM dip is given by the HBT contribution
$\left[\Delta
S_{\text{dip}}^{(\text{max})}\right]=2e^2f\mathcal{RT}\overline{N}_{\text{tot}}$ where
$\overline{N}_{\text{tot}}=\overline{N}_e+\overline{N}_h$ represents to
average total number of excitations emitted per period. 
Using the above notations, the difference between the maximum 
dip and the actual dip is then equal to
\begin{subequations}
	\begin{align}
	\left[\Delta
		S_{\text{dip}}^{(\text{max})}\right]&-\left[\Delta S_{\text{dip}}\right]
	=\nonumber \\
		&2e^2f\mathcal{RT}\left[(\Delta N_e)_\mathrm{w}^2+(\Delta N_e)_\mathrm{w}^2
	-2\,\mathrm{Cov}(N_e,N_h)_\mathrm{w}\right]\,.
	\end{align}
\end{subequations}
The r.h.s is therefore
directly proportional to the
fluctuation $(\Delta Q)^2_\text{w}$ of the excess charge emitted per period by the source.
In units of $-e$, the excess charge operator is given by
\begin{equation}
	\widehat{Q}=\int_\mathbb{R}
	:c^\dagger(\omega)\,c(\omega):\,
	\md\omega\,.
\end{equation}
where the fermionic normal ordering is relative to the reference Fermi
sea at chemical potential $\mu=0$.
Consequently, the ratio of the dip to its absolute upper
bound 
is given by:
\begin{equation}
    \label{eq/HOM/HOM-dip/rescaled}
	\frac{\left[\Delta
	S_{\text{dip}}\right]}{\left[\Delta
	S_{\text{dip}}^{(\text{max})}\right]} =
	1-\frac{(\Delta
	Q)^2_{\text{w}}}{\overline{N}_{\mathrm{tot}}}\,.
\end{equation}
If the many-body state does satisfy Wick's
theorem, which is the case whenever interactions can be neglected, then
having a maximally deep HOM dip corresponds to the actual vanishing charge fluctuations.

\section{Many-body state and Floquet scattering theory}
\label{appendix/manybody}

\subsection{The Floquet-Bloch many-body state}
\label{sec/many-body/Floquet-Bloch}

In this appendix, we discuss the connexion between our approach and the 
$T$-periodic single-electron scattering theory that transforms the 
Fermi sea at chemical potential $\nu=0$ into a pure many-body state.
This corresponds to writing down the explicit form of 
the many-body operator 
$\mathcal{S}$ whose action corresponds to single-particle scattering 
\begin{equation}
	\psi_{\text{out}}(t)
	=
	\mathcal{S} \psi_{\text{in}}(t) \mathcal{S}^\dagger
	=
	\int S(t,t') \psi_{\text{in}}(t') \, \md t\,
\end{equation}
where $S(t,t')$ denotes the $T$-periodic single-particle scattering matrix in the
time-domain representation. Without lack of generality, we shall
consider here the case of an ac Floquet source, that is 
a single particle scattering operator $\mathcal{S}$ leading to a vanishing
average dc-current.

Finding an expression for $\mathcal{S}$ is important for two reasons:
first it gives insights on the action of Floquet sources on the
incoming equilibrium state at the many-body level, then is enables us to
connect the form of the many-body operator to the Floquet-Bloch
spectrum and eigenstates and electron-home coherences between them. 

\subsubsection{Two modes}
\label{sec/many-body/two-modes}

Let us first discuss the simple two-mode case where only
one electron mode $\varphi_e$ and one
hole mode $\varphi_h$ are considered. 
At zero
temperature, the incoming hole mode $\varphi_h$ is filled and contains
exactly one electron whereas the incoming electron mode $\varphi_e$ is
empty. The Floquet source will scatter the mode $\varphi_h$ into a
linear combination of $\varphi_e$ and $\varphi_h$.

We will show that the operator~$\mathcal{S}$ can be written as
\begin{align}
	\mathcal{S} &= \mathcal{S}_\text{d} \mathcal{S}_{\text{p}}
	\\
	\mathcal{S}_{\text{d}}
	&=
	\exp\left(
		\lambda \psi^\dagger[\varphi_e] \psi[\varphi_h]
		-
		\lambda^* \psi^\dagger[\varphi_h] \psi[\varphi_e]
	\right)
	\\
	\mathcal{S}_{\text{p}}
	&=	\exp\left(
		\mi \left(
			\theta_e (\psi^\dagger \psi)[\varphi_e]
			+
			\theta_h (\psi^\dagger \psi)[\varphi_h]
		\right)
	\right)\,.
\end{align}
In this decomposition, $\mathcal{S}_\text{d}$ is a displacement-like
operator, with the complex parameter $\lambda$. It corresponds to the
scattering processes of between the electron and hole modes.
$\mathcal{S}_{\text{p}}$ is a phase-shifting operator, that 
independently the phases of the incoming electron and hole modes.
Since we can reabsorb the phase of $\lambda$ by changing the relative
phase between the wavefunctions $\varphi_e$ and $\varphi_h$, we will
consider $\lambda \in \mathbb{R}^+$.

Let us first focus on the displacement-like operator.
The exponential can be expanded using
the following identity
\begin{multline}
	-\left(
		\psi^\dagger[\varphi_e] \psi[\varphi_h]
		-
		\psi^\dagger[\varphi_h] \psi[\varphi_e]
	\right)^2
	=
	\\
	n_e (1-n_h) + n_h (1-n_e)
	=
	\Pi_{\text{odd}},
\end{multline}
where $n_{e/h} = (\psi^\dagger \psi)[\varphi_{e/h}]$ denotes the number operator for
the corresponding $\varphi_{e/h}$ mode and $\Pi_{\text{odd}}$ is the
projector on the one-particle sector.
If we also introduce
the orthogonal projector $\Pi_{\text{even}}$ that projects on the
zero or two-particle sector, a simple expression
for the many-body scattering operator follows:
\begin{align}
	\label{eq:manybody:scattering}
\mathcal{S}_{\text{d}}
	=&
	\Pi_{\text{even}}
	+
	\Pi_{\text{odd}}
	\left[
		\cos \lambda \right.
		\nonumber \\
		&+\left.
		\sin \lambda
		\left(
			\psi^\dagger[\varphi_e] \psi[\varphi_h]
			-
			\psi^\dagger[\varphi_h] \psi[\varphi_e]
		\right)
	\right].
\end{align}
Applying the operator $\mathcal{S}$ to the creation operators leads to
\begin{align}
	\mathcal{S} \psi^\dagger[\varphi_e] \mathcal{S}
	&=
	\cos \lambda \, \me^{\mi \theta_e} \psi^\dagger[\varphi_e]
	-
	\sin \lambda \, \me^{\mi \theta_h} \psi^\dagger[\varphi_h]
	\nonumber
	\\
	\mathcal{S} \psi^\dagger[\varphi_h] \mathcal{S}
	&=
	\cos \lambda \, \me^{\mi \theta_h} \psi^\dagger[\varphi_h]
	+
	\sin \lambda \, \me^{\mi \theta_e}
	\psi^\dagger[\varphi_e]\,.
\end{align}
The outgoing creation operators are thus linear combinations of the
incoming creation operators. The corresponding linear operator involved is indeed
unitary and any unitary operator can be brought in that form by tuning
the phase between the wavefunctions $\varphi_e$ and $\varphi_h$.

\subsubsection{The many-mode case}

To understand the full many-body case, we first 
recast our Floquet-Bloch analysis in terms of
the scattering operator. For this, we split the
scattering operator into two parts. 
The first one
rearranges electrons and holes independently and is
described by a unitary matrix $\me^{\mi (\Theta^{(h)}+\Theta^{(e)})}$,
where $\Theta^{(h)}$ and $\Theta^{(e)}$ are Hermitian matrices acting on
the hole and electron subspaces respectively. These operators generalize
the phases $\theta_h$ and $\theta_e$ of the previously discussed
two-mode example.
Their action on the Fermi sea is, as we shall see,
to add a global phase to the many-body state. 
The second part involves 
a two-mode scattering process, in which each pair of modes is scattered
according such that
\begin{align}
	\mathcal{S} \psi^\dagger[\varphi_e] \mathcal{S}
	&=
	u \psi^\dagger[\varphi_e]
	+
	v \psi^\dagger[\varphi_h]
	\nonumber
	\\
	\mathcal{S} \psi^\dagger[\varphi_h] \mathcal{S}
	&=
	u \psi^\dagger[\varphi_h]
	-
	v \psi^\dagger[\varphi_e]\,.
\end{align}
with $u, v \in \mathbb{R}_+$. The mathematical details for
such a decomposition of general unitary operators can be found in
\cref{chap:unitmatrep}.

The final result of this procedure is an expression of the full many-body
scattering operator as a product of uncoupled elementary two-mode
operators, with a prefactor that scatters electron and hole subspaces
independently:
\begin{widetext}
\begin{subequations}
	\label{eq:manybodyop:full}
	\begin{align}
		\label{eq:manybodyop:full:1}
		\mathcal{S}
		&=
		\exp\Bigg(
			\sum_{a \in \mathbb{N}}
			\int_{0}^{2 \pi f}
			\lambda_a(\nu)
			\Big(
				\psi^\dagger[\psi^{(e)}_{a,\nu}]
				\psi[\psi^{(h)}_{a,\nu}]
				-
				\psi^\dagger[\psi^{(h)}_{a,\nu}]
				\psi[\psi^{(e)}_{a,\nu}]
			\Big)
			\, \frac{\md \nu}{2 \pi}
		\Bigg)
		\\
		\label{eq:manybodyop:full:2}
		&\times \exp\Bigg(
			\mi
			\sum_{a,b \in \mathbb{N}}
			\int_{0}^{2 \pi f}
			\Big(
				\Theta_{ab}^{(e)}(\nu)
				\psi^\dagger[\psi^{(e)}_{a,\nu}] \psi[\psi^{(e)}_{b,\nu}]
				+
				\Theta_{ab}^{(h)}(\nu)
				\psi^\dagger[\psi^{(h)}_{a,\nu}] \psi[\psi^{(h)}_{b,\nu}]
			\Big)
			\, \frac{\md \nu}{2 \pi}
		\Bigg)
		.
	\end{align}
\end{subequations}
\end{widetext}
Furthermore, when the bands are flat, we can reorganize a combination of
Floquet-Bloch modes as a combination of electronic atoms of signals
directly at the many-body level. Of course the choice of Floquet-Wannier
functions in the electron quadrant will constrain the choice of
Floquet-Wannier functions in the hole quadrant. Namely, we have
\begin{equation}
	\label{eq:blochtowannier:manybody}
	\int_{0}^{2 \pi f}
	\psi^\dagger\left[\psi^{(e)}_{a,\nu}\right]
	\psi\left[\psi^{(h)}_{a,\nu}\right]
	\, \frac{\md \nu}{2 \pi}
	=
	\sum_{l \in \mathbb{Z}}
	\psi^\dagger\left[\varphi^{(e)}_{a,l}\right]
	\psi\left[\varphi^{(h)}_{a,l}\right].
\end{equation}

\subsection{Splitting unitary matrices}
\label{chap:unitmatrep}

In this appendix, we will introduce a decomposition for unitary matrices
useful when we partition equally the Hilbert space in two. In what
remains, we will consider a matrix $S \in \mathrm{U}(2n)$, acting on a
Hilbert space $\mathcal{H} = \mathcal{H}_+ \otimes \mathcal{H}_-$, where
$\dim \mathcal{H}_+ = \dim \mathcal{H}_- = n$. The goal here is to show
that there exists an orthogonal change of basis $P = P_- P_+$
that acts independently on $\mathcal{H}_+$ and $\mathcal{H}_-$ in which
we can write
\begin{equation}
\renewcommand{\arraystretch}{1.5}
	P S P^\dagger
	=
	\left(\begin{array}{@{}c|c@{}}
		u & v \\ \hline
		v & -u
	\end{array}\right)
	\me^{\mi (\Theta_- + \Theta_+)}
\end{equation}
where $u, v \in \mathcal{M}_{n}$ are positive real diagonal matrices,
$\Theta_\pm$ are Hermitian matrices of size $n$.
The first block-column corresponds to the Hilbert space $\mathcal{H}_-$
and the second one corresponds to $\mathcal{H}_+$.

Generically, one can write the $S$ matrix as
\begin{equation}
\renewcommand{\arraystretch}{1.5}
	S
	=
	\left(\begin{array}{@{}c|c@{}}
		S_{--} & S_{-+} \\ \hline
		S_{+-} & S_{++}
	\end{array}\right).
\end{equation}
For the sake of simplicity, we will consider that each submatrix is
invertible. Other cases would correspond to either fully scattered modes
or fully reflected modes, which can be separated from the start without
much problems.

We will first introduce the polar decomposition of $S_{--} = H_{--}
\me^{\mi\theta_-}$, where $\theta_-$ is Hermitian and $H_{--}$ is a positive
semi-definite Hermitian matrix. This allows us to rewrite $S$ as
\begin{equation}
\renewcommand{\arraystretch}{1.5}
	S
	=
	\left(\begin{array}{@{}c|c@{}}
		H_{--} & S_{-+} \\ \hline
		S_{+-}\me^{-\mi \theta_-} & S_{++}
	\end{array}\right) \me^{\mi \theta_-}.
\end{equation}
$H_{--}$ being positive semi-definite, we can write it as $H_{--} =
P_-^\dagger u P_-$, where $u$ is a diagonal, real-valued, positive matrix.
This leads us to
\begin{equation}
\renewcommand{\arraystretch}{1.5}
	P_- S P_-^\dagger
	=
	\left(\begin{array}{@{}c|c@{}}
		u & S'_{-+} \\ \hline
		S'_{+-} & S_{++}
	\end{array}\right) \me^{\mi \Theta_-}.
\end{equation}
where $\Theta_- = P_- \theta_- P_-^\dagger$, $S'_{+-} = P_- S_{+-}
\me^{-\mi \theta_-}$ and $S'_{-+} = S_{-+} P_-^\dagger$. The first matrix
of the rhs must be unitary. Since $u$ is diagonal, it imposes that each
column of $S'_{+-}$ is orthogonal to each other. As such, we can rewrite
this matrix as a product of a unitary matrix and a diagonal positive
real matrix, $S'_{+-} = P_+^\dagger v$. Noting $P = P_- P_+$, we
have shown
\begin{equation}
\renewcommand{\arraystretch}{1.5}
	P S P^\dagger
	=
	\left(\begin{array}{@{}c|c@{}}
		u & S''_{-+} \\ \hline
		v & S'_{++}
	\end{array}\right) \me^{\mi \Theta_-}.
\end{equation}
where $S''_{-+} = S'_{-+} P_+^\dagger$, $S'_{++} = P_+ S_{++}
P_+^\dagger$.

We can now use the hermitian properties of unitary matrices to build
explicitly the constraints between $u$, $v$, $S'_{++}$ and $S''_{-+}$.
The orthogonality constraint gives us
\begin{equation}
	S'_{++} = - (u/v) S''_{-+}
\end{equation}
where $u/v$ is the diagonal matrix formed by $u v^{-1}$. Conversely, the
normalization conditions give us
\begin{equation}
	{S'_{++}}^\dagger (\mathbbm{1}_n + (u/v)^2) S'_{++} = \mathbbm{1}
\end{equation}
Since $u^2 + v^2 = \mathbbm{1}$, this shows that $v^{-1} S'_{++} =
\me^{\mi \Theta_+}$, where $\Theta_+$ is an Hermitian matrix. Putting
everything together, we have
\begin{equation}
\renewcommand{\arraystretch}{1.5}
	P S P^\dagger
	=
	\left(\begin{array}{@{}c|c@{}}
		u & v \\ \hline
		v & -u
	\end{array}\right) \me^{\mi (\Theta_- +
	\Theta_+)}.
\end{equation}
This is the property we wanted to show.

\subsection{Many-body state at non-zero temperature}

At non-zero temperature, all terms of
\cref{eq:manybodyop:full} will play a role.
The contribution \cref{eq:manybodyop:full:2}, arising from the
separate rearrangement of electron and hole modes will have a
non-trivial contribution to the total state. This contribution may
scatter electrons deep into the Fermi sea compared to the thermal scale
into the thermal fluctuations. It will as well scatter holes from the thermal
fluctuations deeper into the Fermi sea. It is also possible to rearrange
wavefunctions inside the thermal band. Similar processes appear in the
electron subspace. Notably, this term will explicitly couple
different bands. The contribution \cref{eq:manybodyop:full:1} will also
act differently, since sectors of even parities are expected if one of
the Floquet-Bloch waves possesses thermal fluctuations at this point.
We expect that the atoms of signal, as well as their respective
coherences to be modified by this term. 

Remarkably, 
the description in terms of Floquet-Bloch waves at zero temperature
allows us to give a many-body description up to the two Hermitian
operators $\Theta^{(e)}$ and $\Theta^{(h)}$. This is interesting since
it gives a way to see which processes will occur when ``heating'' an
ideal single-electron source. Our approach may thus lead to new insights
on the effect of non-zero temperatures on electronic correlations studied in Refs.
\cite{Moskalets-2017-1,Moskalets-2018-3}.

\section{The purity indicator}
\label{appendix/purity}

Wick's theorem is valid whenever the many-body state $\rho$ of the
electron fluid is Gaussian
\begin{equation}
	\rho=\frac{\me^{-\psi^\dagger\cdot \mathbf{K}\cdot \psi}}{Z_K}
\end{equation}
where $Z_\mathbf{K}=\mathrm{Tr}(\me^{-\psi^\dagger\cdot \mathbf{K}\cdot
\psi})$ is the
corresponding partition
function. The $\mathbf{K}$ operator is related to the single-electron coherence
through 
\begin{equation}
	\label{eq/purity/K2G}
	\mathbf{G}^{(e)}=(\mathbf{1}+\me^{\mathbf{K}})^{-1}
\end{equation}
The many-body ``unregularized'' purity indicator
$\mathbb{P}^{(\text{un})}_\rho=\mathrm{Tr}(\rho^2)$ can then
formally rewritten as a quotient on infinite dimensional determinants
over the single-particle space of states $\mathcal{H}_{\mathrm{1p}}$.
Using \cref{eq/purity/K2G}, it can then be 
conveniently expressed in terms of the total single-electron coherence:
\begin{subequations}
	\begin{align}
		\mathbb{P}^{(\text{un})}_\rho
		&=\frac{\mathrm{Det}\left[\mathbf{1}+\me^{2\mathbf{K}}\right]}{\mathrm{Det}\left[\mathbf{1}+\me^\mathbf{K}\right]^2}\\
		&=
		\mathrm{Det}\left[\mathbf{1}+2\left((\mathbf{G}^{(e)})^2-\mathbf{G}^{(e)}\right)\right]
	\end{align}
\end{subequations}
We can now use the decomposition of the total single-electron coherence
$\mathbf{G}^{(e)}=\mathbf{\Pi}_h+\Delta_0\mathbf{G}^{(e)}$ where
$\Pi_h$ denotes the projection onto the space of hole excitations and
\begin{equation}
	\Delta_0\mathbf{G}^{(e)}=\begin{pmatrix}
		\mathbf{g}_e & \mathbf{g}_{eh} \\
		\mathbf{g}_{he} & -\mathbf{g}_h
	\end{pmatrix}
\end{equation}
is the excess single-electron coherence with respect to the Fermi sea
to obtain an expression for the purity indicator only in terms of data
that can be reconstructed by the single-electron tomography protocol.
These are
$\mathbf{g}_e$, $\mathbf{g}_h$ and the off diagonal parts $\mathbf{g}_{eh}$ 
and $\mathbf{g}_{he}$. This finally leads to:
\begin{widetext}
\begin{equation}
	\mathbb{P}^{(\text{un})}_\rho=\left|
	\begin{matrix}
		\mathbf{1}-2\left[\mathbf{g}_e(\mathbf{1}-\mathbf{g}_e)-\mathbf{g}_{eh}\mathbf{g}_{he}\right] 
		&
		2\left[\mathbf{g}_{eh}\mathbf{g}_{h}-\mathbf{g}_{e}\mathbf{g}_{eh}\right]
		\\
2\left[\mathbf{g}_{h}\mathbf{g}_{he}-\mathbf{g}_{he}\mathbf{g}_{e}\right]
		& \mathbf{1}-2\left[\mathbf{g}_h(\mathbf{1}-\mathbf{g}_h)-\mathbf{g}_{he}\mathbf{g}_{eh}\right] 
	\end{matrix}
	\right|
\end{equation}
\end{widetext}
The final step involves using the fact that, for a $T$-periodic
sources, all these operators are block diagonal with respect to the 
decomposition of the single-particle state into subspaces indexed by the
quasi-energy $\nu\in[0,2\pi f[$. Let us consider a block diagonal 
1-particle
operator $\mathbf{M}$ block-diagonal with respect to the quasi-energy $\nu$. Its determinant 
can be approximated by discretizing the first Floquet-Brillouin zone
$[0,2\pi f[$:
\begin{equation}
		\ln\left[\prod_{n=0}^{N-1}
		\mathrm{Det}\left(\mathbf{M}_{\frac{2\pi
		nf}{N}}\right)\right] 
		\simeq N\int_0^{2\pi f}
		\mathrm{Tr}(\ln(\mathbf{M}_\nu))\,\frac{\md\nu}{2\pi f}
\end{equation}
in which $\mathbf{M}_\nu$ precisely denotes the restriction of
$\mathbf{M}$ to the subspace of single-particle state with quasi-energy
$\nu$. Such a discretization corresponds to juxtaposing $N$ periods of
duration $T$ and considering states with periodic boundary conditions on
this interval, thereby introducing a formal IR
regularizaton.

This finally gives us the following compact expression for the many-body
purity indicator:
\begin{widetext}
\begin{equation}\label{purity}
	\mathbb{P}_\rho =  
    \exp\left[
        \int_0^{2\pi f} \ln\left( 1-2A(\nu)(1-A(\nu))-2B(\nu)(1-B(\nu))
        \right) \frac{d\nu}{2\pi f} \right]
\end{equation}
\end{widetext}
in which 
\begin{align}
A(\nu) &= g^{(ee)}(\nu)(1-g^{(hh)}(\nu))-|g^{(eh)}(\nu)|^2 \\ B(\nu) &=
g^{(hh)}(\nu)(1-g^{(ee)}(\nu))-|g^{(eh)}(\nu)|^2
\end{align} 
are computed in terms of the eigenvalues $g^{(ee)}(\nu)$ and
$g^{(hh)}(\nu)$ obtained from our diagonalization algorithm (see
\cref{sec/Floquet}) and of the
corresponding electron/hole coherences $g^{(eh)}(\nu)$. Discussion of
the conditions for
unit purity can be found in Ref. \cite{Bisognin-2019-2}: it
corresponds to a pure many-body state of the form 
\begin{equation}
|\Psi\rangle= \prod_{0\leq \nu<2\pi f} \left(u(\nu)+
    v(\nu)\psi^\dagger[\varphi_\nu^{(e)}]\psi[\varphi_\nu^{(h)}]\right)|F\rangle
\end{equation}
where $|u(\nu)|^2+|v(\nu)|^2=1$ for all $0\leq \nu<2\pi f$.

\section{Mesoscopic capacitor: case of a square drive}
\label{appendix/capacitor/square-drive}

In the case of a square drive used to demonstrate
single-electron emission by the mesoscopic capacitor 
\cite{Feve-2007-1}, the $T$-periodic voltage drive
is defined by $V_g(t)=-V/2$ for $-T/2\leq t<0$ and
$V_g(t)=V/2$ for $0<t<T/2$. 

\subsection{Electron/hole entanglement}
\label{appendix/capacitor/square-drive/entropy}

\begin{figure}
\centering
\includegraphics{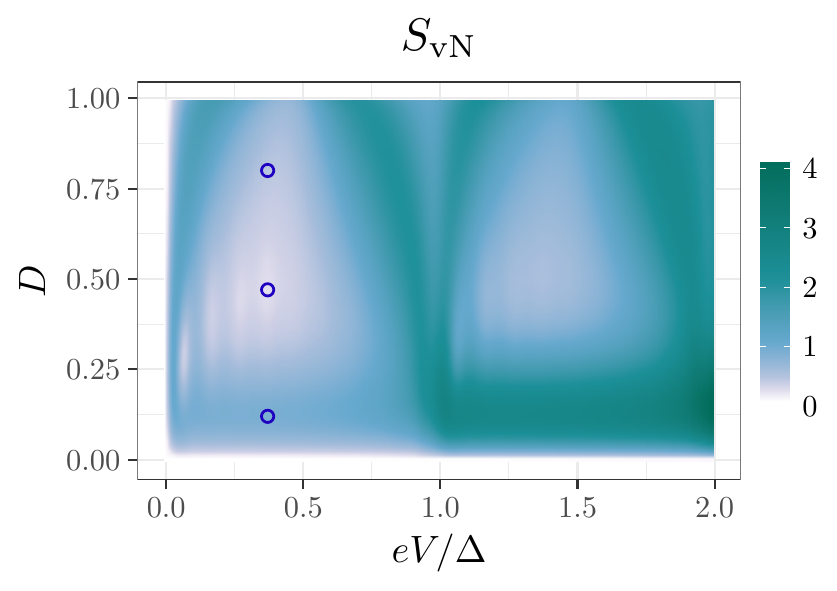}
\caption{\label{fig:capa-meso:examples:square:entropy-map} Density 
plot of the electron/hole
entanglement entropy at zero temperature for the mesoscopic capacitor
operated with a square drive of frequency $f$ such that $\Delta/hf=20$ 
as a function of $eV/\Delta$ and $D$.
}
\end{figure}

\Cref{fig:capa-meso:examples:square:entropy-map} presents a density plot of
the entropy defined by \cref{eq:eh-entanglement:entropy} as a function of $D$ and
$eV/\Delta$ at fixed $\Delta/hf=20$. There are shallow zones with minima in
each square $eV/\Delta\in ]n,n+1]$ ($n\in\mathbb{N}$) and $0<D\leq 1$.
In the single-electron sector, a global minimum can be found at
$eV_{\text{opt}}/\Delta \approx \num{0.37}$ and $D_{\text{opt}}
\approx \num{0.47}$ 
and the corresponding entropy is
$\SI{0.20}{\bit}$. As we shall see, this is the regime where the
mesoscopic capacitor behaves almost ideally, emitting exactly one 
electron and one hole excitation per period.

There is also a minimum in the second square where $1<eV/\Delta\leq 2$ but
the zone is further from zero. In this zone, three electrons are emitted
during the first half period and three holes during the other one, due to the
fact that at zero voltage, there is a level at the Fermi energy. It is
not surprising that in this zone the deviation from the ideal regime is
greater than in the previous case, since we expect a generation of more
electron/hole pairs.

A surprising feature are the substructures that appear within each
shallow zone. At the time of this writing, we do not yet understand this
fact. Further numerical exploration will be necessary, especially to
see if the ratio $\Delta/hf$ plays a role in these substructures.

\begin{figure}
\centering
	\includegraphics{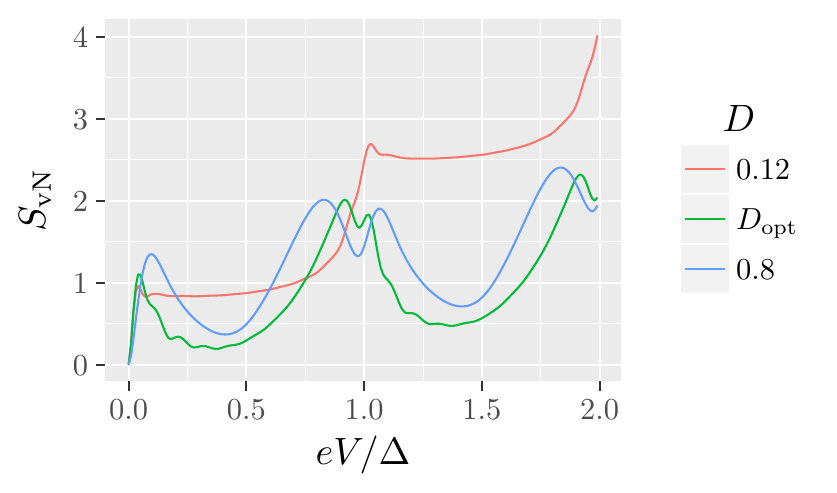}
	\caption{\label{fig:capa-meso:examples:square:entropy-cuts} Cuts of
	the entropy $S_{\text{vN}}$ for a square voltage drive depicted on
	\cref{fig:capa-meso:examples:square:entropy-map} 
	as functions of $eV/\Delta$ for $D=0.12$, $D=D_{\text{opt}}$ and
	$D=0.8$.}
\end{figure}

\begin{figure}
	\centering
	\includegraphics[width=8.6cm]{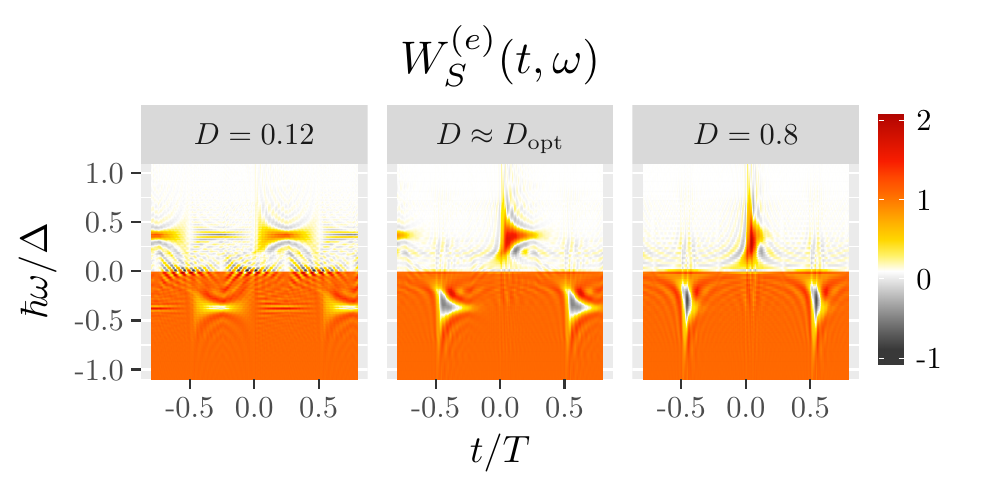}
	\caption{\label{fig:capa-meso:examples:square:Wigner} Density plot of
	the full Wigner distribution function $W_S^{(e)}(t,\omega)$ for the
	square-voltage driven mesoscopic capacitor as a
	function of $t/T$ and $\hbar\omega/\Delta$ for the three selected
	points appearing on \cref{fig:capa-meso:examples:square:entropy-map}.}
\end{figure}

In order to understand more precisely the electron/hole entanglement
properties described by this plot, we have chosen specific points
for which we will push the analysis further. The corresponding
electronic Wigner distribution functions are plotted on
\cref{fig:capa-meso:examples:square:Wigner}.

\subsection{The Floquet-Bloch spectrum}
\label{appendix/capacitor/square-drive/spectrum}

Let us review the Floquet-Bloch spectrum for the three points that are
marked in \cref{fig:capa-meso:examples:square:entropy-map}. This figure
presents the corresponding bands as functions of the adimensionned
quasi-energy $\omega/2\pi f$ and 
orders them according to their averages, the $a=0$ band being one
with the highest average.

The middle panel
corresponds to the absolute minimum of the entropy and therefore to the
best operating point as a single-electron source. Only one band gives
eigenvalues close to one and it is flat. All the other bands are really
close to zero as expected.

Opening the dot ($D=0.8$, right panel) leads to flat bands as expected since at
$D=1$ it is really what is expected but we note that the eigenvalues are
almost unity and that the $a=1$ band has value $0.02$, thus showing that we
are departing from the ideal single-electron regime. 

\begin{figure}
	\centering
	\includegraphics{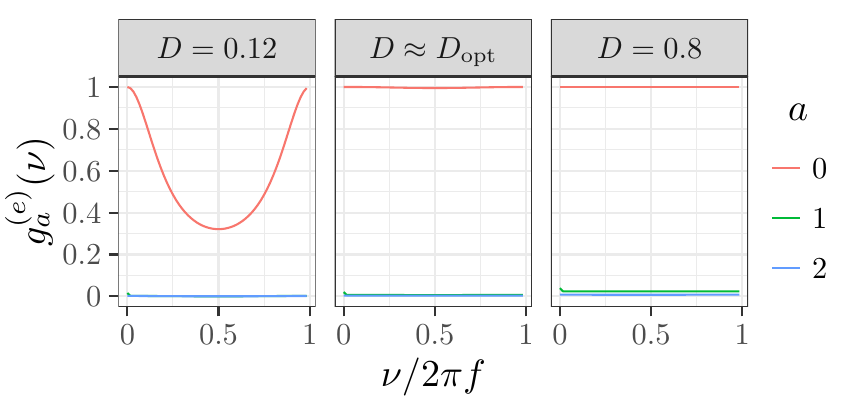}
	\caption{\label{fig:capa-meso:examples:square:spectrum}
	Floquet-Bloch spectrum for the three selected
    points appearing on
	\cref{fig:capa-meso:examples:square:entropy-map}, in the case of a
	square-voltage mesoscopic capacitor.
	Only the first three bands are represented, all the other ones
	being even closer to zero. 
	}
\end{figure}

Closing the dot ($D=0.12$, left panel)
mostly changes the shape of the $a=0$
band which shows some curvature. 
Its average is equal to $0.57$ which
shows that strong electron/hole coherences are expected. 
\subsection{Electronic atoms of signals and coherences}
\label{appendix/capacitor/square-drive/wannier}

\begin{figure}
	\centering
	\includegraphics[width=8.6cm]{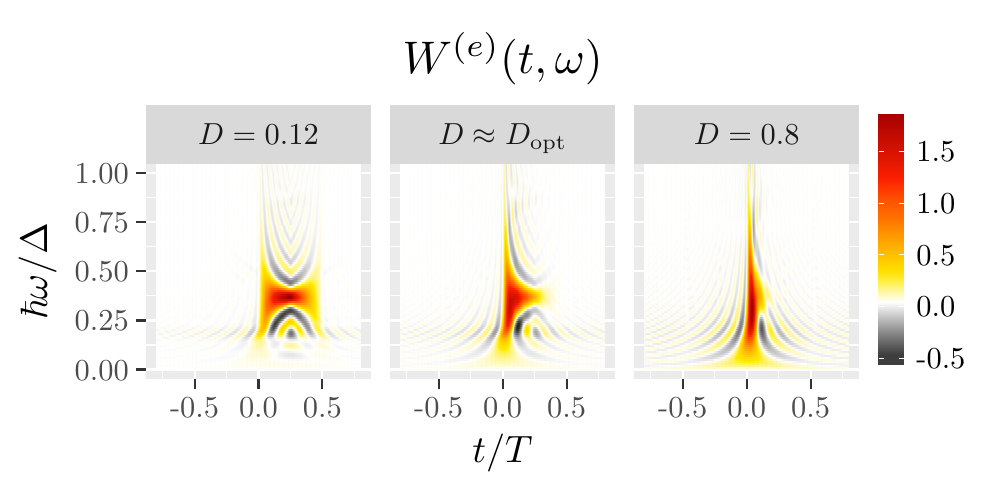}
	\caption{
\label{fig:capa-meso:examples:square:wannier}
Wigner distribution functions for the Floquet-Wannier electronic atoms of
signal corresponding to the $a=0$ Floquet-Bloch band represented as a
function of $t/T$ and $\hbar\omega/\Delta$ for the three operating
points of \cref{fig:capa-meso:examples:square:entropy-map} in the case
	of a square voltage drive.
	}
\end{figure}

In order to get a clearer view of the electronic state emitted by the
source, let us now extract the corresponding electronic atoms of signal.
\Cref{fig:capa-meso:examples:square:wannier} presents the electronic
atoms of signal associated with the $a=0$ Floquet-Bloch band for the
three operating points considered before.

As expected, the duration of each wavepacket increases with decreasing
$D$ reflecting the fact that the escaping time from the dot is longer at
low QPC transparency. At the optimal value $D_{\text{opt}}$, 
we expect the source to emit a wavepacket of the form
\begin{equation}
\label{eq:lorentzian}
\widetilde{\varphi}_{e}(\omega)= 
\frac{\mathcal{N}_{\mathrm{e}}\,\heaviside(\omega)}{\omega-\omega_{\mathrm{e}}-i\gamma_{\mathrm{e}}/2},
\end{equation}
where $\mathcal{N}_{\mathrm{e}}$ ensures normalization and
$\gamma_{\mathrm{e}}$ denotes the electron
escape rate from the quantum dot
which is given by
$\gamma_{\mathrm{e}}=D\Delta/h(1-D/2)$ \cite{Mahe-2008-1,Nigg-2008-1}.
We note that for $D=0.12$, the electronic
wavepacket remains limited to the firsf half 
period $0\lesssim t\lesssim
T/2$. At very low $D$, we expect this wavepacket to be the projection on
the space of single-particle states with positive energy of the dual of
the Martin-Landauer wavepacket, that is of an electronic wavefunction
constant on a time interval. 

\begin{figure}
	\centering
	\includegraphics{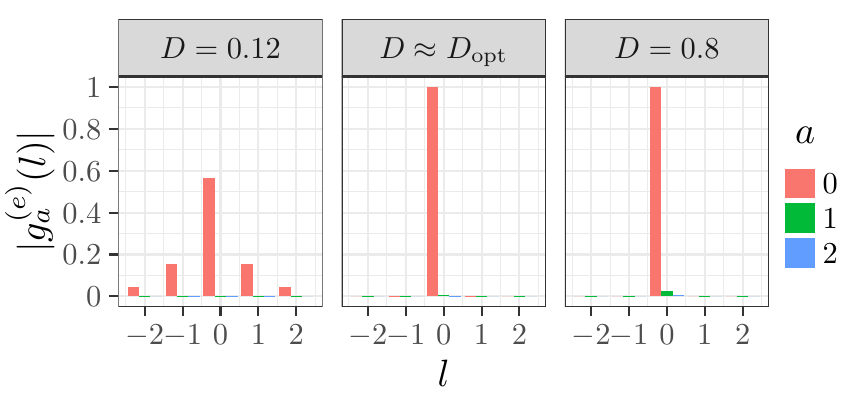}
	\caption{
\label{fig:capa-meso:examples:square:e-coherences}
Temporal coherences $p_n(\Delta l)$ between the electronic atoms
of signal of the $a=0$, $1$ and $2$ Floquet-Bloch bands given by
\cref{eq:Wannier:pa-ee} as a function of $\Delta l$ for the three
operating points of \cref{fig:capa-meso:examples:square:entropy-map} in
	the case of a square drive.
	}
\end{figure}

Since the bands are flat for $D=D_{\text{opt}}$ and $D=0.8$, no
inter-period coherence is expected as can be seen from the middle and
right panels of \cref{fig:capa-meso:examples:square:e-coherences}.
However, when
closing the dot
($D\approx 0.12$, left panel), inter-period coherences for the electronic
excitations start to unfold, an expected consequence of the
delocalization of the emitted electronic excitations over more than a
half period. This shows that the electronic coherence time is given by
the electronic escape time which, in this
case, exceeds the duration of an electronic atom of signal.

\section{Wavefunctions within a Leviton train}
\label{appendix/levitonoids}

We define a Levitonoid as a normalized wavefunction $\psi(t)$ such that
their time translations by multiples of $T$ are mutually orthogonal and
$\sum_{l\in \mathbb{Z}}
\psi(t-lT)\psi^*(t'-lT)$ is the excess electronic first-order coherence
generated by a $T$-periodic train of Lorentzian pulses. As we will see, this
wavefunction is not unique but, in the case of a $T$-periodic Leviton
train, an analytical expressions for
the minimally-spread Levitonoids can be obtained.

\subsection{The Moskalets atoms of signal}
\label{appendix/levitonoids/Moskalets}

In a recent work \cite{Moskalets-2015-1}, Moskalets has identified one
possible Levitonoid as:
\begin{equation}
	\psi(t)
	=
	\sqrt{\frac{\tau_0}{\pi}}
	\frac{1}{t - \mi \tau_0}
	\prod_{n=1}^{\infty}\frac{t+nT+\mi \tau_0}{t+nT-\mi\tau_0}
\end{equation}
where $T = 1/f$ is the period and $\tau_0$ is the typical time width of the
excitation. This wavefunction has a spatial extention given by $\tau_0$.

To discuss its energy content, les us use
the identity
\begin{equation}
	\Gamma(z) = \frac{1}{z}
		\prod_{n=1}^{\infty} \frac{
			\left(1 + \frac{1}{n}\right)^z
		}{
			1+\frac{z}{n}
		}
\end{equation}
to rewrite the infinite product as a ratio of $\Gamma$ functions, up to a
global phase
\begin{subequations}
\begin{align}
	\frac{\Gamma((t-\mi\tau_0)/T)}{\Gamma((t+\mi\tau_0)/T)}
	=
	\frac{t+\mi\tau_0}{t-\mi\tau_0}
	&\prod_{n=1}^{\infty}
	\left( 1 + \frac{1}{n} \right)^{-2 \mi \tau_0} \\
	&\prod_{n=1}^{\infty}
	\frac{t + n T + \mi \tau_0}{t + n T - \mi \tau_0}
\end{align}
\end{subequations}
Then, up to a global phase factor, we can rewrite:
\begin{equation}
	\psi(t)
	=
	\sqrt{\frac{\tau_0}{\pi}} \frac{1}{t+\mi\tau_0}
	\frac{\Gamma((t - \mi\tau_0)/T)}{\Gamma((t + \mi \tau_0)/T)}\,.
\end{equation}
Fourier transforming this wavepacket gives, up
to global phase factor:
\begin{align}
	\psi(\omega) = \frac{1}{\sqrt{\mathcal{N}}}
	\heaviside(\omega)
	\left(
		2 \sin \frac{\nu T}{2}
	\right)^{2 \mi \tau_0 / T}
	\me^{-\omega_{\text{int}} \tau_0}
\end{align}
where $\omega = \omega_{\text{int}} + \nu$, with $\nu \in [0, 2 \pi f[$ and thus
$\omega_{\text{int}} = 2 \pi f \lfloor \omega/2\pi f \rfloor$. Here, $\mathcal{N} =
f/v_F(1-\me^{-4\pi \tau_0/T})$ is a normalization factor and $\heaviside$
the Heaviside step function.
We can then rewrite the wavefunction as a real part and a periodic
phase:
\begin{equation}
	\psi(\omega) = \frac{1}{\sqrt{\mathcal{N}}}
	\heaviside(\omega)
	\me^{\mi \theta(\omega)}
	\me^{-\omega_{\text{int}} \tau_0}
	\label{eq:wavefunction:misha}
\end{equation}
with the phase satisfying the condition $\theta(\omega + 2 \pi f) =
\theta(\omega)$. This expression shows that the electronic distribution
function of this Levitonoid is the staircase approximation of an exponential decay,
with step widths given by $2\pi f$ as expected from $T$-periodicity.
Note that the electronic distribution function
does not depend on the phase $\theta(\omega)$.

\subsection{The minimally-spread atoms of signal}
\label{appendix/levitonoids/minimally-spread}

The Moskalets Levitonoids having a spreading $\tau_0$, they are naturally
expected to be among the minimally-spread atoms of signals when
$f\tau_0\ll 1$, that is when the Leviton spacing is large compared to
their duration. But in the opposite limit $f\tau_0\gtrsim 1$, this is
certainly not the case. Let us search for other Levitonoids that could
be spreat over the period $T$ and clarify
the relation between our algorithm and Moskalets work
\cite{Moskalets-2015-1}.

If a quantum electrical current has a time-reversal symmetry, which is
the case for a Leviton train, then there must be a set of
Wannier wavefunctions that possess this symmetry. Consequently,
there is a set of real valued Wannier functions in the frequency domain:
$\varphi_{\text{Lev}}(\omega) \in
\mathbb{R}$. Assuming that $\varphi_{\text{Lev}}(\omega) \ge 0$,
the time-spreading minimization problem becomes trivial and we find that, up to
time-translation by $T$, the minimal wavefunctions are the ones that
possess the time-reversal symmetry.

In the case of Levitonoids, 
\cref{eq:wavefunction:misha} shows that 
the wavefunctions can be
written as the product of a real part and a phase part, the phase part
being periodic in time. Thus the minimization of
\cref{eq:Wannier:spreading} is realized when the minimal wavefunction
has a constant phase. Setting this global phase to zero, we
have the following wavefunction:
\begin{equation}
	\label{eq:app:levitonoid:minimal}
	\varphi_{\text{Lev}}(\omega) = \frac{1}{\sqrt{\mathcal{N}}}
	\heaviside(\omega)
	\me^{-\omega_{\text{int}} \tau_0}
\end{equation}
which is time-reversal invariant. In this case, the current
of one pulse is different from the current of one Leviton of duration
$\tau_0$. The
time-domain expression for this wavepacket is:
\begin{equation}
	\varphi_{\text{Lev}}(t) = \frac{\mi}{\sqrt{\mathcal{N'}}}
	\frac{1}{t}
	\frac{1 - \me^{-2 \mi \pi f t}}{1-\me^{-2\pi f(\tau_0 + \mi t)}}
\end{equation}
The corresponding average current $i_{\text{Lev}}(t)=-v_F|\varphi(t)|^2$ (in
units of $-e$)
is then
\begin{equation}
	i_{\text{Lev}}(t) \propto \frac{\mathrm{sinc}^2(\pi f t)
		}{
			1 - \cos(2 \pi f t)/\cosh(2 \pi f \tau_0)
		}\,.
\end{equation}
The overlap between this wavepacket and a unique Leviton $\varphi_1$ is given by
\begin{equation}
	\left|
		\Braket{\varphi_{\text{Lev}} | \varphi_{1}}
	\right|^2
	=
	\frac{1}{\pi f \tau_0}
	\frac{1 - \me^{-2 \pi f \tau_0}}{1 + \me^{-2 \pi f \tau_0}}.
\end{equation}
The behavior when $f\tau_0\ll 1$ is approached by
\begin{equation}
	\left|
		\Braket{\varphi_{\text{Lev}} | \varphi_{1}}
	\right|^2
	\simeq
	1 - \frac{(\pi f \tau_0)^2}{6}
\end{equation}
making the Leviton approximation a fairly good approximation in this
case. When $f \tau_0 \gg 1$, we have a rather slow decay:
\begin{equation}
	\left|
		\Braket{\varphi_{\text{Lev}} | \varphi_{1}}
	\right|^2
	\simeq
	\frac{1}{\pi f \tau_0}\,.
\end{equation}

\subsection{Obtention from Martin-Landauer's wavepackets}
\label{appendix/levitonoids/fromML}

As noticed in \cite[Appendix 5]{Dubois-2013-2}, at zero temperature, the
effect of a $T$-eriodic classical drive $V_d$ is to reshuffle the Martin-Landauer wavepackets 
associated with a given period through a unitary transformation
of the following form:
\begin{equation}
\label{eq:app:levitonoid:voltage-on-ML}
	\ket{\text{ML}_{n,l}}\mapsto
	\sum_{k\in\mathbb{Z}}p_k[V_d]\ket{\text{ML}_{n+k,l}}
\end{equation}
in which $\ket{\text{ML}_{n,l}}$ denotes the time translated to period $l$
of the single-particle state defined by \cref{eq/Floquet/examples/Martin-Landauer} and
$p_k[V_d]$ is the photoassisted transition amplitude associated with
this drive. In the case
of a Leviton train, the photoassisted amplitudes $p_k$ are known and
given by~\cite{Dubois-2013-2}
\begin{subequations}
\label{eq:app:levitonoid:amplitudes}
\begin{align}
p_{k<-1} &=0\\
p_{-1}&=\me^{-2\pi f\tau_0}\\
p_{k\geq 0}&=(1-\me^{-4\pi f\tau_0})\,\me^{-2\pi kf\tau_0}
\end{align}
\end{subequations}
Consequently, introducing $\beta=\me^{-2\pi f\tau_0}$ for compacity, for
all states with the Fermi sea ($n\leq -1$):
\begin{subequations}
\begin{align}
\label{eq:app:levitonoid:action-on-ML}
	\ket{\text{ML}_{-n,l}} &\mapsto 
	-\beta\ket{\text{ML}_{-(n+1),l}}+(1-\beta^2)\sum_{k=1}^n\beta^{n-k}\ket{\text{ML}_{-k,l}}\\
	&+(1-\beta^2)\beta^n\sum_{q=0}^{+\infty}\beta^q\ket{\text{ML}_{q,l}}
\end{align}
\end{subequations}
in which we have separated what remains into the Fermi sea (fist line) from what
emerges from the Fermi sea (second line). 
This shows that, as expected, the projection of the single-particle
scattering from the space generated by the Martin-Landauer wavepackets of negative
energy and given period onto the space of state generated by the ones of
positive energy is of rank $1$. 
This is sufficient to obtain an
electronic atom of signal for the Leviton
associated with the period $l$ is
\begin{equation}
\label{eq:app:levitonoid:Lev1}
	\ket{\text{Lev}_l}=\sqrt{1-\beta^2}\,\sum_{p=0}^{+\infty}\beta^p\ket{\text{ML}_{p,l}}.
\end{equation}
Since the Martin-Landauer wavepacket $\text{ML}_{n,l=0}$ has a non-zero constant
wavefunction over $2\pi nf\leq \omega <2\pi (n+1)f$ and zero for
$\omega\geq 2\pi (n+1)f$ or $\omega <2\pi nf$, the wavefunction of 
$\ket{\text{Lev}_1}$ in the frequency domain is exactly given by
\cref{eq:app:levitonoid:minimal} therefore showing that it is indeed the
minimally-spread Levitonoid!

\section{Random emission}
\label{appendix/random}

\subsection{Expression in the frequency domain}
\label{appendix/random/frequency-domain}

Let us start form the excess coherence of a random train of $-e$
charge Lorentzian pluse
\begin{multline}
	\Delta \mathcal{G}^{(e)}
	(t+\tau/2, t-\tau/2)
	= \\
	\mathcal{G}^{(e)}_F(\tau)
	\frac{
		\cos (2 \pi \theta_p(f\tau)) - \cos \left(2 \pi f (
			\tau/2 - \mi \tau_0
		)\right)
	}{
		\cos \left(2 \pi f (\tau / 2 - \mi \tau_0)\right)
		-\cos(2\pi ft)
	}\,.
\end{multline}
This expression is periodic in $t$ with a period $T = 1/f$. 
The $n$-th harmonics $A_n(\tau)$ of its Fourier series expansion
is then given by
\begin{align}
	A_n(\tau)
	&=
	-\mi \mathcal{G}_F(\tau)
	\frac{\cos (2 \pi \theta_p(f \tau)) -
		\cos(2 \pi f (\tau/2 - \mi \tau_0))
	}{
		\sin(2 \pi f (\tau/2 - \mi \tau_0))
	}
	\nonumber \\ &\times
	\me^{-2 \pi |n| f \tau_0} \me^{- \mi \pi |n| f \tau}\,.
\end{align}
Performing the Fourier transform along the variable
$\tau$ leads to the energy representation of the first-order
coherence. Let us notice that the Fourier transform is found to be zero
when $\omega < \pi |n|$, meaning that the first-order coherence is
non-zero only in the electronic quadrant, as expected. By looking at the
electronic quadrant, we derive
\begin{equation}
	\mathcal{G}^{(e)}_{++, n}(\omega)
	=\heaviside(\omega-\pi f|n|)\,
	\me^{-2 \omega \tau_0} \mathcal{F}_p(\omega + \pi n f)
	\label{eq:randomlev:coherence:1}
\end{equation}
where $\mathcal{F}_p(\omega)$ is a $2\pi f$-periodic real-valued function defined by
the sum
\begin{equation}
	\mathcal{F}_p(\omega)
	=
	\frac{\mi}{\pi}
	\sum_{k \in \mathbb{Z}}
	\frac{\cos(\chi_p(k)) - 1}{k + 2 \mi f \tau_0}
\,	\me^{\mi \omega k T}
\end{equation}
where $\chi_p(x)=2\pi \theta_p(2\mi f \tau_0+k)-\pi k$.
In order to evaluate numerically this expression, we will decompose
$\mathcal{F}_p =
\mathcal{F}_p^{(\text{sing})} + \mathcal{F}_p^{\text{(reg})}$, where
$\mathcal{F}_p^{(\text{sing})}$ contains
singularities due to the slow decay ($\sim 1/k$) at infinity. We have
\begin{align}
	\mathcal{F}_p^{(\text{sing})}(\omega)
	&=
	\frac{\mi}{\pi}
	\sum_{k \in \mathbb{Z}}
	\frac{\cosh(4\pi p \tau_0) - 1}{k + 2 \mi f \tau_0}
	\,\me^{\mi \omega k T}
	\\
	\mathcal{F}_p^{(\text{reg})}(\omega)
	&=
	\frac{\mi}{\pi}
	\sum_{k \in \mathbb{Z}}
	\frac{\cos(\chi_p(k)) - \cosh(4\pi p \tau_0)}{k + 2 \mi f \tau_0}
	\,\me^{\mi\,\omega k T}\,.
\end{align}
The singular part then contributes to the single-elecron coherence by
\begin{equation}
	\mathcal{G}^{(e,\text{sing})}_{+,n}(\omega)
	=4\,\heaviside(\omega-\pi f|n|)\,
	\frac{\sinh^2(2 \pi f p \tau_0)}{\me^{4 \pi f \tau_0} - 1}
	\,\me^{-2 \omega_n \tau_0}
\end{equation}
where $\omega_n = 2 \pi f \lfloor \omega/ 2 \pi f\rfloor$ if $n$ is even and
$\omega_n = 2 \pi f \lfloor \omega/2\pi f - 1/2\rfloor + \pi f$ if $n$ is odd.
The regular part can then be evaluated numerically by direct summation.
To be more precise, we can bound the truncation error on the regular
part, when using $K$ terms on the sum. Using an asymptotic expansion, we
find that the error scales as
\begin{equation}
	\epsilon = 16 p (1-p) (f\tau_0)^2
	\frac{\sinh(4 \pi p f \tau_0)}{K}\,.
\end{equation}
On top of the polynomial scaling, we notice an exponential scaling in
$p$ and in $f \tau_0$. Actually, as we will see later, we can use a
symmetry in $p \leftrightarrow 1-p$, to ensure that all the computations
are done at $p \le 1/2$. With that, it is reasonable to compute the sum
for $\tau \lesssim 1$.

When $p = 1$, the regular contribution cancels out, and we find the
usual expression for a Leviton train. When $p \to 0$, it is the singular
contribution that disappears first (scaling as $p^2$) leaving only the
regular contribution (scaling as $p$). When $p\ll \min(1,f\tau_0)$,
$\mathcal{F}_p(\omega) = 4 \pi fp 
\tau_0$ is constant. Each harmonics of the first-order coherence is thus
an exponential decay
\begin{equation}
	\mathcal{G}^{(e)}_{++,n}(\omega)
	=
	4 \pi f p\,\heaviside(\omega-\pi |n|f)\, \tau_0\, \me^{-2\omega\tau_0}\,.
\end{equation}

\subsection{Electronic atoms of signal}
\label{sec:randomlev:atomsofsignals}

The electronic atoms of signal describing the random train's first
order coherence are obtained by following the method
presented in \cref{sec/Floquet}. The projection of the single-electron coherence
on the electronic quadrant has non-zero matrix elements only for
$\omega_\pm=\nu+2\pi n_\pm f$ in which $0\leq \nu<2\pi f$ and 
$n_\pm$ are positive integers. This corresponds to 
$\omega=(\omega_++\omega_-)/2=\nu+\pi f(n_++n_-)$ and
$\Omega=\omega_+-\omega_-=2\pi (n_+-n_-)f$. Then, 
using
\cref{eq:randomlev:coherence:1} and the periodicity of
$\mathcal{F}_p(\Omega)$ in $\Omega\rightarrow \Omega+2\pi f$, the excess
first order
electronic coherence can be rewritten as
\begin{subequations}
	\begin{align}
		\Delta_0\mathbf{G}^{(e)}&=\int_0^{2\pi f}\me^{-2\nu \tau_0}\mathcal{F}_p(\nu)
		\mathbf{M}(\nu)
    \frac{\md \nu}{2\pi f}
	\label{eq:randomlev:coherence:2}\\
\mathbf{M}(\nu)
		&=\sum_{n_\pm\in\mathbb{N}}
	\me^{-2 \pi (n_+ + n_-) f \tau_0}\ket{\nu+2\pi fn_+}\bra{\nu+2\pi fn_-}
	\end{align}
\end{subequations}
We thus have to diagonalize the operator $\mathbf{M}(\nu)$ for each 
$0\leq \nu<2\pi f$. We have already see how to diagonalize it:
for each $\nu$, this operator has rank one and its eigenvactor is
the one obtained in
\cref{eq:app:levitonoid:Lev1}. This immediately shows
that we have the same Floquet-Bloch states (only one band here) and therefore the same
electronic atoms of signals than for the periodic Leviton train ($p=1$,
\cref{appendix/levitonoids/minimally-spread,appendix/levitonoids/fromML}).

\begin{figure}
	\includegraphics{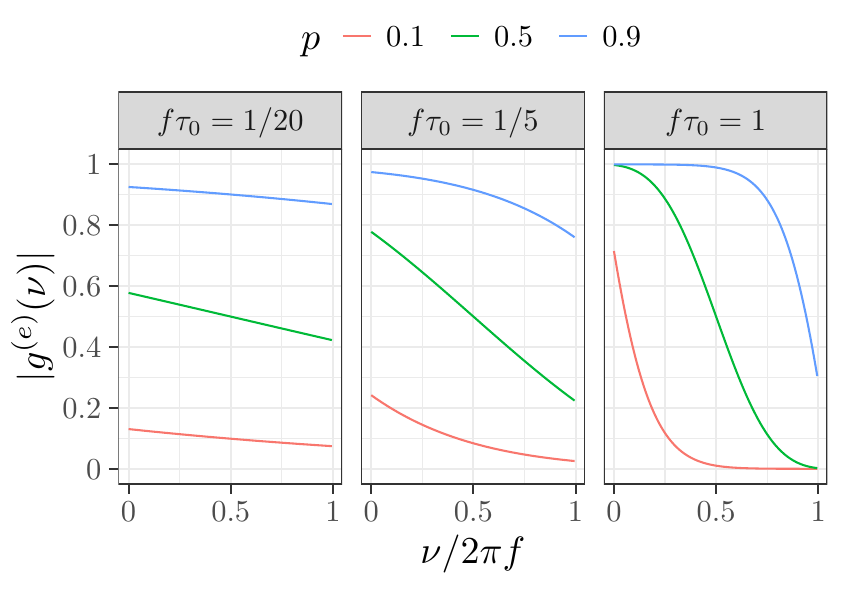}
	\caption{\label{figure/random/spectrum}
Floquet-Bloch 
spectrum for the random train of Lorentzian pulses (it has only one band) for width $f\tau_0=1/10$, $1/5$ and
$1$ for three different values of $p$ ($1/10$, $1/2$ and $9/10$). Bands
are more and more flat when decreasing $f\tau_0$ as expected since, in
this limit, one recovers the random emission of
a single electronic atom of signal per period with probability $p$. When
$p$ goes to unity, we see the band getting closer to a flat band at
value one, which corresponds to the $T$-periodic train of Levitons.
	}
\end{figure}

Only the eigenvalue is
modified by randomness:
\begin{equation}
\label{eq/random/spectrum}
	g^{(e)}(\nu)=\left(1-\me^{-4\pi
	f \tau_0}\right)^{-1}\,\me^{-2\nu\tau_0}\mathcal{F}_p(\nu)\,.
\end{equation}
When $p=1$, the periodic Leviton train with its flat
band with $g^{(e)}(\nu=1)$ is recovered but for $p<1$, the band is not flat anymore.
This means that for $p<1$, the emission probability of emission of the
Levitonoid is lower than unity but this also leads
to inter-period coherences between the Levitonoids. A numerical
evaluation of the r/h/s of \cref{eq/random/spectrum} is shown on
\cref{figure/random/spectrum} which confirms these features. 

Finally, using this last expression, we can rewrite the zeroth-harmonic
of the first order coherence as
\begin{equation}
	\mathcal{G}^{(e)}_0(\omega)
	=
	(1-\me^{-4 \pi f \tau_0})\,
	\me^{-2\pi f\lfloor\frac{\omega}{2\pi f}\rfloor \tau_0}
	g^{(e)}(\omega)\,.
\end{equation}
By performing the inverse Fourier transform on this expression, we can
thus express the coherences between the time-shifted Levitonoids as
$g^{(e)}(l) = T A_0(-l T)$. Thus, we find
\begin{align}
	g^{(e)}(l = 0)
	&= p \\
	g^{(e)}(l \neq 0)
	&= \frac{\mi}{2\pi l}
		\frac{\cosh(2 \pi f \tau_0) - (-1)^l \cos(2 \pi \theta_p(-l))}
			{\sinh 2 \pi f \tau_0}
\end{align}
This expression satisfies $g^{(e)}_{p}(l \neq 0) =
-g^{(e)}_{1-p}(l)^*$ which leads to the symmetry property
\begin{equation}
	g^{(e)}_{p}(\omega)
	=
	1 - g^{(e)}_{1-p}(-\omega)\,.
\end{equation}
We use this symmetry to perform all the numerical computations at $p \le
1/2$.

\subsection{Resumming inter-period coherences}
\label{appendix/random/glattlions}

In \cref{sec:randomlev:atomsofsignals}, we have seen that it was
possible to decompose the signal on Levitonoids. In a sense, Levitonoids
are proper atoms of signals, because they do not overlap when they are
time-shifted by an integer number of periods. However, when $p < 1$,
coherences appear between time-shifted Levitonoids.

Remarkably, is it possible to express the excess single-electron
coherence $\mathbf{G}^{(e)}_{R_p}$ in terms of wavefunctions associated with each
period (and obtained by applying the translation operator
$\mathbf{T}_T)$, each of them emitted with
probability $p$, without any inter-period coherence:
\begin{equation}
	\Delta_0\mathbf{G}^{(e)}
	=
	p
	\sum_{l \in \mathbb{Z}}
	\ket{\text{Gla}_{p,l}} \bra{\text{Gla}_{p,l}}
\end{equation}
in which $\ket{\text{Gla}_{p,l}}=\mathbf{T}_T^l\ket{\text{Gla}_p}$ is
obtained by a time translation by $l$ periods from 
\begin{equation}
	\label{eq/Glattlions/products}
	\ket{\text{Gla}_p}
	=\int_0^{+\infty} %
	\sqrt{\frac{1-\me^{-4 \pi f \tau_0}}{p f}}
	\me^{-\omega_{\text{int}}\tau_0}
	\sqrt{g^{(e)}(\omega)}\,\ket{\omega}\,\md\omega\,
\end{equation}
in which $\omega_{\text{int}}=2\pi f\lfloor \omega/2\pi f\rfloor$.
We will
call such a single-particle state a $p$-Glattlion in reference to
\cite{Glattli-2018-1}. However, 
the overlap of adjacent $p$-Glattlions is non-vanishing and is related
to the electronic coherences between two different periods in the
minimally-spread Levitonoids:
\begin{equation}
	\label{eq/Glattlions/scalar-products}
	p\,
	\braket{\text{Gla}_{p, 0} | \text{Gla}_{p,l}}
	=
	\int_0^{2\pi f} g^{(e)}(\nu)\, \me^{\mi \nu l T} \frac{\md \nu}{2 \pi f}
	= g^{(e)}(l)
\end{equation}
In other words, the coherences between the Levitonoids are given by the
overlap between time-shifted $p$-Glattlions which, therefore, cannot be
considered as electronic atoms of signals. Despite this non vanishing
overlap, it is
quite remarkable that the excess single-electron coherence for
the random train of Levitons can be rewritten in such a simple
form.

Finally, it's worth noting that the 1-Glattlion is nothing else than the
minimal Levitonoid. On the other hand, when $p \to 0$, a $p$-Glattlion
becomes close to
isolated Levitons. This is related to the fact that when $p$ is
small, pulses are emitted as if they were isolated: the probability for
a single pulse to be separated by a distance at least $k$ from the
previous and the next one is $(1-p)^{2k}$ which goes to unity when
$p\rightarrow 0$.
As such, they do
not exhibit Pauli exclusion principle between pulses.

\bibliography{biblio/bigbib,biblio/livres}

\end{document}